\documentclass[traditabstract]{aa} 
\usepackage{epsfig,amsmath}
\usepackage[dvips]{color}
\usepackage{verbatim}
\usepackage{graphicx}
\usepackage{txfonts}
\usepackage{natbib}
\bibpunct{(}{)}{;}{a}{}{,}
\usepackage{hyperref}
\usepackage{rotating}
\usepackage{multicols,multirow}

\definecolor{Black}{named}{Black}
\definecolor{Red}{named}{Red}
\definecolor{Green}{named}{Green}
\definecolor{Blue}{named}{Blue}

\begin{document}

\title{The ESO UVES advanced data products quasar sample - II. Cosmological evolution of the neutral gas mass density}
\author{Tayyaba Zafar\inst{1}
\and C\'{e}line P\'{e}roux\inst{1}
\and Attila Popping\inst{2}
\and Bruno Milliard\inst{1}
\and Jean-Michel Deharveng\inst{1}
\and Stephan Frank\inst{1,3}}

\institute{Aix Marseille Universit\'e, CNRS, LAM (Laboratoire d'Astrophysique de Marseille) UMR 7326, 13388, Marseille, France.
\and International Centre for Radio Astronomy Research (ICRAR), The University of Western Australia, 35 Stirling Hwy, Crawley, WA 6009, Australia.
\and Department of Astronomy, Ohio State University, 140 West 18th Avenue, Columbus, OH 43210-1173, USA.}

\titlerunning{Cosmological Evolution of the Neutral Gas Mass Density}
\authorrunning{T. Zafar et al.}

\offprints{tayyaba.zafar@oamp.fr}

\date{Received  / Accepted }

\abstract
{Damped absorbers, seen in the spectra of background quasars, are unique probes to select {H}\,{\sc i}-rich galaxies. These galaxies allow one to estimate the neutral gas mass over cosmological scales. The neutral gas mass is a possible indicator of gas consumption as star formation proceeds. The damped Ly$\alpha$ absorbers (DLAs; $N_{{\rm H}\,{\sc \rm I}}\ge2\times10^{20}$ cm$^{-2}$) and sub-DLAs ($10^{19}\le N_{{\rm H}\,{\sc \rm I}}\le2\times10^{20}$ cm$^{-2}$) are believed to contain a large fraction of neutral gas mass in the Universe. In paper I of the series, we presented the results of a search for DLAs and sub-DLAs in the European Southern Observatory (ESO) Ultraviolet Visual Echelle Spectrograph (UVES) advanced data products dataset of 250 quasars. Here we use an unbiased subsample of sub-DLAs from this dataset to derive their statistical properties. We built a subset of 122 quasars ranging from $1.5<z_{\rm em}<5.0$, suitable for statistical analysis. The statistical sample was analyzed in conjunction with other sub-DLA samples from the literature. This resulted in a combined sample of 89 sub-DLAs over a redshift path of $\Delta z=193$. We derived the redshift evolution of the number density and the line density for sub-DLAs and compared then with the Lyman-limit systems (LLSs) and DLA measurements from the literature. The results indicate that these three classes of absorbers are evolving in the redshift interval $1.0<z<5.0$. Thanks to the ESO UVES advanced data products data, we were able to determine the column density distribution, $f_{{\rm H}\,{\sc \rm I}}(N,z)$, down to the sub-DLA limit. The flattening of $f_{{\rm H}\,{\sc \rm I}}(N,z)$ in the sub-DLA regime is present in the observations. The redshift evolution of $f_{{\rm H}\,{\sc \rm I}}(N,z)$ down to log $N_{{\rm H}\,{\sc \rm I}}=19.0$ cm$^{-2}$ is also presented, indicating that there are more sub-DLAs at high redshift than at low redshift. $f_{{\rm H}\,{\sc \rm I}}(N,z)$ was also used to determine the neutral gas mass density, $\Omega_{\rm g}$, at $1.5<z<5.0$. The complete sample shows that sub-DLAs contribute 8--20\% to the total $\Omega_{\rm g}$ from $1.5<z<5.0$. In agreement with previous studies, no evolution of $\Omega_{\rm g}$ was observed from low to high redshift (i.e., $1.5<z<5.0$), suggesting that star formation alone cannot explain this non-evolution and replenishment of gas and that recombination of ionized gas is needed.}

\maketitle{}

\section{Introduction\label{intro}}
Baryons account for only a small fraction of mass in the Universe \citep{fukugita04}. In neutral and molecular phase they are reservoirs of gas from which stars form over cosmological scales. A way of measuring the assembly of galaxies is to probe the rate at which they convert their gas into stars. Therefore, to understand this, it is important to determine the cosmological mass density of neutral gas ($\Omega_{\rm g}$; \citealt{ma94} \citealt{klypin95}. The observations of {H}\,{\sc i}-rich galaxies, selected by the imprint they leave on a background quasar, allow one to measure the neutral gas mass density, $\Omega_{\rm g}$. The quasar absorbers are divided into several classes according to the number of atoms along the line of sight: the damped Ly$\alpha$ absorbers (DLAs) with $N_{{\rm H}\,{\sc \rm I}}\ge2\times10^{20}$ cm$^{-2}$, the sub-DLAs with $10^{19} \le N_{{\rm H}\,{\sc \rm I}}\le2\times10^{20}$ cm$^{-2}$, the Lyman-limit systems (LLSs) with $1.6\times10^{17} \le N_{{\rm H}\,{\sc \rm I}} \le 10^{19}$ cm$^{-2}$ and the Ly$\alpha$ forest with $N_{{\rm H}\,{\sc \rm I}}\le1.6\times10^{17}$ cm$^{-2}$. 

Ly$\alpha$ absorbers provide a useful probe of the distribution of baryonic matter in the Universe over cosmological scales. The {H}\,{\sc i} column density distribution function, $f_{{\rm H}\,{\sc \rm I}}(N,z)$, for different classes of absorbers is determined by counting the number of Ly$\alpha$ absorbers in the spectra of background quasars \citep[e.g.,][]{wolfe95,storrie00,prochaska05,noterdaeme09,noterdaeme12b}. Current measurements of the $f_{{\rm H}\,{\sc \rm I}}(N,z)$ distribution span almost ten orders of magnitude in column density, ranging from $N_{{\rm H}\,{\sc \rm I}} \sim 10^{12}$ -- $10^{22}$. The $f_{{\rm H}\,{\sc \rm I}}(N,z)$ distribution for quasar absorbers is analogous to the luminosity function of galaxy surveys. This function is then used to derive the amount of neutral gas mass as a function of redshift. 

The contribution of sub-DLAs to the total neutral gas mass is the source of intense discussion in recent years \citep[see][]{viegas95,peroux05,guimares09}. The contribution of sub-DLAs cannot be neglected in an accurate determination of the total neutral gas mass. The gas in LLSs/sub-DLAs is located in extended haloes, whereas the gas in DLAs is located in dense and compact regions \citep{peroux03,voort12}. The evolution of $\Omega_{\rm g}$ has been open to debate \citep{peroux05,zwaan05,rao06,lah07,noterdaeme09,martin10,noterdaeme12b,braun12}. At $z=0$, the neutral gas is traced by the hyperfine 21-cm emission of atomic hydrogen, but, the limited sensitivity of current generation of radio telescopes prevents detection beyond $z$$\sim$0.2 (\citealt{lah07}, see also \citealt{chang10}). 


In this work we take advantage of the advanced data products archival data of UVES \citep{dekker00} mounted at the ESO-Very Large Telescope (VLT). The excellent spectral resolution of UVES allows measuring of the {H}\,{\sc i} column density with high precision. In \citet[][hereafter paper I]{zafar12}, the details of the ESO UVES advanced data products dataset together with the new $N_{{\rm H}\,{\sc \rm I}}$ measurements of DLAs/sub-DLAs have been provided. Because the database is not a blind search for sub-DLAs, we build here an unbiased statistical sample. Our main aim is to include absorbers down to log $N_{{\rm H}\,{\sc \rm I}}\ge19.0$ to determine the shape of the $f_{{\rm H}\,{\sc \rm I}}(N,z)$ in the sub-DLA regime. The current sample allows us to determine the $f_{{\rm H}\,{\sc \rm I}}(N,z)$ redshift evolution. Quantities such as the number density and line density for different classes of quasar absorbers are also determined. Moreover, the incidence of the sub-DLAs together with the DLAs are used to determine the total gas mass density.

In \S 2 we define our dataset. The `statistical'  and combined samples are defined in \S 3 and 4. In \S 5 we present our results for $f_{{\rm H}\,{\sc \rm I}}(N,z)$ and $\Omega_{\rm g}$, followed by a discussion. Conclusions are provided in \S 6. All log values and expressions correspond to log base 10. Throughout the paper we adopt the $\Lambda$CDM cosmology with the cosmological parameters $\Omega_\Lambda=0.7$, $\Omega_m=0.3$, and $H_0=70$ km s$^{-1}$ Mpc$^{-1}$ (e.g., \citealt{spergel03}).

\section{EUADP sample}
Details of the building of the quasar sample can be found in paper I and are briefly summarized here. The sample consists of 250 quasar spectra ranging from $0.191\leq z_{\rm em}\leq6.311$ taken with the VLT instrument UVES. The lowest resolving power of the UVES is $\sim$ 41,400 when a slit of $1''$ is used. The echelle spectrograph covers the wavelength range 300--500 nm (BLUE) and 420--1100 nm (RED) by use of different standard and non-standard settings. The data were downloaded from the ESO UVES advanced data products (EUADP) facility. The downloaded data were processed by the ESO UVES pipeline (version 3.2) within the \texttt{MIDAS} environment with the best available calibration data. The different spectra of each quasar were then merged after interpolating to a common frame. Throughout the paper, this sample of quasars is coined the ``EUADP" sample. 

To derive a complete census of DLAs/sub-DLAs, an automated search accompanied by a visual search was made to search for DLAs and sub-DLAs. The algorithm consists of building an equivalent-width spectrum over 400-pixel-wide boxes for each quasar blueward of the Ly$\alpha$ emission line of the quasar (see paper I). This search resulted in 93 DLAs and 57 sub-DLAs whose Ly$\alpha$ absorption lines are covered by our data. A careful search in the literature indicated that 87 of these DLAs and 44 of these sub-DLAs have been reported before. The $N_{{\rm H}\,{\sc \rm I}}$ measurements for the remaining 6 DLAs and 13 sub-DLAs are reported in paper I. These DLA/sub-DLAs are confirmed by the detection of lines from higher-order Lyman series and associated metal-lines. Their {H}\,{\sc i} column densities were determined using Voigt-profile fitting with \texttt{FITLYMAN} package in \texttt{MIDAS} \citep{fontana}. In addition, another 47 DLAs/sub-DLAs along the lines of sight of EUADP quasars have been found for which Ly$\alpha$ absorption lines are not covered by our data. This yields a total of 197 DLAs/sub-DLAs along the lines-of-sight of 250 EUADP quasars.

\section{Defining the statistical EUADP sample}
The EUADP data originally come from different observing programs where quasars were targeted by several groups for a variety of studies with different goals, many of which include the analysis of a known DLA. We have followed a strict and careful process to select quasar lines-of-sight and sub-DLAs that can be used for statistical analysis. This process, detailed below, led to a subset of the EUADP sample dubbed the ``statistical" EUADP sample.

\subsection{Survey sensitivity for the statistical EUADP sample}
The first step was defining the redshift path for each  quasar. As common is practice, we defined a redshift interval $\Delta z = z_{\rm max} - z_{\rm min}$ to obtain the redshift path over which we could reliably determine the presence of damped absorbers. The lowest redshift ($z_{\rm min}$) is either the lowest redshift at which the signal-to-noise (S/N) ratio ensures the detection of absorption features at the sub-DLA threshold of $\rm EW_{\rm rest}=2.5$\,$\AA$, or where an intervening LLS is present and thus prohibits the detection of any absorption feature at wavelengths shorter than the restframe Lyman break (912\,$\AA$). The higher redshift value of the two is adopted as $z_{\rm min}$ of the quasar. The highest redshift ($z_{\rm max}$) is $3000$ km s$^{-1}$ blueward of the quasar emission redshift. In addition, some of the UVES-combined spectra have distinct spectral gaps due to no-overlapping settings, which were excluded from the redshift path calculation. The resulting values are provided for each line of sight in Table \ref{sampletable}.

\addtocounter{table}{+1}
\begin{table}
\caption{Summary of the number of quasars and sub-DLAs for samples included in this study. We used a subset of the EUADP sample, dubbed the statistical EUADP sample, combined with previous sub-DLA studies.}     
\label{statsample} 
\centering 
\setlength{\tabcolsep}{1pt}
\begin{tabular}{@{} l c c c c@{}}

\hline
Sample & Ref. & $\Delta z$ & QSOs  & Sub-DLAs  \\
 description & & & included  & included \\
 \hline
EUADP & Paper I  & $\cdots$ &  250 & 57   \\
\vspace{0.2cm}
Statistical EUADP & This work &  88.2 & 122  & 27 \\
\multirow{2}{*}{Previous sub-DLA} & \citet{peroux03b,peroux05} & \multirow{2}{*}{104.9} & \multirow{2}{*}{116}  & \multirow{2}{*}{62} \\
 & \citet{omeara07} & \\
{\it Combined sub-DLA} & This work &  193.1 & 238 & 89  \\
 \hline
\end{tabular}
\end{table}

Furthermore, we ignored quasars with $z_{\rm em}<1.51$ because the UVES spectral coverage does not extend blueward to wavelengths required for detecting the Ly$\alpha$ line. As a result of the S/N cut, the quasars above $z_{\rm em}>5.0$ were excluded because the density of absorbers becomes high enough beyond that redshift to affect our ability to reliably determine the absence or presence of sub-DLAs. In addition, we examined the abstract of each observing proposal to determine if any knowledge of the sightline characteristics were known prior to the execution of the proposal that may represent a bias. We excluded all sightlines that were observed for studying the Ly$\alpha$ forest and the intergalactic medium (IGM). This includes $\ion{He}{ii}$ Gunn-Peterson effect, low column density distribution, and the tomography of the IGM as traced by quasar pairs or groups. Indeed, these sightlines might be biased toward having no high $N_{{\rm H}\,{\sc \rm I}}$ system, precisely the type of feature we are looking for in the present study. We chose the conservative approach to exclude all these redshift paths and sub-DLAs from our analysis. Broad-absorption-line (BALs) quasars are often excluded from the search for quasar absorbers. Indeed, the gas associated with the quasar itself can be confused with DLAs. The EUADP sample contains a few BAL quasars, but the absorption features are weak and cannot be confused with DLAs. In the appendix, we describe the reason for including or excluding each individual $z_{\rm em}$ $>$ 1.5 quasar sightline and sub-DLA. This leaves us with a statistical EUADP sample of 122 quasars from the original 250 EUADP sample (see Table \ref{statsample}).
 
    \begin{figure}
   \centering
      {\includegraphics[width=\columnwidth,clip=]{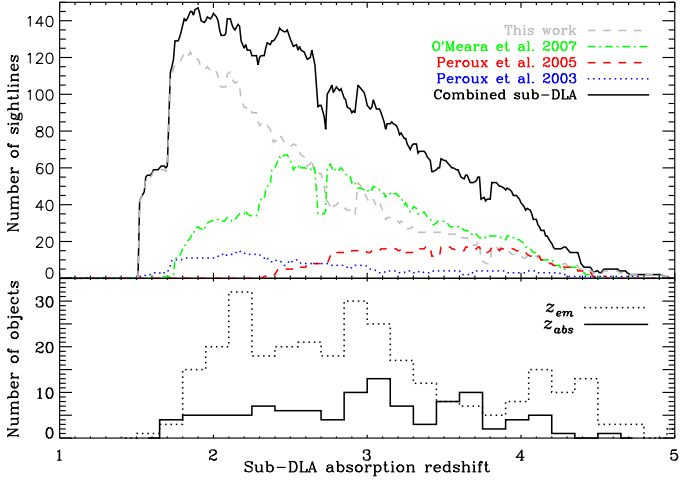}} \\
      \caption{\underline{Upper panel}: Redshift sensitivity function, $g(z)$, as a function of sub-DLA redshift for the statistical EUADP (gray), \citet{omeara07} MIKE $+$ ESI (green), \citet{peroux05} (red), \citet{peroux03b} (blue), and {\it combined sub-DLA} sample (black). \underline{Lower panel}: The distribution of quasar emission and absorption redshift for the {\it combined sub-DLA} sample is illustrated by the black dotted and black solid lines, respectively.}
         \label{gzhist}
   \end{figure}

Where the archival quasar spectra were targeted for the study of dark-matter halos, black-hole masses, stellar masses, and quasar luminosities. where an absorber is not the interest of the study, we included the full redshift path. A large percentage ($\sim$45\%) of the quasars in the EUADP sample were targeted because there was a DLA along their line of sight (studies for  deuterium, molecules, or metal abundances), though many lines of sight have been targeted for a specific known DLA without taking into consideration a sub-DLA along the same line of sight. When an archival quasar spectra was targeted for studying a DLA/sub-DLA, we removed the Ly$\alpha$ regions of a few angstroms in the trough (according to the column densities) from the redshift path, and the Ly$\beta$ regions for DLAs with log $N_{{\rm H}\,{\sc \rm I}}\gtrsim20.90$. The working hypothesis was that the presence of a targeted DLA/sub-DLA does not effect the incidence of other systems. This methodology follows the one applied by others in the statistical analysis of archival data \citep{peroux03b,ribaudo11}.

\vspace{.5cm}
The cumulative number of lines of sight along which an absorber at any given redshift could have been detected in Ly$\alpha$, i.e., redshift sensitivity $g(z)$, is defined as:
\begin{eqnarray}
g(z) = \sum H(z_i^{\rm max} - z)H(z - z_{i}^{\rm min}),
\end{eqnarray}

where $H$ is the Heaviside step function. The plot of $g(z)$ versus the sub-DLA redshift for the statistical EUADP sample is shown in Fig. \ref{gzhist}. The sharp feature in $g(z)$ at $z\approx2.82$ is caused by a spectral gap between the BLUE 390 and RED 580 settings. Another feature at $z\approx3.8$ is due to a spectral gap in the range of 5756--5834\,$\AA$ in the RED 580 setting. The total redshift path of our statistical EUADP sample is $\Delta z=88.2$. For a comparison, the Magellan Inamori Kyocera Echelle (MIKE) and the Echellette Spectrograph and Imager (ESI) sub-DLA sample from \citet{omeara07} and UVES samples from \citet{peroux03b,peroux05} are also shown in Fig. \ref{gzhist}. Their redshift paths are $\Delta z = 91.8$, 16.6, and 23.9 respectively. The EUADP sample includes a larger number of quasars at lower redshift than other samples.

\subsection{Absorber selection for the statistical EUADP sample}
We defined our statistical EUADP sample as the catalog of sub-DLAs of the EUADP sample from paper I restricted to the redshift path defined above. The complete log of DLAs/sub-DLAs, redshifts, and column densities together with references are given in Table \ref{sampletable}. Only sub-DLAs shown in boldface in Table \ref{sampletable} were included in the statistical EUADP sample. The resulting statistical EUADP sample consists of 27 sub-DLAs found by surveying the 122 quasars' lines of sight.

\section{Defining the combined sub-DLA sample} 
\begin{figure}
   \centering
      {\includegraphics[height=3.5cm,width=8.8cm,bb=58 617 539 778,clip=]{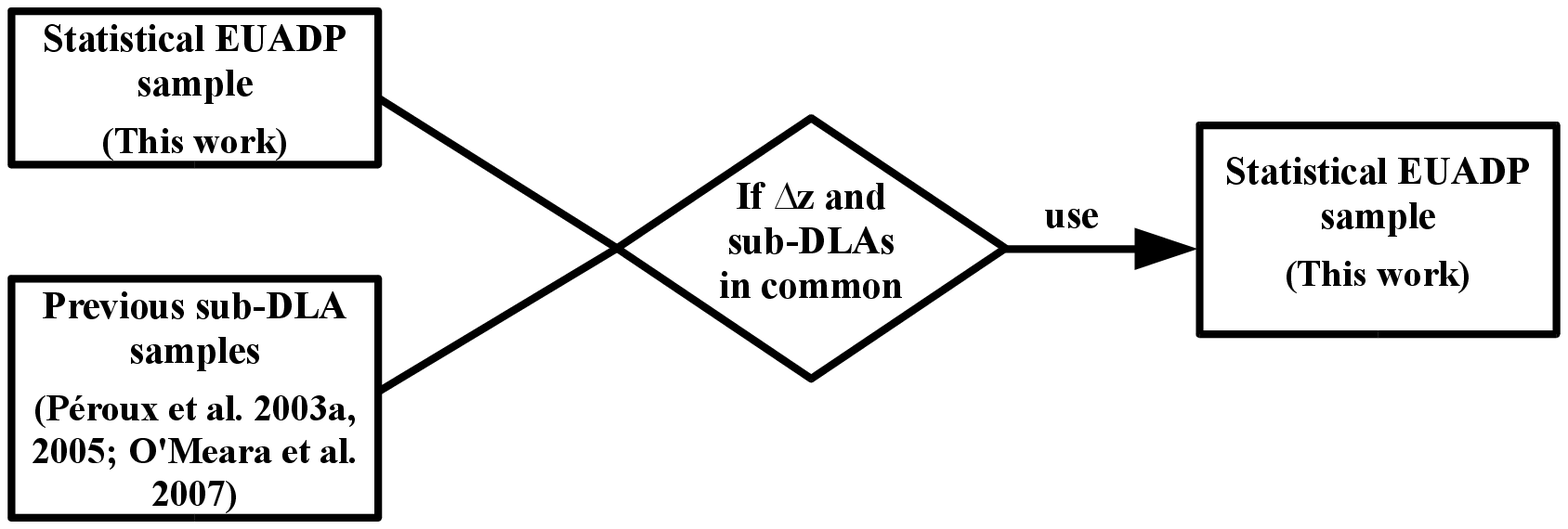}} \\
      \caption{Flowchart describing the building of the {\it combined sub-DLA} sample. For quasars in common with previous sub-DLA samples and the work presented here, we used $N_{{\rm H}\,{\sc \rm I}}$ and the redshift path from this work.}
         \label{flowchart}
   \end{figure}

For the remained of the paper, the statistical EUADP sample is used in combination with previous sub-DLA studies. For the analysis of sub-DLAs, we included the statistical EUADP sample and the results of the high-resolution surveys by \citet{peroux03b,peroux05} and \citet{omeara07} in our statistics. When combining these samples, we encountered cases where the same quasars were targeted in two independent surveys. For these we used $\Delta z$ and the sub-DLA from this work. For example, because the EUADP sample is based on UVES archival data, {\it by construction} it includes the data from \cite{peroux05} which are based on a PI UVES project aiming at finding new sub-DLAs at high redshift. The method used prevents any multiple counting in such cases. The procedure of defining the {\it combined sub-DLA} sample is illustrated in Fig. \ref{flowchart}. In Fig. \ref{gzhist}, $g(z)$ for the {\it combined sub-DLA}, \citet{peroux05}, \citet{peroux03}, and \citet{omeara07} samples are shown. The total redshift path for the {\it combined sub-DLA} sample is $\Delta z=193.1$.

\begin{figure}
   \centering
      {\includegraphics[width=\columnwidth,clip=]{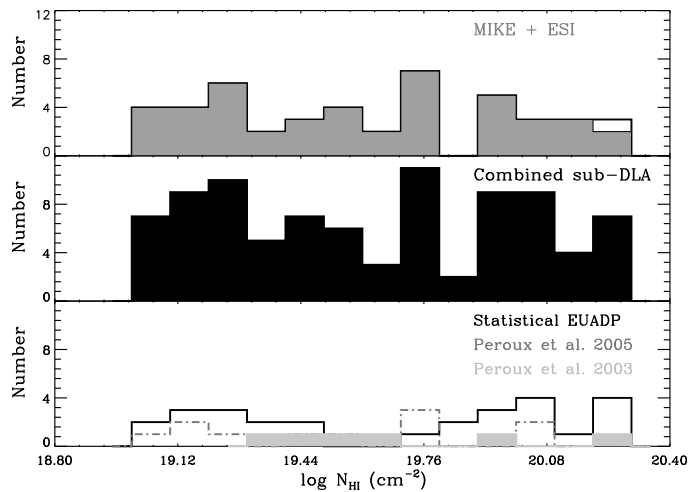}} \\
      \caption{Histograms showing the number of sub-DLAs as a function of log $N_{{\rm H}\,{\sc \rm I}}$. \underline{Top panel}: MIKE $+$ ESI sample from \citet{omeara07}. The filled area represents the new sub-DLAs in MIKE $+$ ESI sample and the unfilled area illustrates the ones overlapping with the {\it combined sub-DLA} sample. \underline{Middle panel}: {\it Combined sub-DLA} sample. \underline{Bottom panel}: The solid black, dash-dotted gray and filled gray histograms show the contribution of sub-DLAs to the {\it combined sub-DLA} sample from the statistical EUADP, \citet{peroux05}, and \citet{peroux03b} samples respectively.}
         \label{nhhist}
   \end{figure}

The resulting {\it combined sub-DLA} sample consists of 89 sub-DLAs found by surveying 238 quasars (see Table \ref{statsample}). The column density distributions of the {\it combined sub-DLA} sample, \citet{peroux05}, \citet{peroux03b}, and MIKE $+$ ESI samples from \citet{omeara07} are shown in Fig. \ref{nhhist}.

\begin{table*}
\caption{Summary of the number density ($d(n)/dz$) and line number density ($d(n)/dX$) of DLAs and sub-DLAs. The DLA results are taken from the \citet{peroux03} sample. The $d(n)/dz$ and $d(n)/dX$ for sub-DLAs is measured from the contribution of both DLAs and sub-DLAs. The final row shows the total redshift path and absorption path length for both DLAs and sub-DLAs.}      
\label{nzresults} 
\centering
\setlength{\tabcolsep}{6pt}
\begin{tabular}{ l c c c c c c c c c c c c}
\hline
\multicolumn{1}{l}{log $N_{{\rm H}\,{\sc \rm I}}$} & \multicolumn{6}{c}{$>19.0$} & \multicolumn{6}{c}{$>20.3$}  \\ 
$z$ range	& $\langle z \rangle$ & $dz$ & $dX$ & \# & $d(n)/dz$ & $d(n)/dX$ & $\langle z \rangle$ & $dz$ & $dX$ & \# & $d(n)/dz$ & $d(n)/dX$  \\
 \hline
1.51--2.00 & 1.80 & 29.9 & 87.3 & 4 & $0.35\pm0.12$ & $0.12\pm0.04$ & 1.84 & 95.6 & 279.4 & 19 & $0.22\pm0.05$ & $0.08\pm0.02$ \\  
2.00--2.50 &  2.26 & 49.2 & 156.8  & 11 & $0.43\pm0.10$ & $0.14\pm0.03$ & 2.27 & 125.1 & 396.6 & 26 &$0.21\pm0.04$ & $0.07\pm0.01$ \\  
2.50--3.00 &  2.76 & 47.1 & 162.8 & 24 & $0.72\pm0.13$ & $0.21\pm0.04$ & 2.73 & 87.2 & 301.0 & 18 & $0.21\pm0.05$ & $0.06\pm0.01$ \\
3.00--3.50 & 3.21 & 33.7 & 124.5 & 23 & $0.96\pm0.18$ & $0.26\pm0.05$ & 3.25 & 64.8 & 239.6 &  18 & $0.28\pm0.07$ & $0.08\pm0.02$ \\
3.50--4.00 & 3.72 & 23.3 & 91.7  & 19 & $1.20\pm0.24$ & $0.30\pm0.06$ & 3.77 & 49.5 & 194.6 & 19 & $0.38\pm0.09$ & $0.10\pm0.02$ \\
4.00--5.00 & 4.18 & 9.9 & 41.0 & 8 & $1.25\pm0.38$ & $0.30\pm0.09$ & 4.20 & 23.1 & 95.1 & 10 & $0.43\pm0.14$ & $0.10\pm0.03$ \\
1.51--5.00 & $\cdots$ & 193.1 & 664.1 & 89 & $\cdots$ & $\cdots$ & $\cdots$ & 445.4 & 1506.9 & $\cdots$ & $\cdots$ & $\cdots$  \\
\hline
\end{tabular}
\end{table*}

\section{Statistical sample properties\label{sample}}

\subsection{Number density}   \label{nztext} 
The number density of quasar absorbers, a quantity directly observable, is the number of absorbers per unit redshift, i.e., $d(n)/dz$ also represented as $n(z)$, 
\begin{eqnarray}
d(n)/dz = n(z) = \frac{m}{dz},
\end{eqnarray}
where $m$ is the total number of absorbers in a given redshift bin and $dz$ is computed by summing the redshift paths in this bin.

   \begin{figure}
   \centering
      {\includegraphics[width=\columnwidth,clip=]{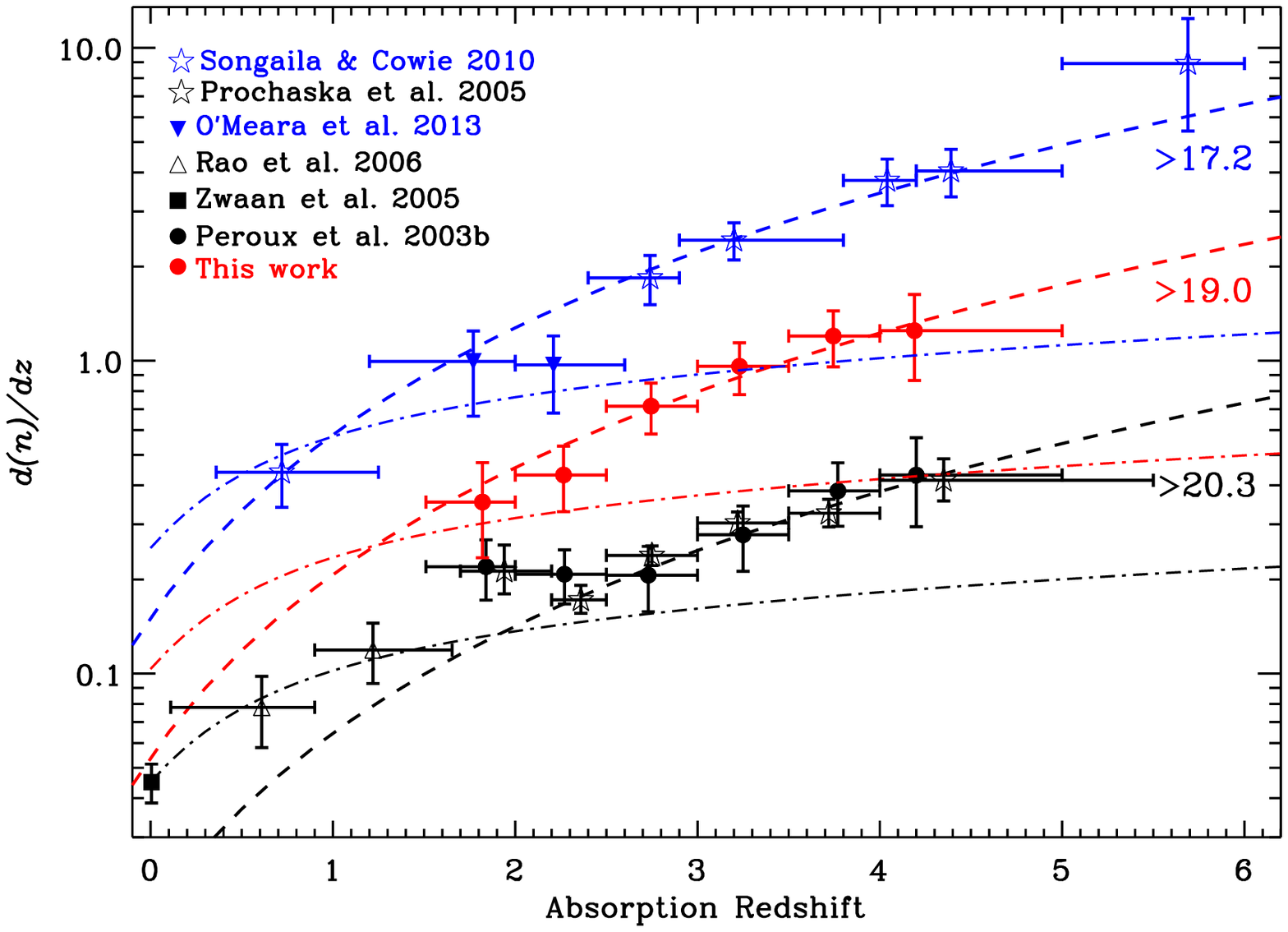}} \\
      \caption{Number density ($d(n)/dz$) of quasar absorbers as a function of redshift. $d(n)/dz$ for DLAs:  \citet[][black circles]{peroux03}, \citet[][black open stars]{prochaska05}, \citet[][black open triangles]{rao06}, and \citet[][black square]{zwaan05b} are shown. $d(n)/dz$ for sub-DLAs: the {\it combined sub-DLA} sample (red circles), and LLSs: \citet[][blue open stars]{songaila10} and \citet[][blue inverted triangles]{omeara12} are shown. The size of the horizontal bars represents the redshift bin sizes, and the vertical error bars are the 1$\sigma$ Poisson uncertainties in the number of absorbers. The blue dashed line is the simple evolution fit to the LLS from \citet{songaila10}. The red and black dashed lines are the LLS evolution fit scaled by a factor of 2.8 and 9.0 for the difference in $d(n)/dz$ of sub-DLAs and DLAs with respect to LLSs (in the range $2.5<z<3.0$), respectively. The black dot-dashed curve represents a non-evolving population of non-zero $\Lambda$-Universe. The red and blue dot-dashed curves are non-evolution curves scaled to a factor of 2.3 and 5.6, respectively (see text for more detail).}
         \label{nz}
   \end{figure}

The combined sample allows one to determine the number density of sub-DLAs and compare it with those for the DLAs and LLSs from the literature. Fig. \ref{nz} shows the number density per unit redshift for DLAs (\citealt{peroux03}, \citealt{prochaska05}, \citealt{rao06}, and \citealt{zwaan05b}), sub-DLAs (the {\it combined sub-DLA} sample), and LLSs (\citealt{songaila10} and \citealt{omeara12}). At low-redshift, results for DLAs come from the \citet{rao06} survey of quasars targeted because of known {Mg}\,{\sc ii} absorbers. The \citet{zwaan05b} results come from 21-cm emission of atomic hydrogen of $z$ $=$ 0 analogues to DLAs. At low redshifts, the results of \citet{omeara12} for LLS come from {\it Hubble Space Telescope} survey. The binning for our data was chosen with an increment of one half in redshift except for $4.0<z<5.0$ because of low number statistics. Vertical error bars reflect 1$\sigma$ Poisson uncertainty in the number of absorbers. All results are tabulated in Table \ref{nzresults}. 

\citet{songaila10} found that $d(n)/dz$ of LLSs can be well fitted by a simple evolution law of the form $d(n)/dz=n_{3.5}\left[(1+z)/4.5\right]^\gamma$ with $n_{3.5}=2.80\pm0.33$ and $\gamma=1.94^{+0.36}_{-0.32}$ for the entire redshift range from $0<z<6$, where $n_{3.5}$ is the value of $n(z)$ at $z=3.5$. In Fig. \ref{nz} we plot this simple evolution fit for LLS from \citet{songaila10}, which provides a reasonable fit to the data. For sub-DLAs and DLAs, the maximum-likelihood fit for the LLS was scaled according to a factor of difference in number density of the sub-DLA (by a factor of 2.8) and DLA (by a factor of 9.0) in the redshift range $2.5<z<3.0$. The redshift range was chosen arbitrarily. This shows that the number density of the absorbers per unit redshift is increasing with increasing redshift for all three classes of absorbers. The number density evolution inferred for these three classes of quasar absorbers is generally steeper with increasing redshift \citep[see][]{songaila10}.

  \begin{figure}
   \centering
      {\includegraphics[width=\columnwidth,clip=]{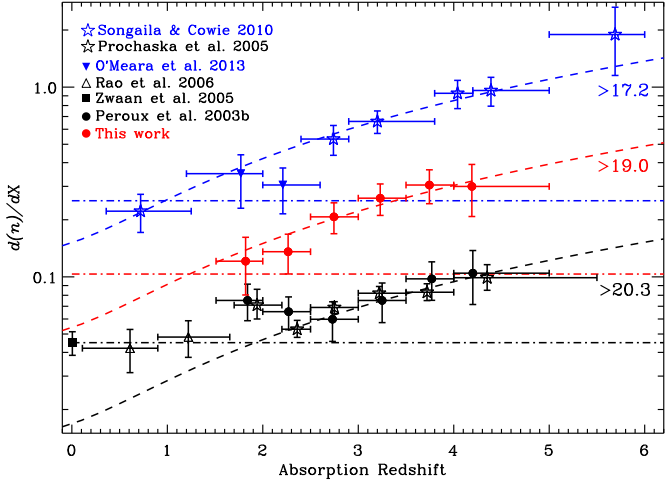}} \\
      \caption{Line density ($d(n)/dX$) of quasar absorbers as a function of redshift. $d(n)/dX$ for DLAs, sub-DLAs and LLSs is shown as black, red, and blue data points, respectively. The symbols have the same meaning as in Fig. \ref{nz}.}
         \label{lx}
   \end{figure}

Assuming no-evolution in a non-zero $\Lambda$-Universe, the number density can be expressed as \citep[see e.g.,][]{peroux05}
\begin{eqnarray}
n(z) = n_0(1+z)^2\left(\frac{H(z)}{H_0}\right)^{-1},
\end{eqnarray}
where $H(z)$ is the Hubble parameter given as
\begin{eqnarray}
H(z)= {H_0} \left[ \Omega_m (1+z)^3 +\Omega_\Lambda \right]^{1/2}.
\end{eqnarray}
As derived by \citet{zwaan05b}, $n_0$ was taken to be $n(z=0)=0.045\pm0.006$. The non-evolution curve for DLAs is shown in Fig. \ref{nz}. The non-evolution curve for LLS was scaled according to a factor of difference in number density of the LLS (5.6) in the redshift range $0.26<z<1.06$. For sub-DLAs, the non-evolution curve was arbitrarily scaled (by a factor of 2.3) in the middle of the DLA and LLS curves because there are no results available at low redshift. Departure from non-evolution beyond $z$$\sim$1 is evident for the three classes of quasar absorbers. The DLAs seems to behave differently at $z$ $<$ 2, which might be due to the different data surveys and methods.

\subsection{Line density}
The cosmological line density of quasar absorbers is the number of absorbers per unit absorption path length $dX$ i.e., $d(n)/dX$. The line density of quasar absorbers, also termed $l(X)$, is the product of the comoving number density and the cross section of absorbers \citep[see][]{prochaska05}. The values for quasar absorbers were calculated in the discrete limit by using 
\begin{eqnarray}
d(n)/dX = l(X) = \frac{m}{dX}.
\end{eqnarray}
This can also be expressed as $l(X)dX=n(z)dz$, where the absorption distance \citep{bahcall69} is defined as 
\begin{equation}
dX = \int \frac{H_0}{H(z)}(1+z)^2 dz .
\end{equation}

In Fig. \ref{lx} the line densities for DLAs (\citealt{peroux03}, \citealt{prochaska05}, \citealt{rao06}, and \citealt{zwaan05b}), sub-DLAs ({\it combined sub-DLA} sample), and LLSs (\citealt{songaila10} and \citealt{omeara12}) versus redshift are presented. The binning and error calculation procedure is the same as defined in Sec. \ref{nztext}. The results are tabulated in Table \ref{nzresults}. 

The simple evolution fit to the line density for the LLS from \citet{songaila10} is also plotted in Fig. \ref{lx}. For DLAs and sub-DLAs the fit is scaled in a similar manner as explained in Sec. \ref{nztext}. Clearly the line density for all three classes of absorbers increases with redshift. The rise in $d(n)/dX$ reflects an increase in the comoving number density or the cross-section of the absorber. In Fig. \ref{lx} the line density non-evolution curve for DLAs is shown and for LLSs and sub-DLAs is scaled in a similar manner as explained in Sec. \ref{nztext}. The flatness of the non-evolution curve reflects no variation in the covering fraction of absorbers on the sky with increasing redshift. There is a clear departure from non-evolution beyond $z$$\sim$1 for all three classes of quasar absorbers.

   \begin{figure}
   \centering
      {\includegraphics[width=\columnwidth,clip=]{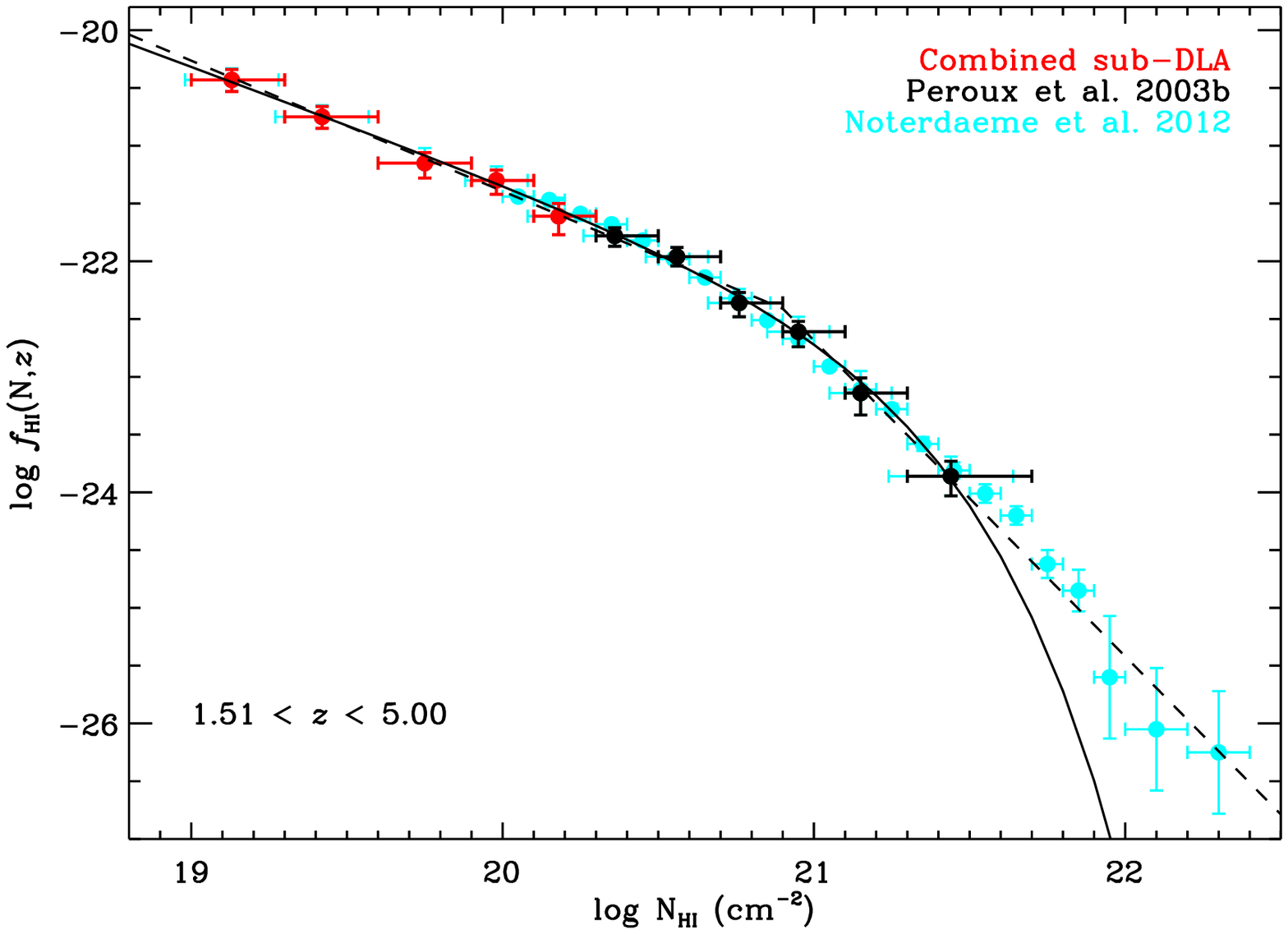}} 
      \caption{$N_{{\rm H}\,{\sc \rm I}}$ frequency distribution of DLAs \citep{peroux03} and sub-DLAs ({\it combined sub-DLA} sample) in a range of $1.51$ $<$ $z$ $<$ $5.00$ is shown as black and red circles, respectively. The $\Gamma$-function and double power-law fit to these DLA and sub-DLA results are shown as solid and dashed black lines, respectively. The cyan data points represent the results from \citet{noterdaeme12b} over two orders of magnitude in $N_{{\rm H}\,{\sc \rm I}}$. The departure at high column densities (log $N_{{\rm H}\,{\sc \rm I}}$ $>$ 21.0) from the $\Gamma$-function shape is due to low number statistics at higher column density in the former studies.}
      \label{fnalluves}
   \end{figure}

\subsection{Column density distribution}
The distribution of {H}\,{\sc i} column densities for all classes of quasar absorbers has been investigated in several studies \citep[e.g.,][]{peroux05,ribaudo11,noterdaeme12b}. The differential column density distribution, $f_{{\rm H}\,{\sc \rm I}}(N,z)$, describes the evolution of quasar absorbers as a function of atomic column density and redshift,
\begin{eqnarray}
f_{{\rm H}\,{\sc \rm I}}(N,z)dNdX = \frac{m}{\Delta N\sum_{i=1}^{n}\Delta X_i}dNdX ,
\end{eqnarray}
where m is the number of quasar absorbers in a range of column densities $(N,N+dN)$ obtained from the observations of $n$ quasar spectra with the total absorption distance coverage $\sum_{i=1}^{n}\Delta X_i$. In Fig. \ref{fnalluves}, we present the $f_{{\rm H}\,{\sc \rm I}}(N,z)$ of the {\it combined sub-DLA} sample while the values are tabulated in Table \ref{fnresults}. The horizontal bars indicate the bin sizes plotted at the mean column density for each bin, and the vertical error bars are Poissonian statistical errors. For a comparison with the DLA regime, the results of the \citet{peroux03} samples are also shown with the same redshift bins. In recent SDSS samples, \citet{noterdaeme09,noterdaeme12b} observed absorbers down to log $N_{{\rm H}\,{\sc \rm I}}=20.0$. The results of \citet{noterdaeme12b} are also plotted in Fig. \ref{fnalluves}. Empirically, $f_{{\rm H}\,{\sc \rm I}}(N,z)$ were fitted by a power-law for various $N_{{\rm H}\,{\sc \rm I}}$ regimes \citep[e.g.,][]{tytler87,rao06,omeara07}. Early determinations of $f_{{\rm H}\,{\sc \rm I}}(N,z)$ for a broad column density range were defined with a single power-law \citep{tytler87}, but the improved quality of observations shows that a single power-law does not represent the observations well over a broad column density range. For this reason, a $\Gamma$-function (power-law with an exponential turn-over), analogous to the Schechter function, was used to reasonably describe the observations \citep[e.g.,][]{storrie00,peroux03,noterdaeme09}. In addition, the integral of this functional form used to compute $\Omega_{\rm g}$ converges, as opposed to a power law. The $\Gamma$-function is defined as
\begin{eqnarray}
f_{{\rm H}\,{\sc \rm I}}(N,z) = k_g\left( \frac{N}{N_g} \right)^{\alpha_g}e^{(-N/N_g)} .
\end{eqnarray}

\begin{table}
\begin{minipage}[t]{\columnwidth}
\caption{Summary of the column density distribution of the {\it combined sub-DLA} sample.}     
\label{fnresults} 
\centering 
\setlength{\tabcolsep}{8pt}
\begin{tabular}{ l c c c }
\hline
 log $N_{{\rm H}\,{\sc \rm I}}$ &  & log & log \\
 range & \# & $\langle N_{{\rm H}\,{\sc \rm I}} \rangle$ & $f_{{\rm H}\,{\sc \rm I}}(N,z)$  \\
 \hline
19.00--19.30 & 23 & 19.13 & $-20.43^{+0.09}_{-0.10}$ \\    
19.30--19.60 & 21 & 19.42 & $-20.75^{+0.09}_{-0.10}$  \\   
19.60--19.90 & 16 & 19.75 & $-21.15^{+0.09}_{-0.13}$ \\
19.90--20.10 & 18 & 20.06 & $-21.30^{+0.09}_{-0.12}$ \\ 
\vspace{0.05cm}
20.10--20.30 & 11 & 20.06 & $-21.61^{+0.11}_{-0.16}$ \\ 
\hline
\end{tabular}
\end{minipage}
\end{table}

\begin{table}
\begin{minipage}[t]{\columnwidth}
\caption{Best-fit parameters of the $\Gamma$-function fits (see Eq. [6]) to the $N_{{\rm H}\,{\sc \rm I}}$ frequency distribution for different redshift bins.}      
\label{fnfit} 
\centering 
\setlength{\tabcolsep}{4pt}
\begin{tabular}{ l c c c c}
\hline
$z$ range & log $k_g$ & log $N_g$ & $\alpha_g$ & $\chi_\nu^2$\\
 \hline
1.51--5.00 & -22.33 & 21.02 & -1.00 & 0.71 \\
1.51--3.10 & -22.30 & 21.08 & -0.95 & 1.16 \\
3.10--5.00 & -22.05 & 20.91 & -1.00 & 0.76 \\
 \hline
\end{tabular}
\end{minipage}
\end{table}

\begin{table}
\caption{Summary of the column density distribution of the {\it combined sub-DLA} sample in two redshift bins.}    
\label{fnzresults} 
\centering
\setlength{\tabcolsep}{5pt}
\begin{tabular}{ l c c c c c c}
\hline
\multicolumn{1}{l}{$z$ range} & \multicolumn{3}{c}{1.51--3.10} & & \multicolumn{2}{c}{3.10--5.00}  \\ 
 log $N_{{\rm H}\,{\sc \rm I}}$ &  & log  & log &  & log  & log \\
 range & \# & $\langle N_{{\rm H}\,{\sc \rm I}} \rangle$ & $f_{{\rm H}\,{\sc \rm I}}(N,z)$ & \# & $\langle N_{{\rm H}\,{\sc \rm I}} \rangle$ & $f_{{\rm H}\,{\sc \rm I}}(N,z)$  \\
  \hline
19.00--19.40 & 14 & 19.20 & $-20.64^{+0.11}_{-0.13}$ & 17 & 19.17 & $-20.29^{+0.10}_{-0.12}$ \\ 
19.40--19.90 & 14 & 19.60 & $-21.16^{+0.10}_{-0.13}$ & 15 & 19.65  & $-20.88^{+0.10}_{-0.13}$ \\
\vspace{0.05cm}
19.90--20.30 & 16 & 20.08 & $-21.48^{+0.10}_{-0.12}$ & 13 & 20.03  & $-21.30^{+0.10}_{-0.14}$ \\
\hline
\end{tabular}
\end{table}

  \begin{figure*}
   \centering
      {\includegraphics[height=8cm,width=16.5cm,clip=]{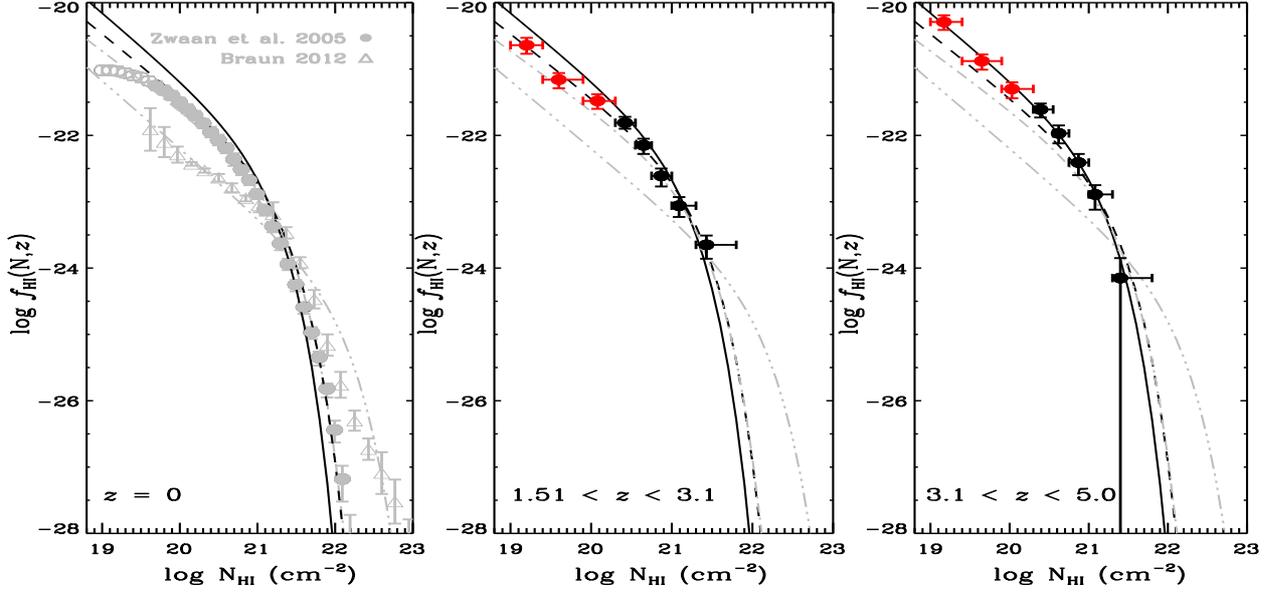}} 
      \caption{{\it Middle and right panels}: The differential column density distribution for two redshift ranges. The horizontal error bars are the bin sizes and the vertical error bars represent the Poissonian statistical errors on the column density distribution. The bins in red correspond to the {\it combined sub-DLA} sample results and bins in black correspond to the \citet{peroux03} results. The black dashed and solid lines represent the $\Gamma$-function fit to the $f_{{\rm H}\,{\sc \rm I}}(N,z)$ in 3.1 $<$ $z$ $<$ 5.0 and 1.51 $<$ $z$ $<$ 3.0 redshift bins, respectively. {\it Left Panel}: $f_{{\rm H}\,{\sc \rm I}}(N,z)$ at $z=0$ from \citet[][gray open triangles]{braun12} and \citet[][gray filled circles]{zwaan05b}. Open gray circles correspond to the measurements below the sensitivity limit of \citet{zwaan05b}. The gray triple dot-dashed and dot-dashed line is $\Gamma$-function fit to \citet{braun12} and \citet{zwaan05b} data, respectively, above their sensitivity limit.}
      \label{fnuvesz}
   \end{figure*}

We fit the data with the $\Gamma$-function for the total redshift range. The best-fit values of the parameters are provided in Table \ref{fnfit} together with the reduced $\chi^2$ ($\chi_\nu^2$) values that define the goodness-of-fit. We also fit the data with a double power-law resulting in the best-fit values of $k_{\rm d}=-22.41$, $N_{\rm d}=20.90$ (fixed), $\alpha_{\rm d1}=-1.13$, and $\alpha_{\rm d2}=-2.74$ \citep[see also][]{noterdaeme09}.
As apparent in Fig. \ref{fnalluves}, the column density distribution flattens in the sub-DLA regime. This flattening of the distribution above log $N_{{\rm H}\,{\sc \rm I}}=18.0$ cm$^{-2}$ is indeed expected \citep[e.g.,][]{peroux05,omeara07,guimares09}. It is due to less self-shielding in the absorbers because at these column densities, part of the neutral gas of the absorber is ionized by incident UV flux \citep[e.g.,][]{katz96,pontzen08,mcquinn11,altay11}. The distribution steepens around log $N_{{\rm H}\,{\sc \rm I}}=20.3$--21.5 as the absorbing gas becomes fully neutral \citep{altay11,erkal12} and because of the atomic-to-molecular hydrogen transition \citep{schaye01,krumholz09,altay11}. There is a deficit of log $N_{{\rm H}\,{\sc \rm I}}>21.7$ DLAs in the \citet{peroux03} sample because their redshift path is smaller than that of the SDSS surveys \citep{noterdaeme09,noterdaeme12b}, the latter providing a better estimation of $f_{{\rm H}\,{\sc \rm I}}(N,z)$ down to log $N_{{\rm H}\,{\sc \rm I}}\gtrsim22$. 


In a recent study, \citet{erkal12} argued that the atomic-to-molecular hydrogen transitions do not significantly contribute to the high column density redshift-independent turnover. They furthermore argued that the similarity of the $f_{{\rm H}\,{\sc \rm I}}(N,z)$ at $z=3$ \citep[e.g.,][]{noterdaeme09} and $z=0$ \citep{zwaan05b} is due to the similarity of the highest central {H}\,{\sc i} surface densities of high-$z$ and low-$z$ disks. \citet{voort12} advocated that above log $N_{{\rm H}\,{\sc \rm I}}$ = 17 most of the absorbers reside inside haloes and below log $N_{{\rm H}\,{\sc \rm I}}$ = 17 the absorbers reside outside haloes. They also argued that the absorbing gas between column densities 17 $<$ log $N_{{\rm H}\,{\sc \rm I}}$ $<$ 21 has never experienced a virial shock.

   \begin{figure*}
   \centering
      {\includegraphics[width=16cm,height=12cm]{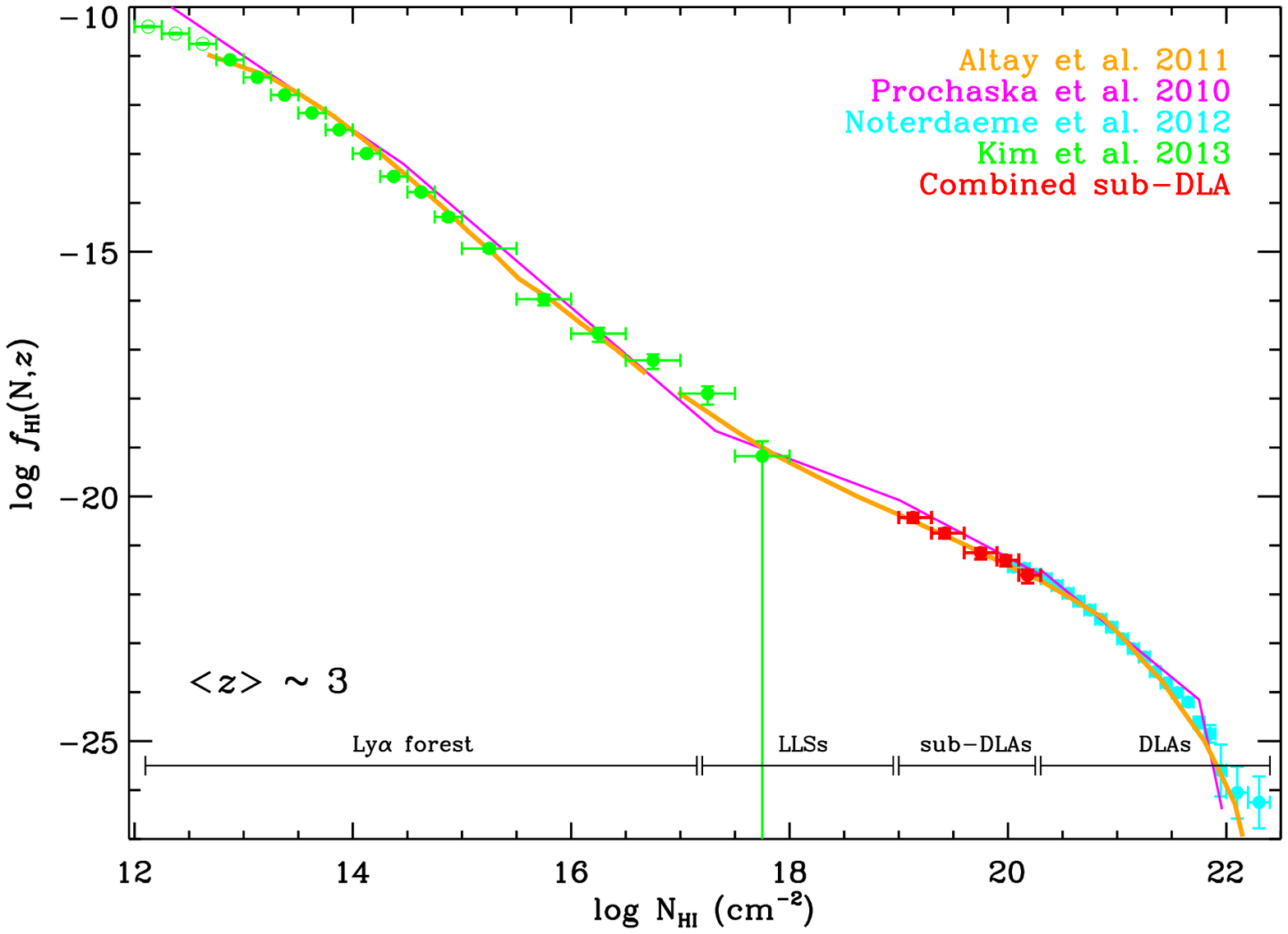}} \\
      \caption{Differential column density distribution ($f_{{\rm H}\,{\sc \rm I}}(N,z)$) plotted against log $N_{{\rm H}\,{\sc \rm I}}$. The red data points indicate $f_{{\rm H}\,{\sc \rm I}}(N,z)$ for sub-DLAs from the {\it combined sub-DLA} sample. The green filled data points represent $f_{{\rm H}\,{\sc \rm I}}(N,z)$ for Ly$\alpha$ forest from \citet{kim13}. The green open circles are column densities from \citet{kim13} affected by incompleteness. The cyan data points represent the results from \citet{noterdaeme12b}. The solid magenta line is the estimate of $f_{N_{{\rm H}\,{\sc \rm I}}}(N,z)$ at $z\sim3.7$ using a series of six power-laws by \citet{prochaska10}. The solid orange line is hydrodynamic/analytical model prediction at $z\sim3$ by \citet{altay11}.}
  \label{fnallrange}
   \end{figure*}


\subsubsection{$f_{{\rm H}\,{\sc \rm I}}(N,z)$ redshift evolution\label{fnevolution}}
We estimated the column density distribution for low and high redshifts. The differential column density distributions for two redshift ranges, $1.51<z<3.1$ and $3.1<z<5.0$, are shown in Fig. \ref{fnuvesz}. These redshift ranges are chosen to divide the number of sub-DLAs equally for each redshift bin. The results of the {\it combined sub-DLA} sample for two redshift bins are tabulated in Table \ref{fnzresults}. Because the $dz$ were available to us, we fit the {\it combined sub-DLA} and \citet{peroux03} results with the $\Gamma$-function for both redshift ranges; the best-fit values of the parameters are summarized in Table \ref{fnfit}. For comparison, the column density distribution at $z=0$ is also shown in Fig. \ref{fnuvesz} from \citet{zwaan05b} and \citet{braun12}. The results of \citet{zwaan05b} come from the survey of 355 {H}\,{\sc i} 21-cm individual maps of nearby galaxies and the \citet{braun12} results come from {H}\,{\sc i} distribution, based on high-resolution maps of M31, M33, and the Large Magellanic Cloud (LMC). The data of \citet{zwaan05b} have a lower resolution and therefore cannot resolve the highest column densities. \citet{braun12} used data at very high-resolution and also applied a correction to the column density based on opacity. In Fig. \ref{fnuvesz}, we overplotted the $\Gamma$-function fit from \citet{zwaan05b} with $\alpha_g=1.24$, log $k_g=-22.91$ and log $N_g=21.20$. We performed a $\Gamma$-function fit to the \citet{braun12} results with $\alpha_g=1.02$, log $k_g=-24.10$ and log $N_g=21.87$.

Fig. \ref{fnuvesz} indicates that the $f_{{\rm H}\,{\sc \rm I}}(N,z)$ evolves with redshift. There is a tendency for a flattening of the distribution function in the sub-DLA regime at lower redshift, indicating that the number of sub-DLAs is larger at higher redshift. This evolution of sub-DLAs might be due to the merging history of absorbers or the ionization of the systems with time.

\subsubsection{Comparison with other studies}
We combined our study with previous studies to place constraints on the $f_{{\rm H}\,{\sc \rm I}}(N,z)$ over ten orders of magnitude in $N_{{\rm H}\,{\sc \rm I}}$. In Fig. \ref{fnallrange}, we show the column density distribution at $\langle z\rangle$ $\sim$ 3. In the DLA regime, the results come from \citet{noterdaeme12b} and \cite{peroux03}. In the sub-DLA regime, the analysis comes from the {\it combined sub-DLA} sample. In the Ly$\alpha$ forest and LLS regimes, the results come from \citet{kim13}, who used UVES observations of 18 quasar spectra at 1.9 $<$ $z$ $<$ 3.2. As can be seen in Fig. \ref{fnallrange}, there is a turnover in the distribution function at log $N_{{\rm H}\,{\sc \rm I}}<12.7$ that is caused by the incompleteness of the sample at these column densities. The complete LLS regime remains uncategorized because the column densities are in the logarithmic part of the curve-of-growth.

In Fig. \ref{fnallrange} a prediction from the hydrodynamical/analytical model of \citet{altay11}, treating self-shielding and formation of molecular hydrogen in post-processing, is also plotted. Their low $N_{{\rm H}\,{\sc \rm I}}$ is obtained by fitting mock spectra with VPFIT and their high $N_{{\rm H}\,{\sc \rm I}}$ by projecting the simulation box onto a plane. A gap between low and high $N_{{\rm H}\,{\sc \rm I}}$ can also be clearly seen in Fig. \ref{fnallrange}. The \citet{altay11} model provides a reasonable representation of the data down to Ly$\alpha$ forest range at $z$ $\sim$ 3 (see also \citealt{voort12}; \citealt{rahmati13a}; \citealt{rahmati13b}).
\subsection{Neutral gas mass density}
\begin{figure}
   \centering
      {\includegraphics[width=\columnwidth,clip=]{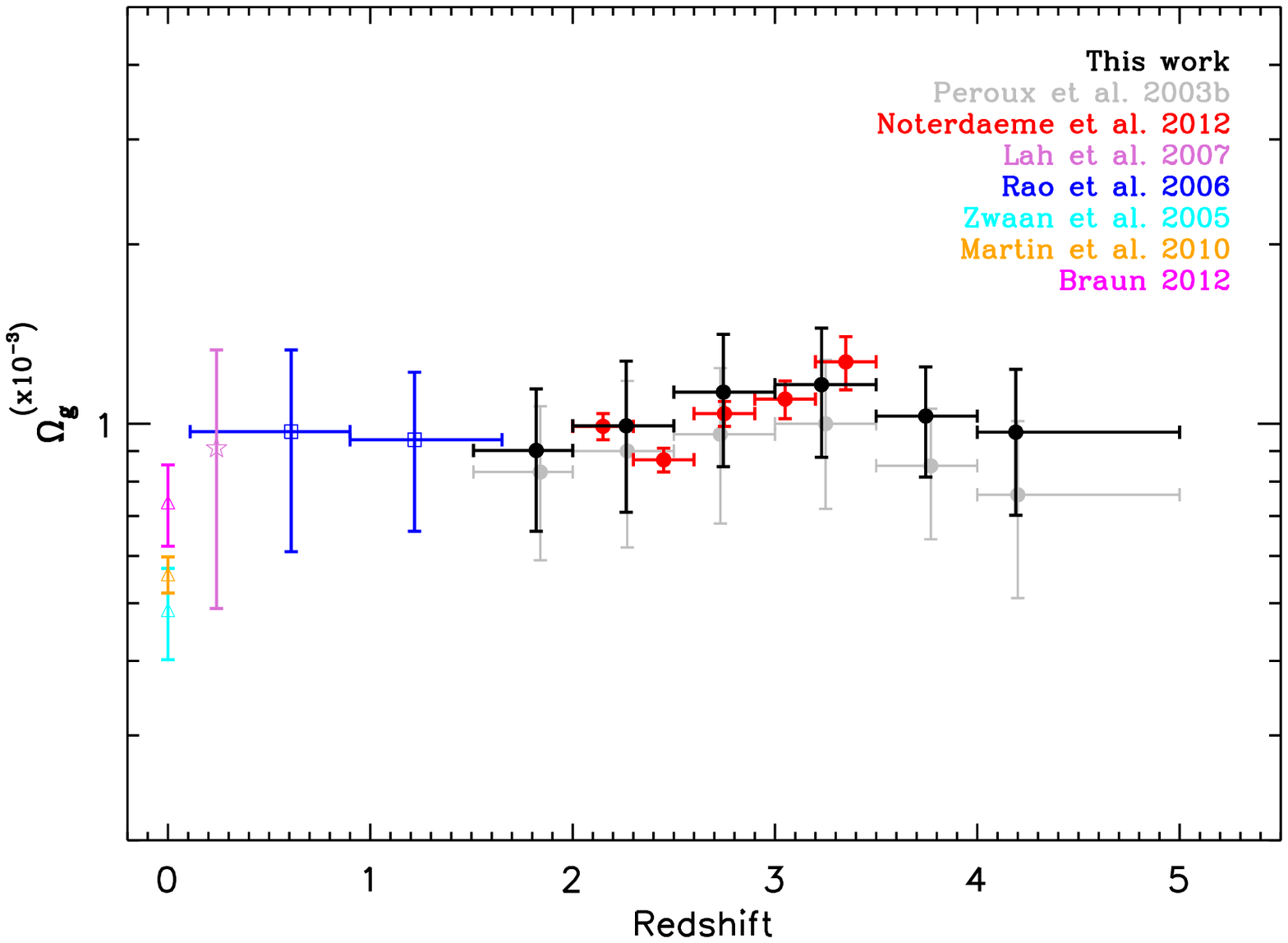}} 
      \caption{Redshift evolution of the neutral gas mass density, $\Omega_{\rm g}$. The black bins correspond to $\Omega_{\rm g}$ measured from DLAs (\citealt{peroux03}) plus sub-DLAs ({\it combined sub-DLA}) contribution. The gray bins show the DLA contribution to the $\Omega_{\rm g}$ from \citet{peroux03} in the same redshift bins. The triangles at $z$ = 0 represent the $\Omega_{\rm g}$ measured in local galaxies by \citet[][cyan]{zwaan05}, \citet[][orange]{martin10}, and \citet[][magenta]{braun12}, respectively. The open purple star at $z$ =0.24 illustrates the {H}\,{\sc i} content measured from the star-forming galaxies by \citet{lah07}. The blue squares at $z<2$ represent the {H}\,{\sc i} measured from {Mg}\,{\sc ii} selected DLAs by \citet{rao06}. The red bins correspond to the $\Omega_{\rm g}$ measured from SDSS by \citet{noterdaeme12b}.} 
      \label{omega}
   \end{figure}

The cosmological mass density of neutral gas, $\Omega_{\rm g}$, observed in high-redshift quasar absorbers is expressed as a fraction of the current critical density \citep[see][]{lanzetta95}:
\begin{eqnarray}
\Omega_{\rm g}(z) = \frac{\mu m_{\rm H} H_0}{c\rho_{\rm crit}} \int_{N_{\rm min}}^{N_{\rm max}} N_{{\rm H}\,{\sc \rm I}} f_{{\rm H}\,{\sc \rm I}}(N,z)dN ,
\end{eqnarray}
where $f_{{\rm H}\,{\sc \rm I}}(N,z)$ is the $N_{{\rm H}\,{\sc \rm I}}$ frequency distribution, $m_{\rm H}$ is the mass of the hydrogen atom, $\rho_{\rm cric}$ is the current critical mass density given as $\rho_{\rm cric}=3H_0^2/8\pi G$, and $\mu=1.3$ is the mean molecular mass of the gas used to incorporate helium as well as hydrogen in the estimate of the neutral gas density. No correction was applied for missing sub-DLAs because of dust \citep{pontzen09}. Setting $N_{\rm min}=1\times10^{19}$ and $N_{\rm max}=2\times10^{20}$ gives the mass density of neutral gas for sub-DLAs, $\Omega_{\rm sub-DLA}$, and setting $N_{\rm min}=2\times10^{20}$ and $N_{\rm max}=\infty$ gives the mass density of neutral gas for DLAs, $\Omega_{\rm DLA}$. It is worth mentioning that $f_{{\rm H}\,{\sc \rm I}}(N,z)$ steepens in the Ly$\alpha$ forest regime (see Fig. \ref{fnallrange}). A significant fraction of gas is photoionized at these column densities, therefore, the Ly$\alpha$ forest contributes less than one percent to $\Omega_{\rm g}$ \citep{kim13}. In the discrete limit, $\Omega_{\rm g}$ is given by
\begin{eqnarray}
\Omega_{\rm g} = \frac{\mu m_{\rm H} H_0}{c\rho_{\rm crit}} \frac{\sum N_{{\rm H}\,{\sc \rm I}}}{\Delta X} ,
\end{eqnarray}
where the sum is calculated for sub-DLAs and DLAs along lines of sight with a total absorption path-length $\Delta X$. In Table \ref{omegaresults} we provide total gas mass density for this work, including contribution from DLAs \citep{peroux03} and sub-DLAs ({\it combined sub-DLA}). The total gas mass density, including {H}\,{\sc i} gas in absorbers up to the canonical sub-DLA definition, is plotted in Fig. \ref{omega}. The binning for our data was selected with an increment of one half in redshift except for 4 $<$ $z$ $<$ 5 because of low number statistics. Vertical error bars correspond to 1$\sigma$ uncertainties and the horizontal error bars indicate bin sizes. It is a difficult task to measure $\Omega_{\rm g}$ at intermediate and low redshifts. For comparison, the {H}\,{\sc i} mass density measured from radio observations of local galaxies using 21-cm emission is plotted in Fig. \ref{omega} (triangles at $z=0)$. These observations come from the Westerbork {H}\,{\sc i} survey of spiral and irregular galaxies (WHISP) project \citep{zwaan05}, the 40\% of the Arecibo legacy fast ALFA (ALFALFA) aurvey \citep{martin10}, and M31, M33, and the LMC \citep{braun12}. The measurement at $z=0.24$ comes from the radio observations of star-forming galaxies \citep{lah07}. \citet{rao06} searched for the Ly$\alpha$ absorption associated to {Mg}\,{\sc ii} systems to measure $\Omega_{\rm g}$ at $0.1<z<1.7$. Their results and the measurements from the radio survey of \citet{lah07} are consistent with high-redshift findings, indicating that {H}\,{\sc i} mass density is roughly constant from $z\sim0.1$--5.0. At $z=0$ the 21-cm radio surveys \citep{zwaan05,martin10,braun12} indicate a decrease in $\Omega_{\rm g}$ by at most a factor of 2 and a constant compatible with uncertainties. The observations of 21-cm radio emission of neutral hydrogen are limited to low redshift \citep{zwaan05,lah07}. More distant galaxies are generally too faint for an individual detection at 21cm. Recently, \citet{chang10} used 21-cm maps to compute $\Omega_{\rm g}$ up to $z=0.8$.


The baryons comprise a small fraction of the critical matter-energy density with $\Omega_b = 0.0455\pm0.0028$ \citep{komatsu11}. Most of the baryons resides in the photoionized Ly$\alpha$ forest ($28\pm11$\%) and shocked-heated warm-hot intergalactic medium (WHIM) at T $\approx10^{5-7}$ K ($25\pm8$\%; see \citealt{shull11}). A small fraction of baryons is locked up in stars in galaxies ($7\pm2$\%), groups, and clusters \citep[][and references therein]{shull11}. Current tracers of WHIM include {O}\,{\sc vi} absorption lines, broad Ly$\alpha$ absorbers, and X-ray absorption lines. Another $5\pm3$\% may reside in condensed gas phase in the circumgalactic medium (CGM) within the virial radius of galaxies \citep{shull11}. The cold gas mass density that is the interest of the current study, contribute around $\lesssim$2\% to the baryon fraction at $z=0$ in the {H}\,{\sc i}, {He}\,{\sc i} and H$_2$ phases . From the study of the multiphase mass densities in the present-day Universe, \citet{shull11} found that one third of the baryons are still missing. The $\Omega_{\rm g}$ at $z=0.1$--5 is $\sim$6 and $\sim$12 times smaller than the current $\Omega_{\ast+\rm CGM}$ and $\Omega_{\rm WHIM}$. $\Omega_{\rm g}$ at $z=0.1$--5.0 is 45 times smaller than $\Omega_b$.

   
\begin{table*}
\begin{minipage}{180mm}
\caption{Redshift distribution of the neutral gas mass density, $\Omega_{\rm g}$, in DLAs from \citet{peroux03} and {\it combined sub-DLA} sample expressed as a function of the current critical density. The total amount of {H}\,{\sc i} gas from DLAs $+$ sub-DLAs is given in boldface. In the last column, we provide the {H}\,{\sc i} mass density for DLA $+$ sub-DLA.}      
\label{omegaresults} 
\centering
\setlength{\tabcolsep}{4pt}
\begin{tabular}{ l c c c c c c c c c c c c c c}
\hline
\multicolumn{1}{l}{log $N_{{\rm H}\,{\sc \rm I}}$ range} & \multicolumn{5}{c}{log $N_{{\rm H}\,{\sc \rm I}}>20.3$} & & \multicolumn{4}{c}{19.0 $<$ log $N_{{\rm H}\,{\sc \rm I}}<20.3$} & \textbf{Total}  \\ 
 & & & & $\Omega_{\rm DLA}$ & \# & & & & $\Omega_{\rm sub-DLA}$  & \# & $\mathbf{\Omega_{\rm g}}$ & $\rho_{\rm g}\times10^8$\\           
$z$ range	& $\langle z \rangle$ & $dz$ & $dX$ & $\times10^{-3}$ & DLAs & $\langle z \rangle$ & $dz$ & $dX$ & $\times10^{-3}$ & sub-DLAs & $\times10^{-3}$ & M$_{\odot}$ Mpc$^{-3}$\\
 \hline
1.51--2.00 & 1.84 & 95.6 & 279.4 & $0.83\pm0.24$ & 21 & 1.80 & 29.9 & 87.3 & $0.07\pm0.03$ & 4 & $\mathbf{0.90\pm0.20}$ & $1.23\pm0.33$\\  
 2.00--2.50 & 2.27 & 125.1 & 396.6 & $0.90\pm0.28$ & 26 & 2.26 & 49.2 & 156.8 & $0.09\pm0.03$ & 11 & $\mathbf{0.99\pm0.29}$ & $1.35\pm0.38$\\  
2.50--3.00 & 2.73 & 87.2 & 301.0 & $0.96\pm0.28$ & 18 &  2.76 & 47.1 & 162.8 & $0.17\pm0.04$ & 24 & $\mathbf{1.13\pm0.28}$ & $1.54\pm0.38$\\
 3.00--3.50 & 3.25 & 64.8 & 239.6 & $1.00\pm0.28$ & 18 & 3.21 & 33.7 & 124.5 & $0.16\pm0.05$ & 23 & $\mathbf{1.16\pm0.29}$ & $1.58\pm0.39$ \\
 3.50--4.00 & 3.77 & 49.5 & 194.6 & $0.75\pm0.21$ & 19 & 3.72 & 23.3 & 91.7 & $0.18\pm0.05$ & 19 &  $\mathbf{1.03\pm0.21}$ & $1.40\pm0.29$ \\
 4.00--5.00 & 4.20 & 	23.2 & 95.6 & $0.76\pm0.25$ & 10 & 4.18 & 9.9 & 41.0 & $0.20\pm0.09$ & 8 & $\mathbf{0.96\pm0.25}$ & $1.31\pm0.36$ \\
\hline
\end{tabular}
\end{minipage}
\end{table*}
   
   
\subsubsection {The contribution of sub-DLAs.}
\citet{peroux03b} first derived the contribution of sub-DLAs to the $\Omega_{\rm g}$ as a function of the $N_{{\rm H}\,{\sc \rm I}}$ column density. They showed that by including sub-DLAs, most of the neutral gas mass in the Universe is accounted for. \citet{noterdaeme09} found that sub-DLAs contribute about 20\% (from $\Gamma$-function extrapolation) to 30\% (from double power-law extrapolation) to the total neutral gas mass. While the high column density systems dominate the contribution to $\Omega_{\rm g}$, the relative contribution of each class of absorbers may evolve. In addition, a nearly complete census of neutral gas mass in the Universe can be obtained by including the contribution of sub-DLAs to $\Omega_{\rm g}$. From a direct measurement, we find here that sub-DLAs contribute between 8--20\% to $\Omega_{\rm g}$ with increasing contribution at high redshift. The sub-DLA contribution increases notably at $z\ge3$, indicating that we are observing the assembly of high column density systems from low column density units. Our findings agree with those of \citet{peroux05} and \citet{noterdaeme09,noterdaeme12b}, suggesting no significant evolution of $\Omega_{\rm g}$ over redshift.

      \begin{figure}
   \centering
  {\includegraphics[width=\columnwidth,height=7.7cm,clip=]{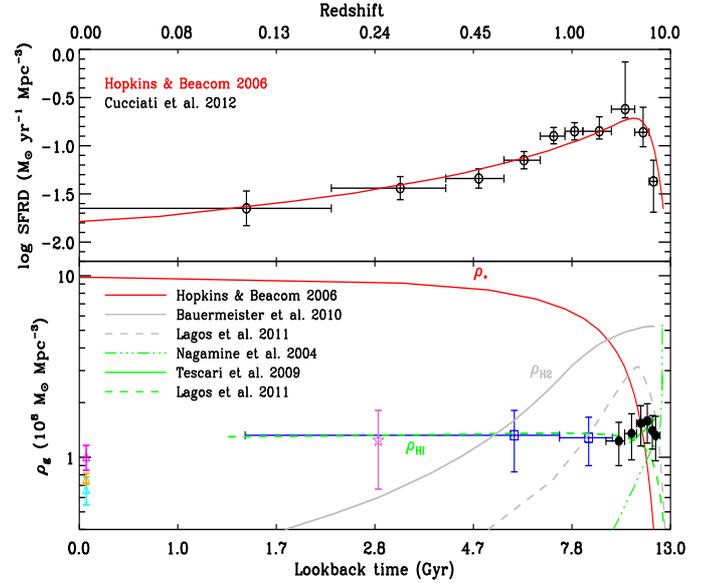}} 
      \caption{{\it Top panel}: Total dust-corrected UV-derived SFR density as a function of lookback time from  the visible multi object-spectrograph (VIMOS)-VLT deep survey (VVDS) Deep $+$ Ultra-Deep sample from \citet{cucciati12} is illustrated as black open circles. The red curve corresponds to the best-fitting parametric form to the SFR density from \citet{hopkins06}. {\it Bottom panel}: The data points have the same meaning as in Fig. \ref{omega}. The green curves represent the prediction of $\rho_{{\rm H}\,{\sc \rm I}}$ from \citet[][green triple dot-dashed line]{nagamine04}, \citet[][green solid line]{tescari09}, and \citet[][green dashed line]{lagos11}. The gray curves correspond to the prediction of $\rho_{\rm H_2}$ from \citet[][solid gray line]{bauermeister10} and \citet[][gray dashed line]{lagos11}. The solid red line illustrates the evolution of the stellar mass density build-up inferred from the star formation history by \citet{hopkins06}.}
      \label{lookbackoml}
   \end{figure}

\subsubsection{Interpretation of the non evolution of $\Omega_{\rm g}$}
Previously, an evolution of $\Omega_{\rm g}$ was expected over cosmic time due to the consumption of neutral gas by star-formation activity. The evolution is also expected because energy is released during the hierarchical build-up of systems from subsystems \citep[e.g.,][]{ledoux98} and/or ejection of gas through galactic winds from the central parts of massive halos into the IGM \citep{fall93}. \citet{prochaska05} and \citet{prochaska09} observed a significant decrease in {H}\,{\sc i} mass between $z=2$--4 from the SDSS surveys. They attributed it to the consumption of neutral gas by star formation activity and/or ejection of gas into the IGM through galactic winds. \citet{zwaan08} and \citet{tescari09} debated the importance of galactic winds for the evolution of $\Omega_{\rm g}$. 

In recent years, new observations have considerably changed the genetal picture of the observable baryons in the Universe \citep[e.g,][]{cole01,fukugita04,danforth08,shull11,noterdaeme12b}. The {H}\,{\sc i} clouds form molecules, and the molecular clouds then cool, fragment, and initiate star formation in galaxies. Therefore DLAs provide the reservoir of neutral gas and serve as a barometer of recent starformation activity. The neutral gas mass density $\rho_{\rm g}=\Omega_{\rm g}\rho_{\rm crit}$ (where $\rho_{\rm crit}=2.78\times 10^{11}$ $h^2$ M$_\odot$ Mpc$^{-3}$) has been measured over most of the age of the Universe (lookback time of about 12 Gyr). The space density of {H}\,{\sc i} appears to evolve surprisingly little from $z=0.1$--5.0. In contrast, the space density of the starformation rate (SFR) in galaxies steadily increases toward $z\simeq2$ by almost an order of magnitude \citep[e.g,][]{cucciati12}. Given the SFR density, it is expected that {H}\,{\sc i} plus H$_2$ gas at high-redshift would be exhausted on timescales of a few Gyr. Fig. \ref{lookbackoml} indicates that there is no evolution of $\rho_{{\rm H}\,{\sc \rm I}}$ over a lookback time of 1.3--12.3 Gyr. This suggests that SFR, ionization of neutral gas, and/or the formation of molecular hydrogen alone cannot explain the non-evolution of the {H}\,{\sc i} space density.

Several surveys have been conducted to study inflows and outflows in distant galaxies via UV absorption lines. Outflows are ubiquitous in almost all distant star-forming galaxies \citep{weiner09,rubin10,steidel10,kacprzak11b,bordoloi11} and their rates aresimilar to or higher than the SFR \citep[e.g.,][]{kornei12}. Recently, cool gas inflows  have been detected in a few galaxies, although with an inflow rate lower than the SFR \citep{sato09,giavalisco11,rubin12,martin12}. Because the outflow rates are generally found to be higher than the SFR, a constant {H}\,{\sc i} space density as we observe seems to indicate a higher accretion on average than observed in the quoted observations.


Introducing a continuous replenishment of gas the account for an appropriate evolution of outflows and gas replenishment rates, i.e., accretion of gas from filaments and streams to the centers of cold dark matter halos, can possibly help in resolving the puzzle of the non-evolution of the space density of {H}\,{\sc i} \citep{hopkins08}. Hydrodynamical simulation indicate that {H}\,{\sc i} gas might be replenished by accretion of matter from the IGM and/or recombination of ionized gas in the walls of supershells \citep{keres05,birnboim07,erb08}.



\section{Conclusions}
We here presented a statistical sample of 122 quasars (1.5 $<$ $z_{\rm em}$ $<$ 5.0) observed with the VLT/UVES. The statistical properties of 27 sub-DLAs along the lines of sight of these quasars were analyzed in combination with the sub-DLA studies of \citet{peroux03b,peroux05} and \citet{omeara07}. Our main findings are summarized as follows:
\begin{itemize}
\item The {\it combined sub-DLA} sample allows one to determine the redshift evolution of the number density and line density. Comparing these with DLAs and LLSs results from the literature, we found that all three classes of absorbers are evolving in the redshift range 1 $<$ $z$ $<$ 5 in the same way.

\item The shape of the column density distribution function down to log $N_{{\rm H}\,{\sc \rm I}}=19.0$ was determined. The redshift evolution of $f_{{\rm H}\,{\sc \rm I}}(N,z)$ was also presented for two redshift bins 1.5 $<$ $z$ $<$ 3.1 and 3.1 $<$ $z$ $<$ 5.0. The shape of $f_{{\rm H}\,{\sc \rm I}}(N,z)$ at high-redshift in the sub-DLA range is steeper, indicating that there are more sub-DLAs at high-redshift than at low-redshift. This evolution suggests that sub-DLAs may merge and/or ionize over cosmic time. 

\item We furthermore used $f_{{\rm H}\,{\sc \rm I}}(N,z)$ to determine the total {H}\,{\sc i} gas mass in the Universe at $1.5<z<5.0$. The complete sample shows that sub-DLAs are significant at all redshifts and their contribution to $\Omega_{\rm g}$ increases from 8--20\% with an increasing fraction at higher redshift.

\item From 0.1 $<$ $z$ $<$ 5.0 {\it no-evolution} of $\Omega_{\rm g}$ was found. In contrast, star formation peaks at $z\sim2.1$. This indicates that star formation aline cannot explain the non-evolution of $\Omega_{\rm g}$. This may suggest that replenishment of gas through accretion of matter from filaments and/or recombination of ionized gas in the walls of supershells may help in sustaining $\Omega_{\rm g}$ as a constant over cosmic time \citep{hopkins08}.
\end{itemize}

\section{Acknowledgments}
This work has been funded within the BINGO! (`history of Baryons: INtergalactic medium/Galaxies cO-evolution') project by the Agence Nationale de la Recherche (ANR) under the allocation ANR-08-BLAN-0316-01. We are thankful to the anonymous referee for his/her constructive comments. We would like to thank the ESO staff for making the UVES Advanced Data Products available to the community.

\addtocounter{table}{-7}
\begin{table}
\begin{minipage}[t]{\columnwidth}
\caption{Complete log of 250 EUADP quasars of which 122 contribute to the statistical EUADP sample. The columns provide the quasar names, emission redshifts, $z_{\rm min}$ (either LLS or S/N ratio threshold of EW$_{\rm rest}=2.5\,\AA$) and $z_{\rm max}$ (3000 km s$^{-1}$ blueward of the Ly$\alpha$ emission line), LLS redshifts, DLA/sub-DLA redshifts, {H}\,{\sc i} column densities of DLAs/sub-DLAs and references for absorbers. The redshift path, $z_{\rm abs}$ and $N_{{\rm H}\,{\sc \rm I}}$ of the absorbers included in our statistical sample are shown in boldface. For the redshift path surveyed, gaps in non-overlapping settings and known DLAs/sub-DLAs were excluded.}
\label{sampletable}
\centering
\setlength{\tabcolsep}{1pt}
\begin{tabular}{@{} l c c c c c c c c c @{}}
\hline
 Quasar &   $z_{\rm em}$ &   $z_{\rm min}$--$z_{\rm max}$ & $z_{LLS}$ &  $z_{\rm abs}$ &  log $N_{{\rm H}\,{\sc \rm I}}$ &  Ref.  \\
 &  & & & & cm$^{-2}$ &  \\
\hline
LBQS 2359-0216B & 2.810 & \textbf{1.542-2.083,} & $\cdots$ & 2.095 & $20.65\pm0.10$  & 1 \\
$\cdots$ & $\cdots$ & \textbf{2.161-2.772} & $\cdots$ & 2.154 & $20.30\pm0.10$  & 1 \\
QSO J0003-2323 & 2.280 & 1.509-2.247 & $\cdots$ & 2.187 & $19.60\pm0.40$ & 2  \\
QSO B0002-422 & 2.760  & 1.509-2.722 & $\cdots$ & $\cdots$ & $\cdots$ &  \\
QSO J0006-6208 & 4.455 & \textbf{3.083-3.188,}  & $\cdots$ & 2.970 & 20.70 & 3 \\
$\cdots$ & $\cdots$ & \textbf{3.216-3.732}, & $\cdots$ & 3.202 & $20.80\pm0.10$ & 4 \\ 
$\cdots$ & $\cdots$ & \textbf{3.798-4.400} & $\cdots$ & 3.776 & 21.00 & 3 \\
$\cdots$ & $\cdots$& $\cdots$ & $\cdots$ & \textbf{4.145} & $\mathbf{19.37\pm0.15}$  & 4 \\
QSO J0008-0958 & 1.950 & \textbf{1.715-1.754,} & $\cdots$ & 1.768 & $20.85\pm0.15$ & 5 \\
$\cdots$ & $\cdots$ & \textbf{1.782-1.921} & $\cdots$ & $\cdots$ & $\cdots$  & \\
QSO J0008-2900 & 2.645 & 1.715-2.609 & $\cdots$ & 2.254 & $20.22\pm0.10$ & 6 \\
QSO J0008-2901 & 2.607 & 1.715-2.571 & $\cdots$ & 2.491 & $19.94\pm0.11$ & 6 \\
QSO B0008+006 & 2.309 &  \textbf{2.093-2.276} & $\cdots$ & $\cdots$ & $\cdots$ &  \\
LBQS 0009-0138 & 1.998 & \textbf{1.715-1.968}  & $\cdots$ & 1.386 & $20.26\pm0.02$ & 7 \\
LBQS 0010-0012 & 2.145 & \textbf{1.509-2.009,}  & $\cdots$ & 2.025 & $20.95\pm0.10$ & 1 \\
$\cdots$ & $\cdots$ & \textbf{2.041-2.114} & $\cdots$ & $\cdots$ & $\cdots$  & \\
LBQS 0013-0029 & 2.087 & \textbf{1.512-1.958,}  & $\cdots$ & 1.968 & $\lesssim19.43$ & 8 \\
$\cdots$ & $\cdots$ & \textbf{1.988-2.056} & $\cdots$ & 1.973 & $20.83\pm0.05$  & 1 \\
LBQS 0018+0026 & 1.244 & 0 & $\cdots$ & $0.520$ & $19.54\pm0.03$ & 7 \\
 $\cdots$ &  $\cdots$ &  $\cdots$ &  $\cdots$ & 0.940 & $19.38\pm0.15$ & 7 \\
QSO B0027-1836 & 2.550 & \textbf{1.553-1.862,}  & 2.400 & 2.402 & $21.75\pm0.10$ & 9  \\
$\cdots$ & $\cdots$ & \textbf{1.877-2.183} & $\cdots$ & $\cdots$ & $\cdots$  & \\
J004054.7-091526 & 4.976 & \textbf{4.534-4.729,} & $\cdots$ & \textbf{4.538} & $\mathbf{20.20\pm0.09}$ & 6 \\
$\cdots$ & $\cdots$ & \textbf{4.751-4.916} & $\cdots$ & 4.740 & $20.39\pm0.11$  & 6 \\
QSO J0041-4936 & 3.240 & \textbf{1.706-2.236,} & $\cdots$ & 2.248 & $20.46\pm0.13$ & 6 \\
$\cdots$ & $\cdots$ & \textbf{2.260-2.718,} & $\cdots$ & $\cdots$ & $\cdots$  & \\
$\cdots$ & $\cdots$ & \textbf{2.800-3.198} & $\cdots$ & $\cdots$ & $\cdots$ &  \\
QSO B0042-2450 & 0.807 & 0 & $\cdots$ & $\cdots$ & $\cdots$  &\\
QSO B0039-407 & 2.478 & \textbf{1.710-2.443} & $\cdots$ & $\cdots$ & $\cdots$ &  \\
QSO B0039-3354 & 2.480 & \textbf{1.710-2.212,}  & $\cdots$ & 2.224 & $20.60\pm0.10$ & 10 \\
$\cdots$ & $\cdots$ & \textbf{2.236-2.445} & $\cdots$ & $\cdots$ & $\cdots$  & \\
LBQS 0041-2638 & 3.053 & 2.090-3.012 & $\cdots$ & $\cdots$ & $\cdots$ &  \\
LBQS 0041-2707 & 2.786  & 2.090-2.748 & $\cdots$ & $\cdots$ & $\cdots$ &  \\
QSO B0042-2656 & 3.358 & 2.217-3.096 & 3.288 & $\cdots$ & $\cdots$ &  \\
LBQS 0042-2930 & 2.388 & \textbf{1.706-1.798,} & $\cdots$ & 1.809 & $20.40\pm0.10$ & 10 \\
$\cdots$ & $\cdots$ & \textbf{1.820-1.926,} & $\cdots$ & 1.936 & $20.50\pm0.10$  & 10 \\
$\cdots$ & $\cdots$ & \textbf{1.946-2.354} & $\cdots$ & $\cdots$ & $\cdots$ &  \\
LBQS 0042-2657 & 2.898 & 2.090-2.859 & $\cdots$ & $\cdots$ & $\cdots$ &  \\
J004612.2-293110 & 1.675 & \textbf{1.550-1.648} & $\cdots$ & $\cdots$ & $\cdots$ &  \\
LBQS 0045-2606 & 1.242 & 0 & $\cdots$ & $\cdots$ & $\cdots$  &\\
QSO B0045-260 &0.486 & 0 & $\cdots$ & $\cdots$ & $\cdots$  &\\
QSO B0046-2616  & 1.410 & 0 & $\cdots$ & $\cdots$ & $\cdots$  &\\
LBQS 0047-2538 & 1.969  & \textbf{1.706-1.939} & $\cdots$ & $\cdots$ & $\cdots$ &  \\
LBQS 0048-2545 & 2.082 & \textbf{1.710-2.051} & $\cdots$ & $\cdots$ & $\cdots$ &  \\
QSO B0018-2608 & 2.249 & \textbf{1.710-2.217} & $\cdots$ & $\cdots$ & $\cdots$ &  \\
LBQS 0049-2535 & 1.528 & 0 & $\cdots$ & $\cdots$ & $\cdots$  &\\
LBQS 0051-2605 & 0.624& 0 & $\cdots$ & $\cdots$ & $\cdots$  &\\
QSO B0055-26 & 3.662 & 1.533-3.615 & 2.377 & $\cdots$ & $\cdots$ &   \\
QSO B0058-292 & 3.093 & \textbf{1.779-2.090,} & 2.695 & 2.671  & $21.10\pm0.10$ & 1 \\
$\cdots$ & $\cdots$ & \textbf{2.105-2.656,} & $\cdots$ & $\cdots$ & $\cdots$  & \\
$\cdots$ & $\cdots$ & \textbf{2.686-3.052} & $\cdots$ & $\cdots$ & $\cdots$  & \\
LBQS 0059-2735 & 1.595 & \textbf{1.550-1.569} & $\cdots$ & $\cdots$ & $\cdots$ &  \\
QSO B0100+1300 & 2.686 & 1.706-2.649 & $\cdots$ & 2.309 & $21.35\pm0.08$ & 1 \\
QSO J0105-1846 & 3.037 & \textbf{1.957-2.327} & 2.941 & 2.370 & $21.00\pm0.08$ & 10 \\
$\cdots$ & $\cdots$ & \textbf{2.410-3.000} & $\cdots$ & \textbf{2.926} & $\mathbf{20.00\pm0.10}$ & 10 \\
QSO B0102-2931 & 2.220 & 1.768-2.189 & $\cdots$ & $\cdots$ & $\cdots$ &  \\
QSO B0103 -260 & 3.365 & 2.426-3.321 & $\cdots$ & $\cdots$ & $\cdots$ &  \\ 
QSO B0109-353 & 2.406 & 1.513-2.372 & $\cdots$ & $\cdots$ & $\cdots$ &  \\
QSO B0112-30 & 2.985 & \textbf{2.930-2.945} & $\cdots$ & 2.418 & $20.50\pm0.08$ & 1 \\
$\cdots$ & $\cdots$ & $\cdots$ & $\cdots$ & 2.702 & $20.30\pm0.10$ & 1 \\
QSO B0117+031 & 1.609 & 0 & $\cdots$ & $\cdots$ & $\cdots$  &\\
QSO J0123-0058 & 1.550 & 0 & $\cdots$ & 1.409 & $20.08\pm0.09$ & 11 \\
 QSO J0124+0044 & 3.834 & \textbf{2.426-3.786} & $\cdots$ & \textbf{2.988} & $\mathbf{19.18\pm0.10}$ & 4 \\
\hline
\end{tabular}
\end{minipage}
\end{table}

\addtocounter{table}{-1}
\begin{table}
\begin{minipage}[t]{\columnwidth}
\caption{continued.}
\centering
\setlength{\tabcolsep}{1pt}
\begin{tabular}{@{} l c c c c c c c c c @{}}
\hline
 Quasar &   $z_{\rm em}$ &  $z_{\rm min}$--$z_{\rm max}$ & $z_{LLS}$ &   $z_{\rm abs}$ &  log $N_{{\rm H}\,{\sc \rm I}}$ &  Ref.   \\
 &  & & & & cm$^{-2}$ &  \\
\hline
 $\cdots$ & $\cdots$ & $\cdots$ & $\cdots$ & \textbf{3.078} & $\mathbf{20.21\pm0.10}$ & 4 \\
QSO B0122-379 & 2.190 & 1.510-2.158 & $\cdots$ & $\cdots$ & $\cdots$ &  \\ 
QSO B0122-005 & 2.278 & \textbf{1.715-1.749,} & $\cdots$ & 1.761 & $20.78\pm0.07$ &  12 \\
$\cdots$ & $\cdots$ & \textbf{1.773-2.245} & $\cdots$ & \textbf{2.010} & $\mathbf{20.04\pm0.07}$ &  12 \\
QSO B0128-2150 & 1.900 & \textbf{1.505-1.838} & $\cdots$ & 1.857 & $20.21\pm0.09$ & 6 \\
QSO B0130-403 & 3.023 & \textbf{1.715-2.983} & $\cdots$ & $\cdots$ & $\cdots$ &  \\
QSO J0133+0400 & 4.154 & \textbf{2.858-3.347,} & 4.140 & \textbf{3.139} & $\mathbf{19.01\pm0.10}$ & 4 \\
$\cdots$& $\cdots$& \textbf{3.350-3.676,} & $\cdots$ & 3.692 & $20.68\pm0.15$ & 4 \\
$\cdots$& $\cdots$& \textbf{3.707-3.765,} & $\cdots$ & 3.773 & $20.42\pm0.10$ & 4 \\
$\cdots$& $\cdots$& \textbf{3.786-4.102} & $\cdots$ & \textbf{3.995} & $\mathbf{19.94\pm0.15}$ & 4 \\
$\cdots$& $\cdots$& $\cdots$ & $\cdots$ & \textbf{3.999} & $\mathbf{19.16\pm0.15}$ & 4 \\
$\cdots$& $\cdots$& $\cdots$ & $\cdots$ & \textbf{4.021} & $\mathbf{19.09\pm0.15}$ & 4 \\
QSO J0134+0051 & 1.522 & 0 & $\cdots$ & 0.842 & $19.93\pm0.13$  & 11 \\
QSO B0135-42	 & 3.970 & \textbf{2.498-3.920} & 3.662 & \textbf{3.101} & $\mathbf{19.81\pm0.10}$ & 4 \\
$\cdots$& $\cdots$ & $\cdots$ & $\cdots$ & \textbf{3.665} & $\mathbf{19.11\pm0.10}$ & 4 \\
QSO J0138-0005 & 1.340 & 0 & $\cdots$ & 0.782 & $19.81\pm0.09$ & 11 \\
QSO J0139-0824 & 3.016 & 1.715-2.662 & $\cdots$ & 2.677 & $20.70\pm0.15$ & 13  \\
QSO J0143-3917 & 1.807 & 1.509-1.779 & $\cdots$ & $\cdots$ & $\cdots$ &  \\
QSO J0153+0009 & 0.837 & 0 & $\cdots$ & 0.771 & $19.70\pm0.09$ & 11 \\
QSO J0153-4311 & 2.789  & 1.588-2.751 & 2.451 & $\cdots$ & $\cdots$ &  \\
QSO J0157-0048 & 1.545 & 0 & $\cdots$ & 1.416 & $19.90\pm0.07$ & 7 \\
QSO B0201+113 & 3.610 & \textbf{2.309-2.661,} & 3.400 & 3.385 & $21.26\pm0.08$ & 14 \\
$\cdots$ & $\cdots$ & \textbf{2.708-3.097} & $\cdots$ & $\cdots$ & $\cdots$  & \\
QSO J0209+0517 & 4.174 & \textbf{2.767-3.658,} & $\cdots$ & 3.666 & $20.47\pm0.10$ & 4 \\
  $\cdots$ & $\cdots$ & \textbf{3.675-3.885,} & $\cdots$ & \textbf{3.707} & $\mathbf{19.24\pm0.10}$ & 4 \\
  $\cdots$ & $\cdots$ & \textbf{3.872-4.122} & $\cdots$ & 3.863 & $20.43\pm0.15$ & 4 \\
J021741.8-370100 & 2.910 & \textbf{1.715-2.417,} & $\cdots$ & 2.429 & $20.62\pm0.08$ & 6 \\
$\cdots$ & $\cdots$ & \textbf{2.441-2.503,} & $\cdots$ & 2.514 & $20.46\pm0.09$ & 6 \\
$\cdots$ & $\cdots$ & \textbf{2.525-2.718} & $\cdots$ & $\cdots$ & $\cdots$  & \\
QSO J0217+0144 & 1.715 & 0 & $\cdots$ & $1.345$ & $19.89\pm0.09$ & 11 \\
QSO B0216+0803 & 2.996  & \textbf{1.521-2.280,} & $\cdots$ & \textbf{1.769} & $\mathbf{20.20\pm0.10}$ & 7 \\
$\cdots$ & $\cdots$ & \textbf{2.306-2.718,} & $\cdots$ & 2.293 & $20.45\pm0.16$ & 15 \\
$\cdots$ & $\cdots$ & \textbf{2.930-2.956} & $\cdots$ & $\cdots$ & $\cdots$  & \\
QSO B0227-369 & 2.115 & \textbf{1.715-2.084} & $\cdots$ & $\cdots$ & $\cdots$ &  \\
QSO B0237-2322 & 2.225 & 1.509-2.193 & $\cdots$ & 1.365 & $19.30\pm0.30$  & 16  \\
$\cdots$ & $\cdots$ & $\cdots$ & $\cdots$ & 1.672 & $19.65\pm0.10$  & 10  \\
QSO J0242+0049 & 2.071 & \textbf{1.715-2.040} & $\cdots$ & $\cdots$ & $\cdots$ &   \\
QSO J0243-0550 & 1.805 & \textbf{1.715-1.777} & $\cdots$ & $\cdots$ & $\cdots$ &   \\
QSO B0241-01  & 4.053 & \textbf{2.767-4.002} & 4.023 & $\cdots$ & $\cdots$ &  \\
QSO B0244-1249 & 2.201 & \textbf{1.706-2.169} & $\cdots$ & \textbf{1.863} & $\mathbf{19.48\pm0.18}$ & 12 \\
QSO B0253+0058 & 1.346 & 0 & $\cdots$ & 0.725 & $20.70\pm0.17$ & 11 \\
QSO B0254-404 & 2.280 & \textbf{1.708-2.035,} & $\cdots$ & 2.046 & $20.45\pm0.08$ & 10 \\
$\cdots$ & $\cdots$ & \textbf{2.057-2.247} & $\cdots$ & $\cdots$ & $\cdots$  & \\
QSO J0300+0048 & 0.314 & 0 & $\cdots$ & $\cdots$ & $\cdots$  &\\
J030640.8-301032 & 2.096 & 1.515-2.065 & $\cdots$ & $\cdots$ & $\cdots$ &  \\
J030643.7-301107 & 2.130 & 1.517-2.099 & $\cdots$ & $\cdots$ & $\cdots$ &  \\
QSO B0307-195A & 2.144 & 1.509-2.112 & $\cdots$ & $\cdots$ & $\cdots$ &  \\
QSO B0307-195B & 2.122 & 1.521-2.091 & $\cdots$ & 1.788 & $19.00\pm0.10$ & 17 \\
J031856.6-060038 & 1.927 & \textbf{1.550-1.899} & $\cdots$ &  $\cdots$ & $\cdots$ &   \\
QSO B0329-385 & 2.423 & 1.520-2.389 & $\cdots$ & $\cdots$ & $\cdots$  &  \\
QSO B0329-2534  & 2.736 & 1.605-2.699 &  2.470 & $\cdots$ & $\cdots$ &  \\
QSO J0332-4455 &  2.679 & \textbf{1.754-2.402,} & 2.671 & 2.411 & $20.15\pm0.07$ & 10 \\
$\cdots$ & $\cdots$ & \textbf{2.420-2.640} & $\cdots$ & 2.656 & $19.82\pm0.05$ & 10 \\
QSO B0335-122 & 3.442  &  \textbf{2.266-3.097} & 3.353 & 3.178 & $20.80\pm0.07$ & 18 \\
QSO J0338+0021 & 5.020 &  \textbf{4.529-4.962} & $\cdots$ & 4.060 & $20.40$ & 3 \\
QSO J0338-0005 & 3.049 & 1.552-2.183 & $\cdots$ & 2.230 & $21.05\pm0.25$ & 19 \\
$\cdots$ & $\cdots$ & $\cdots$ & $\cdots$ & 2.747 & $20.17\pm0.47$ & 19 \\
QSO B0336-017 & 3.197 &  \textbf{2.090-2.990} & $\cdots$ & 3.062 & $21.20\pm0.09$ & 18 \\
QSO B0347-383 & 3.222 &  \textbf{2.017-2.391,} & 3.043 & 3.025 & $20.73\pm0.05$ & 20 \\
$\cdots$ & $\cdots$ & \textbf{2.397-3.014,} & $\cdots$ & $\cdots$ & $\cdots$  & \\
$\cdots$ & $\cdots$ & \textbf{3.039-3.181} & $\cdots$ & $\cdots$ & $\cdots$  & \\
QSO B0347-2111 & 2.944 & 0 & $\cdots$ & 1.947 & $20.30\pm0.10$ & 21 \\
J035320.2-231418 & 1.911 & \textbf{1.715-1.882} & $\cdots$ & $\cdots$ & $\cdots$ &  \\
QSO J0354-2724 & 2.823 & \textbf{1.715-2.714} & $\cdots$ & 1.405 & $20.18\pm0.15$ & 7  \\
J040114.0-395132 & 1.507 & 0 & $\cdots$ & $\cdots$ & $\cdots$  &\\
QSO J0403-1703 & 4.227 &  \textbf{2.932-3.736,} & $\cdots$ & $\cdots$ & $\cdots$ & \\
$\cdots$ & $\cdots$ & \textbf{3.801-4.174} & $\cdots$ & $\cdots$ & $\cdots$ &\\
QSO J0407-4410 & 3.020 & \textbf{1.736-1.899,} & 2.647 & 1.913 & $20.80\pm0.10$ & 10 \\
$\cdots$ & $\cdots$ & \textbf{1.927-1.974,} & $\cdots$ & 2.551 & $21.15\pm0.15$ & 10 \\
$\cdots$ & $\cdots$ & \textbf{2.048-2.526,} & $\cdots$ & 2.595 & $21.05\pm0.10$ & 10 \\
$\cdots$ & $\cdots$ & \textbf{2.566-2.579} & $\cdots$ & 2.621 & $20.45\pm0.10$ & 10 \\
$\cdots$ & $\cdots$ & \textbf{2.612-2.616,} & $\cdots$ & $\cdots$ & $\cdots$ & \\
$\cdots$ & $\cdots$ & \textbf{2.633-2.980} & $\cdots$ & $\cdots$ & $\cdots$ & \\
\hline
\end{tabular}
\end{minipage}
\end{table}

\addtocounter{table}{-1}
\begin{table}
\begin{minipage}[t]{\columnwidth}
\caption{continued.}
\centering
\setlength{\tabcolsep}{1pt}
\begin{tabular}{@{} l c c c c c c c c c @{}}
\hline
 Quasar &   $z_{\rm em}$ &  $z_{\rm min}$--$z_{\rm max}$ & $z_{LLS}$ &   $z_{\rm abs}$ & log $N_{{\rm H}\,{\sc \rm I}}$ &  Ref. \\
 &  & & & & cm$^{-2}$ &  \\
\hline
QSO J0422-3844 & 3.123 & 2.093-3.120 & 3.115 & 3.082 & $19.37\pm0.02$ & 22 \\
QSO J0427-1302 & 2.166 & \textbf{1.710-2.135} & $\cdots$ & 1.408 & $19.04\pm0.04$ & 7 \\
$\cdots$ & $\cdots$ & $\cdots$ & $\cdots$ & 1.562 & $19.35\pm0.10$ & 23 \\
QSO J0430-4855 & 1.940  & 1.517-1.911 & $\cdots$ & $\cdots$ & $\cdots$ &  \\
QSO B0432-440 & 2.649 & \textbf{1.715-2.283,} & $\cdots$ & 2.297 & $20.78\pm0.10$ & 21 \\
$\cdots$ & $\cdots$ & \textbf{2.311-2.613} & $\cdots$ & $\cdots$ & $\cdots$  & \\
QSO B0438-43	 & 2.852 & \textbf{1.551-2.333,} & $\cdots$ & 2.347 & $20.78\pm0.12$ & 21 \\
$\cdots$ & $\cdots$ & \textbf{2.361-2.815} & $\cdots$ & $\cdots$ & $\cdots$  & \\
QSO J0441-4313 & 0.593 & 0 & $\cdots$ & $\cdots$ & $\cdots$  &\\
QSO B0449-1645 & 2.679 & \textbf{1.708-2.643} & $\cdots$ & 1.007 & $20.98\pm0.07$ & 11 \\
QSO B0450-1310B & 2.250 & \textbf{1.517-2.055,} & $\cdots$ & 2.067 & $20.50\pm0.07$ & 10 \\
$\cdots$ & $\cdots$ & \textbf{2.079-2.187} & $\cdots$ & $\cdots$ & $\cdots$  & \\
QSO J0455-4216 & 2.661 & 1.514-2.624 & $\cdots$ & $\cdots$ & $\cdots$ &  \\
PKS 0454-220 & 0.534 & 0 & $\cdots$ & 0.474 & $19.45\pm0.03$ & 11 \\
4C-02.19	 & 2.286 & 1.550-2.253 & $\cdots$ & 2.040 & $21.70\pm0.10$ & 10 \\
QSO B0512-3329 & 1.569 & 0 & $\cdots$ & $0.931$ & $20.49\pm0.08$ & 24 \\
QSO B0515-4414  & 1.713 & 1.517-1.686 & $\cdots$ & 1.151 & $19.88\pm0.05$ & 25 \\
J051939.8-364613 & 1.349 & 0 & $\cdots$ & $\cdots$ & $\cdots$  &\\
QSO B0528-2505 & 2.765 & \textbf{1.889-2.128,} & 2.827 & 2.141 & $20.95\pm0.05$ & 26 \\
$\cdots$ & $\cdots$ & \textbf{2.154-2.714} & $\cdots$ & 2.811 & $21.35\pm0.07$ & 27 \\
QSO J0530+13	 & 2.070 & \textbf{1.861-2.039} & $\cdots$ & $\cdots$ & $\cdots$ &  \\
QSO B0551-36	 &  2.318  & \textbf{1.517-1.949,} & $\cdots$ & 1.962 & $20.70\pm0.08$ & 20 \\
$\cdots$ & $\cdots$ & \textbf{1.975-2.285} & $\cdots$ & $\cdots$ & $\cdots$  & \\
J060008.1-504036	& 3.130 & \textbf{2.063-2.138,} & 3.080 & 2.149 & $20.40\pm0.12$ & 6 \\
$\cdots$ & $\cdots$ & \textbf{2.160-3.089} & $\cdots$ & $\cdots$ & $\cdots$ &  \\
QSO B0606-2219 & 	1.926 & \textbf{1.750-1.898} & $\cdots$ &$\cdots$ & \\
QSO B0642-5038 & 3.090 & \textbf{1.757-2.644,} & 2.670 & 2.659 & $20.95\pm0.08$ & 10 \\
$\cdots$ & $\cdots$ & \textbf{2.932-3.050} & $\cdots$ & $\cdots$ & $\cdots$  & \\
QSO B0736+01 & 0.191 & 0 & $\cdots$ & $\cdots$ & $\cdots$  &\\
QSO B0810+2554 & 1.510 & 0 & $\cdots$ & $\cdots$ & $\cdots$  &\\
QSO B0827+2421 & 0.939 & 0 & $\cdots$ & $0.518$ & $20.30\pm0.04$ & 11 \\
QSO B0841+129 & 2.495 & \textbf{1.707-1.846,} & $\cdots$ & 1.864 & $21.00\pm0.10$ & 10 \\
$\cdots$ & $\cdots$ & \textbf{1.881-2.362,} & $\cdots$ & 2.375 & $21.05\pm0.10$ & 10 \\
$\cdots$ & $\cdots$ & \textbf{2.387-2.466,} & $\cdots$ & 2.476 & $20.80\pm0.10$ & 10 \\
$\cdots$ & $\cdots$ & $\cdots$ & $\cdots$ & 2.506 & $19.00\pm0.15$  & 28 \\
QSO 0908+0603 & 2.793 & 1.715-2.720 & $\cdots$ & $\cdots$ & $\cdots$ & \\
QSO B0913+0715 & 2.785 & \textbf{1.726-2.608,} & 2.629 & 2.618 & $20.35\pm0.10$ & 20 \\
$\cdots$ & $\cdots$ & \textbf{2.628-2.747} & $\cdots$ & $\cdots$ & $\cdots$  & \\
QSO B0919-260 & 2.299 & \textbf{1.715-2.266} & $\cdots$ & $\cdots$ & $\cdots$ &  \\
QSO B0926-0201 & 1.661 & 0 & $\cdots$ & $\cdots$ & $\cdots$  &\\
QSO B0933-333 & 2.910 & \textbf{2.093-2.670} & $\cdots$ & 2.682 & $20.50\pm0.10$ & 20 \\
$\cdots$ & $\cdots$ & \textbf{2.694-2.871} & $\cdots$ & $\cdots$ & $\cdots$  & \\
QSO B0951-0450 & 4.369 & \textbf{2.936-3.846,} & $\cdots$ & \textbf{3.235} & $\mathbf{20.25\pm0.10}$ & 29 \\
$\cdots$ & $\cdots$ & \textbf{3.870-4.158} & $\cdots$ & 3.858 & $20.60\pm0.10$  & 3 \\
$\cdots$ & $\cdots$ & $\cdots$ & $\cdots$ & 4.203 & $20.55\pm0.10$ & 27 \\
QSO B0952+179 & 1.472 & 0 & $\cdots$ & $0.238$ & $21.32\pm0.05$ & 18 \\
QSO B0952-0115 & 4.426  & 3.034-4.370 & 4.203 & 3.476 & $20.04\pm0.07$ & 6  \\
$\cdots$ & $\cdots$ & $\cdots$ & $\cdots$ & 4.024 & $20.55\pm0.10$ & 26 \\
QSO B1005-333 & 1.837 & \textbf{1.731-1.809} & $\cdots$ & $\cdots$ & $\cdots$ &  \\
QSO J1009-0026 & 1.241 & 0 & $\cdots$ & $0.840$ & $20.20\pm0.06$ & 11 \\
$\cdots$ & $\cdots$ & $\cdots$ & $\cdots$ & 0.880 & $19.48\pm0.08$ & 11 \\
LBQS 1026-0045B & 1.530 & 0 & $\cdots$ & $0.632$ & $19.95\pm0.07$ & 7 \\
$\cdots$ & $\cdots$ & $\cdots$ & $\cdots$ & 0.709 & $20.04\pm0.06$ & 7 \\
QSO B1027+0540 & 6.311& 5.581-6.238 & $\cdots$ & $\cdots$ & $\cdots$ &  \\
Q1036-272 & 3.090  & \textbf{2.096-2.779,} & $\cdots$ & 2.792 & $20.65\pm0.13$ & 6 \\
$\cdots$ & $\cdots$ & \textbf{2.805-3.050} & $\cdots$ & $\cdots$ & $\cdots$  & \\
QSO B1036-2257 & 3.130 & \textbf{1.846-2.179,} & 2.790 & \textbf{2.533} & $\mathbf{19.30\pm0.10}$ & 6 \\
$\cdots$ & $\cdots$ & \textbf{2.194-2.714,} & $\cdots$ & 2.777 & $20.93\pm0.05$ & 10 \\
$\cdots$ & $\cdots$ & \textbf{2.793-3.091} & $\cdots$ & $\cdots$ & $\cdots$ &  \\
QSO J1039-2719 & 2.193 & \textbf{1.706-2.132,} & $\cdots$ & 2.139 & $19.90\pm0.05$ & 10 \\
$\cdots$ & $\cdots$ & \textbf{2.146-2.163} & $\cdots$ & $\cdots$ & $\cdots$  & \\
QSO B1038-2712 & 2.331 & \textbf{1.505-2.298} & $\cdots$ & $\cdots$ & $\cdots$ &  \\
QSO B1036-268 & 2.460  & \textbf{1.707-2.425} & $\cdots$ & \textbf{2.235} & $\mathbf{19.96\pm0.09}$ & 6 \\
QSO J1044-0125 & 5.740 & 0 & $\cdots$ & $\cdots$ & $\cdots$  &\\
J104540.7-101813 & 1.261 & 0 & $\cdots$ & $\cdots$ & $\cdots$  &\\
J104642.9+053107 & 2.682 & 1.517-2.182 & $\cdots$ & $\cdots$ & $\cdots$ &  \\
QSO B1044+059 & 1.226 & 0 & $\cdots$ & $\cdots$ & $\cdots$  &\\
QSO B1044+056 & 1.306 & 0 & $\cdots$ & $\cdots$ & $\cdots$  &\\
QSO B1045+056 & 1.230 & 0 & $\cdots$ & $0.951$ & $19.28\pm0.02$ & 11 \\
QSO B1052-0004 & 1.021 & 0 & $\cdots$ & $\cdots$ & $\cdots$  &\\
QSO B1055-301 & 2.523 & \textbf{1.552-1.885,} & $\cdots$ & 1.904 & $21.54\pm0.10$ & 10 \\
$\cdots$ & $\cdots$ & \textbf{1.923-2.488} & $\cdots$ & $\cdots$ & $\cdots$  & \\
QSO B1101-26	 & 2.145 & 2.109-2.117 & $\cdots$ & 1.838 & $19.50\pm0.05$ & 15 \\
\hline
\end{tabular}
\end{minipage}
\end{table}

\addtocounter{table}{-1}
\begin{table}
\begin{minipage}[t]{\columnwidth}
\caption{continued.}
\centering
\setlength{\tabcolsep}{1pt}
\begin{tabular}{@{} l c c c c c c c c c @{}}
\hline
 Quasar &   $z_{\rm em}$ &  $z_{\rm min}$--$z_{\rm max}$ & $z_{LLS}$ &   $z_{\rm abs}$ & log $N_{{\rm H}\,{\sc \rm I}}$ &  Ref. \\
 &  & & & &  cm$^{-2}$ &  \\
\hline
QSO B1104-181 & 2.319 & 1.517-2.286 & $\cdots$ & 1.661 & $20.85\pm0.01$ & 30 \\
QSO J1107+0048 & 1.392 & 0 & $\cdots$ & 0.740 & $21.00\pm0.04$ & 11 \\
QSO B1108-07	 & 3.922 & \textbf{2.619-3.598,} & 3.824 & \textbf{3.482} & $\mathbf{19.95\pm0.07}$ & 1  \\
$\cdots$ & $\cdots$ & \textbf{3.618-3.874} & $\cdots$ & 3.608 & $20.37\pm0.07$ & 1 \\
QSO J1113-1533 & 3.370 &  \textbf{2.241-2.574,} & 3.295 & 3.265 & $21.30\pm0.05$ & 10 \\
$\cdots$ & $\cdots$ & \textbf{2.615-3.248,} & $\cdots$ & $\cdots$ & $\cdots$  & \\
$\cdots$ & $\cdots$ & \textbf{3.282-3.326} & $\cdots$ & $\cdots$ & $\cdots$ & \\
QSO B1114-220 & 2.282 & 0 & $\cdots$ & $\cdots$ & $\cdots$ & \\
QSO B1114-0822 & 4.495 & \textbf{3.180-3.734,} & $\cdots$ & 4.258 & $20.02\pm0.12$ & 6 \\
$\cdots$ & $\cdots$ & \textbf{3.804-4.246,} & $\cdots$ & $\cdots$ & $\cdots$ \\
$\cdots$ & $\cdots$ & \textbf{4.270-4.302} & $\cdots$ & $\cdots$ & $\cdots$  & \\
QSO B1122-168 & 2.400  & 1.706-2.368 & $\cdots$ & 0.682 &  $20.45\pm0.05$ & 31 \\
J112910.9-231628 & 1.019 & 0 & $\cdots$ & $\cdots$ & $\cdots$  &\\
QSO J1142+2654 & 2.630 & 2.095-2.594 & $\cdots$ & $\cdots$ & $\cdots$ & \\
QSO B1145-676 & 0.210 & 0 & $\cdots$ & $\cdots$ & $\cdots$  &\\
QSO B1151+068 & 2.762 & \textbf{1.517-1.757,} & $\cdots$ & 1.775 & $21.30\pm0.08$ & 24 \\
$\cdots$ & $\cdots$ & \textbf{1.793-2.724} & $\cdots$ & $\cdots$ & $\cdots$  & \\
J115538.6+053050 & 3.475 & \textbf{2.266-2.597,} & 3.350 & 2.608 & $20.37\pm0.11$ & 6 \\
$\cdots$ & $\cdots$ & \textbf{2.619-3.258} & $\cdots$ & 3.327 & $21.00\pm0.10$ & 6 \\
QSO J1159+1337 & 3.984 & \textbf{2.932-3.736,} & $\cdots$ & \textbf{3.726} & $\mathbf{20.00\pm0.10}$ & 29 \\
$\cdots$ & $\cdots$ & \textbf{3.800-3.935} & $\cdots$ & $\cdots$ & $\cdots$ \\
QSO B1158-1842 & 2.453  & 1.509-2.418 & $\cdots$ & $\cdots$ & $\cdots$ & \\
QSO B1202-074 & 4.695  & 3.129-4.635 & 4.500 & 4.383 & $20.60\pm0.14$ & 26 \\
J120550.2+020131 & 2.134  & \textbf{1.778-2.102} & $\cdots$ & 1.747 & $20.40\pm0.10$ & 10 \\
QSO B1209+0919 & 3.292 & \textbf{2.137-2.563,} & 3.182 & 2.584 & $21.40\pm0.10$ & 20 \\
$\cdots$ & $\cdots$ & \textbf{2.599-3.251} & $\cdots$ & $\cdots$ & $\cdots$  & \\
LBQS 1209+1046 & 2.193  & \textbf{1.706-2.161} & $\cdots$ & 0.633 & 20.30 & 32 \\
LBQS 1210+1731 & 2.543 & \textbf{1.517-1.878,} & $\cdots$ & 1.892 & $20.70\pm0.08$ & 10 \\
$\cdots$ & $\cdots$ & \textbf{1.906-2.187} & $\cdots$ & $\cdots$ & $\cdots$  & \\
QSO J1215+3309 & 0.616 & 0 & $\cdots$ & $\cdots$ & $\cdots$  &\\
QSO B1220-1800 & 2.160 & \textbf{1.698-2.102} & $\cdots$ & 2.112 & $20.12\pm0.07$ & 10 \\
$\cdots$ & $\cdots$ & \textbf{2.122-2.130} & $\cdots$ & $\cdots$ & $\cdots$  & \\
LBQS 1223+1753 & 2.940  & \textbf{2.044-2.446,} & 2.941 & 2.466 & $21.40\pm0.10$ & 10 \\
$\cdots$ & $\cdots$ & \textbf{2.488-2.901} & $\cdots$ & \textbf{2.557} & $\mathbf{19.32\pm0.15}$ & 33 \\
QSO B1228-113 & 3.528  & \textbf{1.715-2.180,} & $\cdots$ & 2.193 & $20.60\pm0.10$ & 21 \\
$\cdots$ & $\cdots$ & \textbf{2.206-2.718} & $\cdots$ & $\cdots$ & $\cdots$  & \\
$\cdots$ & $\cdots$ & \textbf{2.932-3.485} & $\cdots$ & $\cdots$ & $\cdots$ \\
QSO B1230-101 & 2.394  & \textbf{1.805-1.919,} & $\cdots$ & 1.931 & $20.48\pm0.10$ & 21 \\
$\cdots$ & $\cdots$ & \textbf{1.943-2.361} & $\cdots$ & $\cdots$ & $\cdots$  & \\
LBQS 1232+0815 & 2.570 & \textbf{1.704-2.327,} & $\cdots$ & \textbf{1.720} & $\mathbf{19.48\pm0.13}$ & 6  \\
$\cdots$ & $\cdots$ & \textbf{2.350-2.535} & $\cdots$ & 2.334 & $20.90\pm0.08$ & 34 \\
LBQS 1242+0006 & 2.084  & \textbf{1.706-1.812,} & $\cdots$ & 1.824 & $20.45\pm0.10$ & 10 \\
$\cdots$ & $\cdots$ & \textbf{1.836-2.053} & $\cdots$ & $\cdots$ & $\cdots$  & \\
QSO J1246-0730 & 1.286 & 0 & $\cdots$ & $\cdots$ & $\cdots$  &\\
LBQS 1246-0217 & 2.117 & \textbf{1.842-2.088} & $\cdots$ & 1.781 & $21.45\pm0.15$ & 35 \\
J124957.2-015929 & 3.635 & 2.414-3.589 & 3.550 & $\cdots$ & $\cdots$ & \\
QSO B1249-02 & 1.192 & 0 & $\cdots$ & $\cdots$ & $\cdots$  &\\
QSO B1256-177 & 1.956 & \textbf{1.789-1.926} & $\cdots$ & $\cdots$ & $\cdots$ &  \\
QSO J1306+0356 & 5.999 & 4.511-5.929 & $\cdots$ & $\cdots$ & $\cdots$ &  \\
QSO B1317-0507 & 3.710 & 1.911-3.663 & 2.880 & $\cdots$ & $\cdots$ &  \\
QSO B1318-263 & 2.027 & \textbf{1.715-1.997} & $\cdots$ & $\cdots$ & $\cdots$ &  \\
LBQS 1320-0006 & 1.388 & 0 & $\cdots$ & 0.716 & $20.54\pm0.15$ & 11 \\
QSO B1324-047 & 1.882 & \textbf{1.706-1.853} & $\cdots$ & $\cdots$ & $\cdots$ & \\
QSO J1330-2522 & 3.910 &  2.547-2.714, & $\cdots$ & 2.654 & $19.56\pm0.13$ & 6 \\
$\cdots$ & $\cdots$ & 2.932-3.736, & $\cdots$ & 2.910 & 20.00 & 3 \\
$\cdots$  & $\cdots$  &  3.800-3.861 & $\cdots$ & 3.080 & $19.88\pm0.09$ & 3  \\
QSO B1331+170 & 2.084 & \textbf{1.517-1.718,} & $\cdots$ & 1.776 & $21.15\pm0.07$ & 10 \\
$\cdots$ & $\cdots$ & \textbf{1.801-2.053} & $\cdots$ & $\cdots$ & $\cdots$  & \\
QSO J1342-1355 & 3.212  & \textbf{2.115-3.092} & 3.152 & 3.118 & $20.05\pm0.08$ & 10 \\
QSO J1344-1035 & 2.134 & 1.512-2.103 & $\cdots$ & $\cdots$ & $\cdots$ &  \\
QSO B1347-2457 & 2.578  & 1.512-2.542 & $\cdots$ & $\cdots$ & $\cdots$ & \\
QSO J1356-1101 & 3.006  & \textbf{2.090-2.493} & $\cdots$ & \textbf{2.397} & $\mathbf{19.85\pm0.08}$ & 6  \\
$\cdots$ & $\cdots$ &  \textbf{2.510-2.924} & $\cdots$ & 2.501 & $20.44\pm0.05$ & 27 \\
$\cdots$ & $\cdots$ & $\cdots$ & $\cdots$ & 2.967 & $20.80\pm0.10$ & 10 \\
QSO B1402-012 & 2.522  &  \textbf{2.093-2.487} & $\cdots$ & $\cdots$ & $\cdots$ & \\
QSO B1409+0930 & 2.838 & \textbf{1.761-2.008,} & 2.680 & 2.019 & $20.65\pm0.10$ &  10 \\
$\cdots$ & $\cdots$ & \textbf{2.030-2.448,} & $\cdots$ & 2.456 & $20.53\pm0.08$ & 10 \\
$\cdots$ & $\cdots$ & \textbf{2.463-2.802} & $\cdots$ & \textbf{2.668} & $\mathbf{19.80\pm0.08}$ &  10 \\
QSO B1412-096 & 2.001 &  \textbf{1.715-1.971} & $\cdots$ & $\cdots$ & $\cdots$ & \\
QSO J1421-0643 & 3.689  &  \textbf{2.357-3.096} & 3.474 & 3.448 & $20.40\pm0.10$ & 18 \\
QSO B1424-41 & 1.522 & 0 & $\cdots$ & $\cdots$ & $\cdots$  &\\
QSO B1429-008B & 2.082  &  1.739-2.052 & $\cdots$ & $\cdots$ & $\cdots$ & \\
QSO J1439+1117 & 2.583 &  \textbf{1.715-2.409,} & $\cdots$ & 2.418 & $20.10\pm0.10$ & 36 \\
\hline
\end{tabular}
\end{minipage}
\end{table}

\addtocounter{table}{-1}
\begin{table}
\begin{minipage}[t]{\columnwidth}
\caption{continued.}
\centering
\setlength{\tabcolsep}{1pt}
\begin{tabular}{@{} l c c c c c c c c c @{}}
\hline
 Quasar &   $z_{\rm em}$ &  $z_{\rm min}$--$z_{\rm max}$ & $z_{LLS}$ &   $z_{\rm abs}$ & log $N_{{\rm H}\,{\sc \rm I}}$ &  Ref. \\
 &  & & & &  cm$^{-2}$ &  \\
\hline
$\cdots$ & $\cdots$ &  \textbf{2.427-2.547} & $\cdots$ & $\cdots$ & $\cdots$  & \\
QSO J1443+2724 & 4.443 &  \textbf{2.932-3.736,} & $\cdots$ & 4.224 & $20.95\pm0.08$ & 27 \\
$\cdots$ & $\cdots$ &  \textbf{3.802-4.209,} & $\cdots$ & $\cdots$ & $\cdots$ \\
$\cdots$ & $\cdots$ &  \textbf{4.242-4.384} & $\cdots$ & $\cdots$ & $\cdots$  & \\
LBQS 1444+0126 & 2.210 & \textbf{1.710-2.082,} & $\cdots$ & 2.087 & $20.25\pm0.07$ & 10 \\
$\cdots$ & $\cdots$ & \textbf{2.091-2.178} & $\cdots$ & $\cdots$ & $\cdots$  & \\
QSO B1448-232 & 2.215  & 1.509-2.184 & $\cdots$ & $\cdots$ & $\cdots$ &  \\
J145147.1-151220 & 4.763  & 3.103-4.705 & 4.369 & $\cdots$ & $\cdots$  &  \\
QSO J1453+0029 & 1.297 & 0 & $\cdots$ & $\cdots$ & $\cdots$  &\\
J151352.52+085555 & 2.904 &  \textbf{1.745-2.865} & $\cdots$ & $\cdots$ & $\cdots$ &  \\
QSO J1621-0042 & 3.700 & 2.110-3.655 & 3.138 & 3.104 & $19.70\pm0.20$ & 37 \\
4C 12.59  &  1.792  &  \textbf{1.517-1.764} & $\cdots$ & 0.531 & $20.70\pm0.09$ & 11 \\
$\cdots$ & $\cdots$ & $\cdots$ & $\cdots$ & 0.900 & $19.70\pm0.04$ & 11 \\
QSO J1723+2243 & 4.520 &  \textbf{3.072-3.687,} & $\cdots$ & 3.697 & $20.35\pm0.10$ & 29 \\
$\cdots$ & $\cdots$ &  \textbf{3.707-3.736,} & $\cdots$ & \textbf{4.155} & $\mathbf{19.23\pm0.12}$ & 6 \\
$\cdots$ & $\cdots$ &  \textbf{3.802-4.464} & $\cdots$ & $\cdots$ & $\cdots$ \\
QSO B1730-130 & 0.902 & 0 & $\cdots$ & $\cdots$ & $\cdots$  &\\
QSO B1741-038 & 1.054 & 0 & $\cdots$ & $\cdots$ & $\cdots$  &\\
QSO B1937-1009 & 3.787 &  2.932-3.738 & $\cdots$ & $\cdots$ & $\cdots$ &  \\
QSO B1935-692 & 3.152 &  2.932-3.110 & $\cdots$ & $\cdots$ & $\cdots$ &  \\
QSO B2000-330 & 3.783 &  2.422-3.735 & 3.560 & 3.172 & $19.80\pm0.25$ & 37 \\
$\cdots$ & $\cdots$ & $\cdots$ & $\cdots$ & 3.188 & $19.80\pm0.15$ & 37 \\
$\cdots$ & $\cdots$ & $\cdots$ & $\cdots$ & 3.192 & $19.20\pm0.25$ & 37 \\
QSO J2107-0620 & 0.642 & 0 & $\cdots$ & $\cdots$ & $\cdots$  &\\
LBQS 2113-4345 & 2.053 & 1.510-2.021 & $\cdots$ & $\cdots$ & $\cdots$ &  \\
LBQS 2114-4347 & 2.040 &  1.509-2.010 & $\cdots$ & 1.912 & $19.50\pm0.10$ & 6 \\
J211739.5-433538 & 2.050 & 1.509-2.020 & $\cdots$ & $\cdots$ & $\cdots$ & \\
QSO J2119-3536 & 2.341 & \textbf{1.710-1.992,} & $\cdots$ & 1.996 & $20.10\pm0.07$ & 27 \\
$\cdots$ & $\cdots$ & \textbf{2.000-2.308} & $\cdots$ & $\cdots$ & $\cdots$  & \\
QSO B2126-15 & 3.268 & 1.838-3.225 & 2.783 & 2.638 & $19.25\pm0.15$ & 37 \\
$\cdots$ & $\cdots$ & $\cdots$ & $\cdots$ & 2.769 & $19.20\pm0.15$ & 37 \\
QSO B2129-4653 & 2.230 & 1.706-2.198 & $\cdots$ & $\cdots$ & $\cdots$ & \\
J213314.2-464031 & 2.208  & 1.730-2.176 & $\cdots$ & $\cdots$ & $\cdots$ & \\
LBQS 2132-4321 & 2.420  & \textbf{1.710-1.907,} & $\cdots$ & 1.916 & $20.74\pm0.09$ & 6 \\
$\cdots$ & $\cdots$ & \textbf{1.923-2.383} & $\cdots$ & $\cdots$ & $\cdots$  & \\
LBQS 2138-4427 & 3.170 & \textbf{2.090-2.371,} & $\cdots$ & 2.383 & $20.60\pm0.05$ & 1 \\
$\cdots$ & $\cdots$ & \textbf{2.395-2.837,} & $\cdots$ & 2.852 & $20.98\pm0.05$ & 27 \\
$\cdots$ & $\cdots$ & \textbf{2.867-3.099} & $\cdots$ & $\cdots$ & $\cdots$  & \\
QSO B2139-4433 & 3.220  & 2.932-3.178 & $\cdots$ & $\cdots$ & $\cdots$ & \\
QSO B2149-306 & 2.345 & \textbf{1.715-2.312} & $\cdots$ & $\cdots$ & $\cdots$ & \\
QSO B2204-408 & 3.155 & \textbf{2.932-3.113} & $\cdots$ & $\cdots$ & $\cdots$ &  \\
LBQS 2206-1958A & 2.560 & \textbf{1.517-1.912,} & $\cdots$ & 1.921 & $20.67\pm0.05$ & 30 \\
$\cdots$ & $\cdots$ & \textbf{1.932-2.071,} & $\cdots$ & 2.076 & $20.44\pm0.05$ & 10 \\
$\cdots$ & $\cdots$ & \textbf{2.083-2.526} & $\cdots$ & $\cdots$ & $\cdots$  & \\
QSO J2215-0045 & 1.476 & 0 & $\cdots$ & $\cdots$ & $\cdots$  &\\
QSO J2220-2803 & 2.406 & 1.517-2.373 & $\cdots$ & $\cdots$ & $\cdots$ & \\
QSO B2222-396 & 2.198 & \textbf{1.706-2.114} & $\cdots$ & 2.154 & $20.85\pm0.10$ & 10 \\
QSO J2227-2243 & 1.891 & 1.517-1.863 & $\cdots$ & $\cdots$ & $\cdots$ & \\
QSO B2225-4025 & 2.030 & \textbf{1.716-1.920} & $\cdots$ & 1.965 & $20.65\pm0.10$ & 10 \\
LBQS 2230+0232 & 2.147 & \textbf{1.517-1.849,} & $\cdots$ & 1.864 & $20.90\pm0.10$ & 10 \\
$\cdots$ & $\cdots$ & \textbf{1.879-2.116} & $\cdots$ & $\cdots$ & $\cdots$  & \\
J223851.0-295301 & 2.387  & 1.535-2.356 & $\cdots$ & $\cdots$ & $\cdots$ &  \\
J223922.9-294947 & 1.849  & 1.564-1.821 & $\cdots$ & $\cdots$ & $\cdots$ &  \\
J223938.9-295451 & 1.907 & 1.535-1.878 & $\cdots$ & $\cdots$ & $\cdots$ &  \\
J223941.8-294955 & 2.102 & 1.656-2.071 & $\cdots$ & 1.825 & $19.84\pm0.14$ & 6 \\
J223948.7-294748 & 2.068 & 1.517-2.037 & $\cdots$ & $\cdots$ & $\cdots$ &  \\
J223951.9-294837  & 2.121 & 1.525-2.090 & $\cdots$ & $\cdots$ & $\cdots$ &  \\
QSO B2237-0607 & 4.558 & \textbf{2.894-3.735,} & $\cdots$ & 4.079 & $20.55\pm0.10$ & 4 \\
$\cdots$ & $\cdots$ & \textbf{3.802-4.068,} & $\cdots$ & $\cdots$ & $\cdots$ &  \\
$\cdots$ & $\cdots$ & \textbf{4.090-4.500} & $\cdots$ & $\cdots$ & $\cdots$ &  \\
QSO J2247-1237  & 1.892 & \textbf{1.716-1.863} & $\cdots$ & $\cdots$ & $\cdots$ &  \\
QSO B2311-373 & 2.476  & \textbf{1.706-2.171,} & $\cdots$ & 2.182 & $20.48\pm0.13$ & 10 \\
$\cdots$ & $\cdots$ & \textbf{2.193-2.443} & $\cdots$ & $\cdots$ & $\cdots$  & \\
J232046.7-294406 & 2.401 & 1.509-2.367 & $\cdots$ &  $\cdots$ & $\cdots$ &  \\
J232059.4-295520 &  2.317 & 1.517-2.284 & $\cdots$ & $\cdots$ & $\cdots$ &  \\
J232114.3-294725 & 2.677 & 2.091-2.640 & $\cdots$ & $\cdots$ & $\cdots$ & \\
J232121.2-294350 & 2.184 & 1.620-2.152 & $\cdots$ & $\cdots$ & $\cdots$ &  \\
QSO B2318-1107 & 2.960 & \textbf{1.509-1.618,} & $\cdots$ & 1.629 & $20.52\pm0.14$ & 6 \\
$\cdots$ & $\cdots$ & \textbf{1.640-1.977,} & $\cdots$ & 1.989 & $20.68\pm0.05$ & 10 \\
$\cdots$ & $\cdots$ & \textbf{2.001-2.714,} & $\cdots$ & $\cdots$ & $\cdots$  & \\
$\cdots$ & $\cdots$ & \textbf{2.932-2.920} & $\cdots$ & $\cdots$ & $\cdots$  & \\
QSO J2328+0022 & 1.302 & 0 & $\cdots$ & 0.652 & $20.32\pm0.07$ & 11 \\
QSO B2332-094 & 3.330 & \textbf{2.177-3.045,} & 3.235 & \textbf{2.287} & $\mathbf{20.07\pm0.07}$ & 8  \\
\hline
\end{tabular}
\end{minipage}
\end{table}

\addtocounter{table}{-1}
\begin{table}
\begin{minipage}[t]{\columnwidth}
\caption{continued.}
\centering
\setlength{\tabcolsep}{1pt}
\begin{tabular}{@{} l c c c c c c c c c @{}}
\hline
 Quasar &   $z_{\rm em}$ &  $z_{\rm min}$--$z_{\rm max}$ & $z_{LLS}$ &   $z_{\rm abs}$ & log $N_{{\rm H}\,{\sc \rm I}}$ &  Ref. \\
 &  & & & &  cm$^{-2}$ &  \\
\hline
$\cdots$ & $\cdots$ & \textbf{3.069-3.289} & $\cdots$ & 3.057 & $20.50\pm0.07$ & 10 \\
J233544.2+150118 & 0.791 & 0 & $\cdots$ & $0.680$ & $19.70\pm0.30$ & 11 \\
QSO B2342+3417 & 3.010  & \textbf{2.091-2.891,} & $\cdots$ & 2.909 & $21.10\pm0.10$ & 10 \\
$\cdots$ & $\cdots$ & \textbf{2.927-2.970} & $\cdots$ & \textbf{2.940} & $\mathbf{20.18\pm0.10}$ & 6 \\
QSO J2346+1247 & 2.578 & 1.715-2.538 & $\cdots$ & 2.569 & $20.98\pm0.04$ & 38 \\
QSO B2343+125 & 2.763 & \textbf{1.601-2.185} & 2.465 & 2.431 & $20.40\pm0.07$ & 10 \\
QSO B2345+000 & 2.654 & 1.715-2.617 & $\cdots$ & $\cdots$ & $\cdots$ & \\
QSO B2347-4342 & 2.885  & 1.812-2.846 & 2.750 & $\cdots$ & $\cdots$ &  \\
QSO B2348-0180 & 3.023 & \textbf{1.962-2.032,} & 2.946 & 2.426 & $20.50\pm0.10$ & 27 \\
$\cdots$ & $\cdots$ & \textbf{2.069-2.415,} & $\cdots$ & 2.615 & $21.30\pm0.08$ & 10 \\
$\cdots$ & $\cdots$ & \textbf{2.437-2.596,} & $\cdots$ & $\cdots$ & $\cdots$  & \\
$\cdots$ & $\cdots$ & \textbf{2.634-2.718,} & $\cdots$ & $\cdots$ & $\cdots$  & \\
$\cdots$ & $\cdots$ & \textbf{2.932-2.982} & $\cdots$ & $\cdots$ & $\cdots$ &  \\
QSO B2348-147 & 2.933 & \textbf{1.838-2.264,} & $\cdots$ & 2.279 & $20.56\pm0.08$ & 10 \\
$\cdots$ & $\cdots$ & \textbf{2.288-2.894} & $\cdots$ & $\cdots$ & $\cdots$  & \\
J235534.6-395355 & 1.579 & 0 & $\cdots$ & $\cdots$ & $\cdots$  &\\
J235702.5-004824 & 3.013 & \textbf{2.002-2.973} & 3.000 & $2.479$ & $20.41\pm0.08$ & 6 \\
QSO J2359-1241 & 0.868 & 0 & $\cdots$ & $\cdots$ & $\cdots$  &\\
\hline
\end{tabular}
\end{minipage}
\vspace{0.1cm}
{\bf References:} (1) \citet{ledoux06}; 
(2) \citet{richter05};
(3) \citet{peroux01};
(4) \citet{peroux05};
(5) \citet{kaplan10};
(6) Paper\,I;
(7) \citet{dessauges09};
(8) \citet{ledoux03};
(9) \citet{noterdaeme07};
(10) \citet{fox09};
(11) \citet{rao06};
(12) \citet{ellison09};
(13) \citet{prochaska04};
(14) \citet{ellison01};
(15) \citet{peroux03};
(16) \citet{srianand07};
(17) \citet{odorico02};
(18) \citet{kanekar09};
(19) \citet{noterdaeme09};
(20) \citet{noterdaeme08};
(21) \citet{ellison012};
(22) \citet{carswell96};
(23) \citet{peroux04};
(24) \citet{lopez05};
(25) \citet{quast08};
(26) \citet{prochaska07};
(27) \citet{srianand98};
(28) \citet{fox11};
(29) \citet{guimares09};
(30) \citet{lopez99};
(31) \citet{ledoux02};
(32) \citet{lebrun97};
(33) \citet{dessauges03};
(34) \citet{ivanchik10};
(35) \citet{herbert06};
(36) \citet{noterdaeme08b};
(37) \citet{omeara07};
(38) \citet{rix07}.
\end{table}

\bibliographystyle{aa}
\bibliography{dla.bib}{}

\begin{thebibliography}{104}
\expandafter\ifx\csname natexlab\endcsname\relax\def\natexlab#1{#1}\fi

\bibitem[{{Altay} {et~al.}(2011){Altay}, {Theuns}, {Schaye}, {Crighton}, \&
  {Dalla Vecchia}}]{altay11}
{Altay}, G., {Theuns}, T., {Schaye}, J., {Crighton}, N.~H.~M., \& {Dalla
  Vecchia}, C. 2011, \apjl, 737, L37

\bibitem[{{Bahcall} \& {Peebles}(1969)}]{bahcall69}
{Bahcall}, J.~N. \& {Peebles}, P.~J.~E. 1969, \apjl, 156, L7

\bibitem[{{Bauermeister} {et~al.}(2010){Bauermeister}, {Blitz}, \&
  {Ma}}]{bauermeister10}
{Bauermeister}, A., {Blitz}, L., \& {Ma}, C.-P. 2010, \apj, 717, 323

\bibitem[{{Birnboim} {et~al.}(2007){Birnboim}, {Dekel}, \&
  {Neistein}}]{birnboim07}
{Birnboim}, Y., {Dekel}, A., \& {Neistein}, E. 2007, \mnras, 380, 339

\bibitem[{{Bordoloi} {et~al.}(2011){Bordoloi}, {Lilly}, {Knobel}, {Bolzonella},
  {Kampczyk}, {Carollo}, {Iovino}, {Zucca}, {Contini}, {Kneib}, {Le Fevre},
  {Mainieri}, {Renzini}, {Scodeggio}, {Zamorani}, {Balestra}, {Bardelli},
  {Bongiorno}, {Caputi}, {Cucciati}, {de la Torre}, {de Ravel}, {Garilli},
  {Kova{\v c}}, {Lamareille}, {Le Borgne}, {Le Brun}, {Maier}, {Mignoli},
  {Pello}, {Peng}, {Perez Montero}, {Presotto}, {Scarlata}, {Silverman},
  {Tanaka}, {Tasca}, {Tresse}, {Vergani}, {Barnes}, {Cappi}, {Cimatti},
  {Coppa}, {Diener}, {Franzetti}, {Koekemoer}, {L{\'o}pez-Sanjuan},
  {McCracken}, {Moresco}, {Nair}, {Oesch}, {Pozzetti}, \&
  {Welikala}}]{bordoloi11}
{Bordoloi}, R., {Lilly}, S.~J., {Knobel}, C., {et~al.} 2011, \apj, 743, 10

\bibitem[{{Braun}(2012)}]{braun12}
{Braun}, R. 2012, \apj, 749, 87

\bibitem[{{Carswell} {et~al.}(1996){Carswell}, {Webb}, {Lanzetta}, {Baldwin},
  {Cooke}, {Williger}, {Rauch}, {Irwin}, {Robertson}, \& {Shaver}}]{carswell96}
{Carswell}, R.~F., {Webb}, J.~K., {Lanzetta}, K.~M., {et~al.} 1996, \mnras,
  278, 506

\bibitem[{{Chang} {et~al.}(2010){Chang}, {Pen}, {Bandura}, \&
  {Peterson}}]{chang10}
{Chang}, T.-C., {Pen}, U.-L., {Bandura}, K., \& {Peterson}, J.~B. 2010, \nat,
  466, 463

\bibitem[{{Cole} {et~al.}(2001){Cole}, {Norberg}, {Baugh}, {Frenk},
  {Bland-Hawthorn}, {Bridges}, {Cannon}, {Colless}, {Collins}, {Couch},
  {Cross}, {Dalton}, {De Propris}, {Driver}, {Efstathiou}, {Ellis},
  {Glazebrook}, {Jackson}, {Lahav}, {Lewis}, {Lumsden}, {Maddox}, {Madgwick},
  {Peacock}, {Peterson}, {Sutherland}, \& {Taylor}}]{cole01}
{Cole}, S., {Norberg}, P., {Baugh}, C.~M., {et~al.} 2001, \mnras, 326, 255

\bibitem[{{Coppolani} {et~al.}(2006){Coppolani}, {Petitjean}, {Stoehr},
  {Rollinde}, {Pichon}, {Colombi}, {Haehnelt}, {Carswell}, \&
  {Teyssier}}]{coppolani06}
{Coppolani}, F., {Petitjean}, P., {Stoehr}, F., {et~al.} 2006, \mnras, 370,
  1804

\bibitem[{{Cucciati} {et~al.}(2012){Cucciati}, {Tresse}, {Ilbert}, {Le
  F{\`e}vre}, {Garilli}, {Le Brun}, {Cassata}, {Franzetti}, {Maccagni},
  {Scodeggio}, {Zucca}, {Zamorani}, {Bardelli}, {Bolzonella}, {Bielby},
  {McCracken}, {Zanichelli}, \& {Vergani}}]{cucciati12}
{Cucciati}, O., {Tresse}, L., {Ilbert}, O., {et~al.} 2012, \aap, 539, A31

\bibitem[{{Danforth} \& {Shull}(2008)}]{danforth08}
{Danforth}, C.~W. \& {Shull}, J.~M. 2008, \apj, 679, 194

\bibitem[{{Dekker} {et~al.}(2000){Dekker}, {D'Odorico}, {Kaufer}, {Delabre}, \&
  {Kotzlowski}}]{dekker00}
{Dekker}, H., {D'Odorico}, S., {Kaufer}, A., {Delabre}, B., \& {Kotzlowski}, H.
  2000, in SPIE Conference Series, Vol. 4008, Society of Photo-Optical
  Instrumentation Engineers (SPIE) Conference Series, ed. {M.~Iye \&
  A.~F.~Moorwood}, 534--545

\bibitem[{{Dessauges-Zavadsky} {et~al.}(2009){Dessauges-Zavadsky}, {Ellison},
  \& {Murphy}}]{dessauges09}
{Dessauges-Zavadsky}, M., {Ellison}, S.~L., \& {Murphy}, M.~T. 2009, \mnras,
  396, L61

\bibitem[{{Dessauges-Zavadsky} {et~al.}(2003){Dessauges-Zavadsky},
  {P{\'e}roux}, {Kim}, {D'Odorico}, \& {McMahon}}]{dessauges03}
{Dessauges-Zavadsky}, M., {P{\'e}roux}, C., {Kim}, T.-S., {D'Odorico}, S., \&
  {McMahon}, R.~G. 2003, \mnras, 345, 447

\bibitem[{{D'Odorico} {et~al.}(2002){D'Odorico}, {Petitjean}, \&
  {Cristiani}}]{odorico02}
{D'Odorico}, V., {Petitjean}, P., \& {Cristiani}, S. 2002, \aap, 390, 13

\bibitem[{{Ellison} \& {Lopez}(2009)}]{ellison09}
{Ellison}, S.~L. \& {Lopez}, S. 2009, \mnras, 397, 467

\bibitem[{{Ellison} {et~al.}(2001{\natexlab{a}}){Ellison}, {Pettini},
  {Steidel}, \& {Shapley}}]{ellison01}
{Ellison}, S.~L., {Pettini}, M., {Steidel}, C.~C., \& {Shapley}, A.~E.
  2001{\natexlab{a}}, \apj, 549, 770

\bibitem[{{Ellison} {et~al.}(2001{\natexlab{b}}){Ellison}, {Yan}, {Hook},
  {Pettini}, {Wall}, \& {Shaver}}]{ellison012}
{Ellison}, S.~L., {Yan}, L., {Hook}, I.~M., {et~al.} 2001{\natexlab{b}}, \aap,
  379, 393

\bibitem[{{Erb}(2008)}]{erb08}
{Erb}, D.~K. 2008, \apj, 674, 151

\bibitem[{{Erkal} {et~al.}(2012){Erkal}, {Gnedin}, \& {Kravtsov}}]{erkal12}
{Erkal}, D., {Gnedin}, N.~Y., \& {Kravtsov}, A.~V. 2012, \apj, 761, 54

\bibitem[{{Fall} \& {Pei}(1993)}]{fall93}
{Fall}, S.~M. \& {Pei}, Y.~C. 1993, \apj, 402, 479

\bibitem[{{Fontana} \& {Ballester}(1995)}]{fontana}
{Fontana}, A. \& {Ballester}, P. 1995, The Messenger, 80, 37

\bibitem[{{Fox} {et~al.}(2011){Fox}, {Ledoux}, {Petitjean}, {Srianand}, \&
  {Guimar{\~a}es}}]{fox11}
{Fox}, A.~J., {Ledoux}, C., {Petitjean}, P., {Srianand}, R., \&
  {Guimar{\~a}es}, R. 2011, \aap, 534, A82

\bibitem[{{Fox} {et~al.}(2009){Fox}, {Prochaska}, {Ledoux}, {Petitjean},
  {Wolfe}, \& {Srianand}}]{fox09}
{Fox}, A.~J., {Prochaska}, J.~X., {Ledoux}, C., {et~al.} 2009, \aap, 503, 731

\bibitem[{{Fukugita} \& {Peebles}(2004)}]{fukugita04}
{Fukugita}, M. \& {Peebles}, P.~J.~E. 2004, \apj, 616, 643

\bibitem[{{Giavalisco} {et~al.}(2011){Giavalisco}, {Vanzella}, {Salimbeni},
  {Tripp}, {Dickinson}, {Cassata}, {Renzini}, {Guo}, {Ferguson}, {Nonino},
  {Cimatti}, {Kurk}, {Mignoli}, \& {Tang}}]{giavalisco11}
{Giavalisco}, M., {Vanzella}, E., {Salimbeni}, S., {et~al.} 2011, \apj, 743, 95

\bibitem[{{Guimar{\~a}es} {et~al.}(2009){Guimar{\~a}es}, {Petitjean}, {de
  Carvalho}, {Djorgovski}, {Noterdaeme}, {Castro}, {Poppe}, \&
  {Aghaee}}]{guimares09}
{Guimar{\~a}es}, R., {Petitjean}, P., {de Carvalho}, R.~R., {et~al.} 2009,
  \aap, 508, 133

\bibitem[{{Herbert-Fort} {et~al.}(2006){Herbert-Fort}, {Prochaska},
  {Dessauges-Zavadsky}, {Ellison}, {Howk}, {Wolfe}, \& {Prochter}}]{herbert06}
{Herbert-Fort}, S., {Prochaska}, J.~X., {Dessauges-Zavadsky}, M., {et~al.}
  2006, \pasp, 118, 1077

\bibitem[{{Hopkins} \& {Beacom}(2006)}]{hopkins06}
{Hopkins}, A.~M. \& {Beacom}, J.~F. 2006, \apj, 651, 142

\bibitem[{{Hopkins} {et~al.}(2008){Hopkins}, {McClure-Griffiths}, \&
  {Gaensler}}]{hopkins08}
{Hopkins}, A.~M., {McClure-Griffiths}, N.~M., \& {Gaensler}, B.~M. 2008, \apjl,
  682, L13

\bibitem[{{Ivanchik} {et~al.}(2010){Ivanchik}, {Petitjean}, {Balashev},
  {Srianand}, {Varshalovich}, {Ledoux}, \& {Noterdaeme}}]{ivanchik10}
{Ivanchik}, A.~V., {Petitjean}, P., {Balashev}, S.~A., {et~al.} 2010, \mnras,
  404, 1583

\bibitem[{{Kacprzak} {et~al.}(2011){Kacprzak}, {Churchill}, {Evans}, {Murphy},
  \& {Steidel}}]{kacprzak11b}
{Kacprzak}, G.~G., {Churchill}, C.~W., {Evans}, J.~L., {Murphy}, M.~T., \&
  {Steidel}, C.~C. 2011, \mnras, 416, 3118

\bibitem[{{Kanekar} {et~al.}(2009){Kanekar}, {Smette}, {Briggs}, \&
  {Chengalur}}]{kanekar09}
{Kanekar}, N., {Smette}, A., {Briggs}, F.~H., \& {Chengalur}, J.~N. 2009,
  \apjl, 705, L40

\bibitem[{{Kaplan} {et~al.}(2010){Kaplan}, {Prochaska}, {Herbert-Fort},
  {Ellison}, \& {Dessauges-Zavadsky}}]{kaplan10}
{Kaplan}, K.~F., {Prochaska}, J.~X., {Herbert-Fort}, S., {Ellison}, S.~L., \&
  {Dessauges-Zavadsky}, M. 2010, \pasp, 122, 619

\bibitem[{{Katz} {et~al.}(1996){Katz}, {Weinberg}, {Hernquist}, \&
  {Miralda-Escude}}]{katz96}
{Katz}, N., {Weinberg}, D.~H., {Hernquist}, L., \& {Miralda-Escude}, J. 1996,
  \apjl, 457, L57

\bibitem[{{Kere{\v s}} {et~al.}(2005){Kere{\v s}}, {Katz}, {Weinberg}, \&
  {Dav{\'e}}}]{keres05}
{Kere{\v s}}, D., {Katz}, N., {Weinberg}, D.~H., \& {Dav{\'e}}, R. 2005,
  \mnras, 363, 2

\bibitem[{{Kim} {et~al.}(2013){Kim}, {Partl}, {Carswell}, \&
  {M{\"u}ller}}]{kim13}
{Kim}, T.-S., {Partl}, A.~M., {Carswell}, R.~F., \& {M{\"u}ller}, V. 2013,
  ArXiv e-prints

\bibitem[{{Klypin} {et~al.}(1995){Klypin}, {Borgani}, {Holtzman}, \&
  {Primack}}]{klypin95}
{Klypin}, A., {Borgani}, S., {Holtzman}, J., \& {Primack}, J. 1995, \apj, 444,
  1

\bibitem[{{Komatsu} {et~al.}(2011){Komatsu}, {Smith}, {Dunkley}, {Bennett},
  {Gold}, {Hinshaw}, {Jarosik}, {Larson}, {Nolta}, {Page}, {Spergel},
  {Halpern}, {Hill}, {Kogut}, {Limon}, {Meyer}, {Odegard}, {Tucker}, {Weiland},
  {Wollack}, \& {Wright}}]{komatsu11}
{Komatsu}, E., {Smith}, K.~M., {Dunkley}, J., {et~al.} 2011, \apjs, 192, 18

\bibitem[{{Kornei} {et~al.}(2012){Kornei}, {Shapley}, {Martin}, {Coil}, {Lotz},
  {Schiminovich}, {Bundy}, \& {Noeske}}]{kornei12}
{Kornei}, K.~A., {Shapley}, A.~E., {Martin}, C.~L., {et~al.} 2012, \apj, 758,
  135

\bibitem[{{Krumholz} {et~al.}(2009){Krumholz}, {Ellison}, {Prochaska}, \&
  {Tumlinson}}]{krumholz09}
{Krumholz}, M.~R., {Ellison}, S.~L., {Prochaska}, J.~X., \& {Tumlinson}, J.
  2009, \apjl, 701, L12

\bibitem[{{Lagos} {et~al.}(2011){Lagos}, {Baugh}, {Lacey}, {Benson}, {Kim}, \&
  {Power}}]{lagos11}
{Lagos}, C.~D.~P., {Baugh}, C.~M., {Lacey}, C.~G., {et~al.} 2011, \mnras, 418,
  1649

\bibitem[{{Lah} {et~al.}(2007){Lah}, {Chengalur}, {Briggs}, {Colless}, {de
  Propris}, {Pracy}, {de Blok}, {Fujita}, {Ajiki}, {Shioya}, {Nagao},
  {Murayama}, {Taniguchi}, {Yagi}, \& {Okamura}}]{lah07}
{Lah}, P., {Chengalur}, J.~N., {Briggs}, F.~H., {et~al.} 2007, \mnras, 376,
  1357

\bibitem[{{Lanzetta} {et~al.}(1995){Lanzetta}, {Wolfe}, \&
  {Turnshek}}]{lanzetta95}
{Lanzetta}, K.~M., {Wolfe}, A.~M., \& {Turnshek}, D.~A. 1995, \apj, 440, 435

\bibitem[{{Le Brun} {et~al.}(1997){Le Brun}, {Bergeron}, {Boisse}, \&
  {Deharveng}}]{lebrun97}
{Le Brun}, V., {Bergeron}, J., {Boisse}, P., \& {Deharveng}, J.~M. 1997, \aap,
  321, 733

\bibitem[{{Ledoux} {et~al.}(2002){Ledoux}, {Bergeron}, \&
  {Petitjean}}]{ledoux02}
{Ledoux}, C., {Bergeron}, J., \& {Petitjean}, P. 2002, \aap, 385, 802

\bibitem[{{Ledoux} {et~al.}(1998){Ledoux}, {Petitjean}, {Bergeron}, {Wampler},
  \& {Srianand}}]{ledoux98}
{Ledoux}, C., {Petitjean}, P., {Bergeron}, J., {Wampler}, E.~J., \& {Srianand},
  R. 1998, \aap, 337, 51

\bibitem[{{Ledoux} {et~al.}(2006){Ledoux}, {Petitjean}, {Fynbo}, {M{\o}ller},
  \& {Srianand}}]{ledoux06}
{Ledoux}, C., {Petitjean}, P., {Fynbo}, J.~P.~U., {M{\o}ller}, P., \&
  {Srianand}, R. 2006, \aap, 457, 71

\bibitem[{{Ledoux} {et~al.}(2003){Ledoux}, {Petitjean}, \&
  {Srianand}}]{ledoux03}
{Ledoux}, C., {Petitjean}, P., \& {Srianand}, R. 2003, \mnras, 346, 209

\bibitem[{{Lopez} {et~al.}(2005){Lopez}, {Reimers}, {Gregg}, {Wisotzki},
  {Wucknitz}, \& {Guzman}}]{lopez05}
{Lopez}, S., {Reimers}, D., {Gregg}, M.~D., {et~al.} 2005, \apj, 626, 767

\bibitem[{{Lopez} {et~al.}(1999){Lopez}, {Reimers}, {Rauch}, {Sargent}, \&
  {Smette}}]{lopez99}
{Lopez}, S., {Reimers}, D., {Rauch}, M., {Sargent}, W.~L.~W., \& {Smette}, A.
  1999, \apj, 513, 598

\bibitem[{{Ma} \& {Bertschinger}(1994)}]{ma94}
{Ma}, C.-P. \& {Bertschinger}, E. 1994, \apjl, 434, L5

\bibitem[{{Martin} {et~al.}(2010){Martin}, {Papastergis}, {Giovanelli},
  {Haynes}, {Springob}, \& {Stierwalt}}]{martin10}
{Martin}, A.~M., {Papastergis}, E., {Giovanelli}, R., {et~al.} 2010, \apj, 723,
  1359

\bibitem[{{Martin} {et~al.}(2012){Martin}, {Shapley}, {Coil}, {Kornei},
  {Bundy}, {Weiner}, {Noeske}, \& {Schiminovich}}]{martin12}
{Martin}, C.~L., {Shapley}, A.~E., {Coil}, A.~L., {et~al.} 2012, \apj, 760, 127

\bibitem[{{McQuinn} {et~al.}(2011){McQuinn}, {Oh}, \&
  {Faucher-Gigu{\`e}re}}]{mcquinn11}
{McQuinn}, M., {Oh}, S.~P., \& {Faucher-Gigu{\`e}re}, C.-A. 2011, \apj, 743, 82

\bibitem[{{Nagamine} {et~al.}(2004){Nagamine}, {Springel}, \&
  {Hernquist}}]{nagamine04}
{Nagamine}, K., {Springel}, V., \& {Hernquist}, L. 2004, \mnras, 348, 421

\bibitem[{{Noterdaeme} {et~al.}(2007){Noterdaeme}, {Ledoux}, {Petitjean}, {Le
  Petit}, {Srianand}, \& {Smette}}]{noterdaeme07}
{Noterdaeme}, P., {Ledoux}, C., {Petitjean}, P., {et~al.} 2007, \aap, 474, 393

\bibitem[{{Noterdaeme} {et~al.}(2008{\natexlab{a}}){Noterdaeme}, {Ledoux},
  {Petitjean}, \& {Srianand}}]{noterdaeme08}
{Noterdaeme}, P., {Ledoux}, C., {Petitjean}, P., \& {Srianand}, R.
  2008{\natexlab{a}}, \aap, 481, 327

\bibitem[{{Noterdaeme} {et~al.}(2012){Noterdaeme}, {Petitjean}, {Carithers},
  {P{\^a}ris}, {Font-Ribera}, {Bailey}, {Aubourg}, {Bizyaev}, {Ebelke},
  {Finley}, {Ge}, {Malanushenko}, {Malanushenko}, {Miralda-Escud{\'e}},
  {Myers}, {Oravetz}, {Pan}, {Pieri}, {Ross}, {Schneider}, {Simmons}, \&
  {York}}]{noterdaeme12b}
{Noterdaeme}, P., {Petitjean}, P., {Carithers}, W.~C., {et~al.} 2012, \aap,
  547, L1

\bibitem[{{Noterdaeme} {et~al.}(2009){Noterdaeme}, {Petitjean}, {Ledoux}, \&
  {Srianand}}]{noterdaeme09}
{Noterdaeme}, P., {Petitjean}, P., {Ledoux}, C., \& {Srianand}, R. 2009, \aap,
  505, 1087

\bibitem[{{Noterdaeme} {et~al.}(2008{\natexlab{b}}){Noterdaeme}, {Petitjean},
  {Ledoux}, {Srianand}, \& {Ivanchik}}]{noterdaeme08b}
{Noterdaeme}, P., {Petitjean}, P., {Ledoux}, C., {Srianand}, R., \& {Ivanchik},
  A. 2008{\natexlab{b}}, \aap, 491, 397

\bibitem[{{O'Meara} {et~al.}(2007){O'Meara}, {Prochaska}, {Burles}, {Prochter},
  {Bernstein}, \& {Burgess}}]{omeara07}
{O'Meara}, J.~M., {Prochaska}, J.~X., {Burles}, S., {et~al.} 2007, \apj, 656,
  666

\bibitem[{{O'Meara} {et~al.}(2013){O'Meara}, {Prochaska}, {Worseck}, {Chen}, \&
  {Madau}}]{omeara12}
{O'Meara}, J.~M., {Prochaska}, J.~X., {Worseck}, G., {Chen}, H.-W., \& {Madau},
  P. 2013, \apj, 765, 137

\bibitem[{{P{\'e}roux} {et~al.}(2004){P{\'e}roux}, {Deharveng}, {Le Brun}, \&
  {Cristiani}}]{peroux04}
{P{\'e}roux}, C., {Deharveng}, J.-M., {Le Brun}, V., \& {Cristiani}, S. 2004,
  \mnras, 352, 1291

\bibitem[{{P{\'e}roux} {et~al.}(2003{\natexlab{a}}){P{\'e}roux},
  {Dessauges-Zavadsky}, {D'Odorico}, {Kim}, \& {McMahon}}]{peroux03b}
{P{\'e}roux}, C., {Dessauges-Zavadsky}, M., {D'Odorico}, S., {Kim}, T.-S., \&
  {McMahon}, R.~G. 2003{\natexlab{a}}, \mnras, 345, 480

\bibitem[{{P{\'e}roux} {et~al.}(2005){P{\'e}roux}, {Dessauges-Zavadsky},
  {D'Odorico}, {Sun Kim}, \& {McMahon}}]{peroux05}
{P{\'e}roux}, C., {Dessauges-Zavadsky}, M., {D'Odorico}, S., {Sun Kim}, T., \&
  {McMahon}, R.~G. 2005, \mnras, 363, 479

\bibitem[{{P{\'e}roux} {et~al.}(2003{\natexlab{b}}){P{\'e}roux}, {McMahon},
  {Storrie-Lombardi}, \& {Irwin}}]{peroux03}
{P{\'e}roux}, C., {McMahon}, R.~G., {Storrie-Lombardi}, L.~J., \& {Irwin},
  M.~J. 2003{\natexlab{b}}, \mnras, 346, 1103

\bibitem[{{P{\'e}roux} {et~al.}(2001){P{\'e}roux}, {Storrie-Lombardi},
  {McMahon}, {Irwin}, \& {Hook}}]{peroux01}
{P{\'e}roux}, C., {Storrie-Lombardi}, L.~J., {McMahon}, R.~G., {Irwin}, M., \&
  {Hook}, I.~M. 2001, \aj, 121, 1799

\bibitem[{{Pontzen} {et~al.}(2008){Pontzen}, {Governato}, {Pettini}, {Booth},
  {Stinson}, {Wadsley}, {Brooks}, {Quinn}, \& {Haehnelt}}]{pontzen08}
{Pontzen}, A., {Governato}, F., {Pettini}, M., {et~al.} 2008, \mnras, 390, 1349

\bibitem[{{Pontzen} \& {Pettini}(2009)}]{pontzen09}
{Pontzen}, A. \& {Pettini}, M. 2009, \mnras, 393, 557

\bibitem[{{Prochaska} \& {Herbert-Fort}(2004)}]{prochaska04}
{Prochaska}, J.~X. \& {Herbert-Fort}, S. 2004, \pasp, 116, 622

\bibitem[{{Prochaska} {et~al.}(2005){Prochaska}, {Herbert-Fort}, \&
  {Wolfe}}]{prochaska05}
{Prochaska}, J.~X., {Herbert-Fort}, S., \& {Wolfe}, A.~M. 2005, \apj, 635, 123

\bibitem[{{Prochaska} {et~al.}(2010){Prochaska}, {O'Meara}, \&
  {Worseck}}]{prochaska10}
{Prochaska}, J.~X., {O'Meara}, J.~M., \& {Worseck}, G. 2010, \apj, 718, 392

\bibitem[{{Prochaska} \& {Wolfe}(2009)}]{prochaska09}
{Prochaska}, J.~X. \& {Wolfe}, A.~M. 2009, \apj, 696, 1543

\bibitem[{{Prochaska} {et~al.}(2007){Prochaska}, {Wolfe}, {Howk}, {Gawiser},
  {Burles}, \& {Cooke}}]{prochaska07}
{Prochaska}, J.~X., {Wolfe}, A.~M., {Howk}, J.~C., {et~al.} 2007, \apjs, 171,
  29

\bibitem[{{Quast} {et~al.}(2008){Quast}, {Reimers}, \& {Baade}}]{quast08}
{Quast}, R., {Reimers}, D., \& {Baade}, R. 2008, \aap, 477, 443

\bibitem[{{Rahmati} {et~al.}(2013{\natexlab{a}}){Rahmati}, {Pawlik},
  {Raicevic}, \& {Schaye}}]{rahmati13a}
{Rahmati}, A., {Pawlik}, A.~H., {Raicevic}, M., \& {Schaye}, J.
  2013{\natexlab{a}}, \mnras, 430, 2427

\bibitem[{{Rahmati} {et~al.}(2013{\natexlab{b}}){Rahmati}, {Schaye}, {Pawlik},
  \& {Raicevic}}]{rahmati13b}
{Rahmati}, A., {Schaye}, J., {Pawlik}, A.~H., \& {Raicevic}, M.
  2013{\natexlab{b}}, \mnras, 431, 2261

\bibitem[{{Rao} {et~al.}(2006){Rao}, {Turnshek}, \& {Nestor}}]{rao06}
{Rao}, S.~M., {Turnshek}, D.~A., \& {Nestor}, D.~B. 2006, \apj, 636, 610

\bibitem[{{Ribaudo} {et~al.}(2011){Ribaudo}, {Lehner}, \& {Howk}}]{ribaudo11}
{Ribaudo}, J., {Lehner}, N., \& {Howk}, J.~C. 2011, \apj, 736, 42

\bibitem[{{Richter} {et~al.}(2005){Richter}, {Ledoux}, {Petitjean}, \&
  {Bergeron}}]{richter05}
{Richter}, P., {Ledoux}, C., {Petitjean}, P., \& {Bergeron}, J. 2005, \aap,
  440, 819

\bibitem[{{Rix} {et~al.}(2007){Rix}, {Pettini}, {Steidel}, {Reddy},
  {Adelberger}, {Erb}, \& {Shapley}}]{rix07}
{Rix}, S.~A., {Pettini}, M., {Steidel}, C.~C., {et~al.} 2007, \apj, 670, 15

\bibitem[{{Rubin} {et~al.}(2012){Rubin}, {Prochaska}, {Koo}, \&
  {Phillips}}]{rubin12}
{Rubin}, K.~H.~R., {Prochaska}, J.~X., {Koo}, D.~C., \& {Phillips}, A.~C. 2012,
  \apjl, 747, L26

\bibitem[{{Rubin} {et~al.}(2010){Rubin}, {Weiner}, {Koo}, {Martin},
  {Prochaska}, {Coil}, \& {Newman}}]{rubin10}
{Rubin}, K.~H.~R., {Weiner}, B.~J., {Koo}, D.~C., {et~al.} 2010, \apj, 719,
  1503

\bibitem[{{Sato} {et~al.}(2009){Sato}, {Martin}, {Noeske}, {Koo}, \&
  {Lotz}}]{sato09}
{Sato}, T., {Martin}, C.~L., {Noeske}, K.~G., {Koo}, D.~C., \& {Lotz}, J.~M.
  2009, \apj, 696, 214

\bibitem[{{Schaye}(2001)}]{schaye01}
{Schaye}, J. 2001, \apjl, 562, L95

\bibitem[{{Shull} {et~al.}(2012){Shull}, {Smith}, \& {Danforth}}]{shull11}
{Shull}, J.~M., {Smith}, B.~D., \& {Danforth}, C.~W. 2012, \apj, 759, 23

\bibitem[{{Songaila} \& {Cowie}(2010)}]{songaila10}
{Songaila}, A. \& {Cowie}, L.~L. 2010, \apj, 721, 1448

\bibitem[{{Spergel} {et~al.}(2003){Spergel}, {Verde}, {Peiris}, {Komatsu},
  {Nolta}, {Bennett}, {Halpern}, {Hinshaw}, {Jarosik}, {Kogut}, {Limon},
  {Meyer}, {Page}, {Tucker}, {Weiland}, {Wollack}, \& {Wright}}]{spergel03}
{Spergel}, D.~N., {Verde}, L., {Peiris}, H.~V., {et~al.} 2003, \apjs, 148, 175

\bibitem[{{Srianand} {et~al.}(2007){Srianand}, {Gupta}, \&
  {Petitjean}}]{srianand07}
{Srianand}, R., {Gupta}, N., \& {Petitjean}, P. 2007, \mnras, 375, 584

\bibitem[{{Srianand} \& {Petitjean}(1998)}]{srianand98}
{Srianand}, R. \& {Petitjean}, P. 1998, \aap, 335, 33

\bibitem[{{Steidel} {et~al.}(2010){Steidel}, {Erb}, {Shapley}, {Pettini},
  {Reddy}, {Bogosavljevi{\'c}}, {Rudie}, \& {Rakic}}]{steidel10}
{Steidel}, C.~C., {Erb}, D.~K., {Shapley}, A.~E., {et~al.} 2010, \apj, 717, 289

\bibitem[{{Storrie-Lombardi} \& {Wolfe}(2000)}]{storrie00}
{Storrie-Lombardi}, L.~J. \& {Wolfe}, A.~M. 2000, \apj, 543, 552

\bibitem[{{Tescari} {et~al.}(2009){Tescari}, {Viel}, {Tornatore}, \&
  {Borgani}}]{tescari09}
{Tescari}, E., {Viel}, M., {Tornatore}, L., \& {Borgani}, S. 2009, \mnras, 397,
  411

\bibitem[{{Tytler}(1987)}]{tytler87}
{Tytler}, D. 1987, \apj, 321, 49

\bibitem[{{van de Voort} {et~al.}(2012){van de Voort}, {Schaye}, {Altay}, \&
  {Theuns}}]{voort12}
{van de Voort}, F., {Schaye}, J., {Altay}, G., \& {Theuns}, T. 2012, \mnras,
  2664

\bibitem[{{Viegas}(1995)}]{viegas95}
{Viegas}, S.~M. 1995, \mnras, 276, 268

\bibitem[{{Weiner} {et~al.}(2009){Weiner}, {Coil}, {Prochaska}, {Newman},
  {Cooper}, {Bundy}, {Conselice}, {Dutton}, {Faber}, {Koo}, {Lotz}, {Rieke}, \&
  {Rubin}}]{weiner09}
{Weiner}, B.~J., {Coil}, A.~L., {Prochaska}, J.~X., {et~al.} 2009, \apj, 692,
  187

\bibitem[{{Wolfe} {et~al.}(1995){Wolfe}, {Lanzetta}, {Foltz}, \&
  {Chaffee}}]{wolfe95}
{Wolfe}, A.~M., {Lanzetta}, K.~M., {Foltz}, C.~B., \& {Chaffee}, F.~H. 1995,
  \apj, 454, 698

\bibitem[{{Zafar} {et~al.}(2013){Zafar}, {Popping}, \& {P{\'e}roux}}]{zafar12}
{Zafar}, T., {Popping}, A., \& {P{\'e}roux}, C. 2013, \aap, 556, A140

\bibitem[{{Zwaan} {et~al.}(2008){Zwaan}, {Walter}, {Ryan-Weber}, {Brinks}, {de
  Blok}, \& {Kennicutt}}]{zwaan08}
{Zwaan}, M., {Walter}, F., {Ryan-Weber}, E., {et~al.} 2008, \aj, 136, 2886

\bibitem[{{Zwaan} {et~al.}(2005{\natexlab{a}}){Zwaan}, {Meyer},
  {Staveley-Smith}, \& {Webster}}]{zwaan05}
{Zwaan}, M.~A., {Meyer}, M.~J., {Staveley-Smith}, L., \& {Webster}, R.~L.
  2005{\natexlab{a}}, \mnras, 359, L30

\bibitem[{{Zwaan} {et~al.}(2005{\natexlab{b}}){Zwaan}, {van der Hulst},
  {Briggs}, {Verheijen}, \& {Ryan-Weber}}]{zwaan05b}
{Zwaan}, M.~A., {van der Hulst}, J.~M., {Briggs}, F.~H., {Verheijen}, M.~A.~W.,
  \& {Ryan-Weber}, E.~V. 2005{\natexlab{b}}, \mnras, 364, 1467

\end{thebibliography}

\begin{appendix}
\section{Description of the individual sightlines}
This appendix describes the individual sightlines of all the quasars with $z_{\rm em}>1.5$ presented in Table \ref{sampletable}. For each of them, we have examined the abstract of the observing proposal to determine if any knowledge of the sightline characteristics were known prior to the execution of the proposal, which may represent a bias for our statistical study of sub-DLAs. We provide here a short description of the original observing aims and a description of our decision on whether to include or exclude the redshift path and/or the sub-DLAs along these sightlines.
\begin{itemize}
\item LBQS\,2359-01216B was observed with the aim of measuring the molecular and metal content of known DLAs. The redshift path was included in our statistical sample to search for unknown sub-DLAs.
\item QSO\,J0003-2323 was observed as part of the ESO Large Program for IGM studies. The redshift path and an additional sub-DLA reported by \citet{richter05} were excluded from the statistical sample. 
\item QSO\,B0002-422 was observed as part of the ESO Large Program for IGM studies. The redshift path was excluded from the statistical sample. 
\item QSO\,0006-6208 was targeted in a search for new sub-DLAs. Therefore, the redshift path and a sub-DLA previously reported by \citet{peroux05} were included in the statistical sample.
\item QSO\,J0008-0958 was observed with the aim of measuring the metal content of a known DLA. The redshift path was included in our statistical sample to search for unknown sub-DLAs.
\item QSO\,J0008-2900 was observed in a study of the tomography of the IGM. Therefore, the redshift path and an additional sub-DLA reported in paper\,I were excluded from the statistical sample.
\item QSO\,J0008-2901 was observed in a study of the tomography of the IGM. Therefore, the redshift path and an additional sub-DLA reported in paper\,I were excluded from the statistical sample.
\item QSO\,B0008+006 was observed in a study of its BAL features and therefore the redshift path was included in our statistical sample to search for unknown sub-DLAs.
\item LBQS\,0009-0138 was targeted in a study the metal content of sub-DLAs at 0.7 $<$ $z$ $<$ 1.5. The redshift path was included to search for sub-DLAs at higher redshift.
\item LBQS\,0010-0012 was observed with the aim of measuring the molecular content of a known DLA. The redshift path was included in our statistical sample to search for unknown sub-DLAs.
\item LBQS\,0013-0029 was observed with the aim of measuring the molecular content of a known DLA. The redshift path was included in our statistical sample to search for unknown sub-DLAs. The sub-DLA found by \citet{ledoux03} was not included in our redshift path because it is near the known DLA.
\item QSO\,B0027-1836 was observed in a search for molecules in a known DLA. The redshift path was included in our statistical sample to search for unknown sub-DLAs.
\item J004054.7-091526 was observed in a study of the intrinsic properties of the quasar. Therefore, the redshift path and an additional sub-DLA found by \cite{noterdaeme09} were included. 
\item QSO\,J0041-4936 was observed with the aim of measuring the deuterium content of a known DLA. The redshift path was included in our statistical sample to search for unknown sub-DLAs. 
\item QSO\,B0039-407 was targeted in a study the metal content of $\ion{Mg}{ii}$ absorbers at 0.6 $<$ $z$ $<$ 1.5. The redshift path was included to search for sub-DLAs at higher redshift. 
\item QSO\,B0039-3354 was observed with the aim of measuring the metal content of known DLAs. The redshift path was included in our statistical sample to search for unknown sub-DLAs.
\item LBQS\,0041-2638 was observed in a study of the tomography of the IGM. Therefore, the redshift path was excluded from the statistical sample.
\item LBQS\,0041-2707 was observed in a study of the tomography of the IGM. Therefore, the redshift path was excluded from the statistical sample.
\item QSO\,B0042-2656 was targeted in a study of the tomography of the IGM. Therefore, the redshift path was excluded from the statistical sample.
\item LBQS\,0042-2930 was observed with the aim of measuring the metal content of known DLAs. The redshift path was included in our statistical sample to search for unknown sub-DLAs.
\item LBQS\,B0042-2657 was targeted in a study of the tomography of the IGM. Therefore, the redshift path was excluded from the statistical sample.
\item J004612.2-293110 was targeted in a study of a dark clump at 0.1 $<$ $z$ $<$ 0.3 along the line of sight to the quasar. Therefore, the redshift path was included in our statistical sample to search for unknown sub-DLAs.
\item LBQS\,0047-2538 was observed in a study of the halo of a nearby galaxy along the quasar sightline. Hence, the redshift path was included in our statistical sample to search for unknown sub-DLAs.
\item LBQS\,0048-2545 was observed in a study of the halo of a nearby galaxy along  the quasar sightline. Therefore, the redshift path was included in our statistical sample to search for unknown sub-DLAs. 
\item QSO\,B0018-2608 was targeted in a study of the halo of a nearby galaxy along  the quasar sightline. Hence, the redshift path was included in our statistical sample to search for unknown sub-DLAs. 
\item QSO\,B0055-26 was observed in a study of the IGM. Therefore, the redshift path was excluded from the statistical sample.
\item QSO\,B0058-292  was observed with the aim of measuring the molecular and metal content of a known DLA. The redshift path was included in our statistical sample to search for unknown sub-DLAs.
\item LBQS\,0059-2735 was observed in a study of quasar outflows. Therefore, the redshift path was included in our statistical sample to search for unknown sub-DLAs.
\item QSO\,0100+1300 was targeted in a study of the variation of the fine-structure constant in a quasar with known metal lines. Hence, the redshift path was excluded from the statistical sample. 
\item QSO\,J0105-1846 was observed in a study of the metal content of a known DLA. Therefore the redshift path and an additional sub-DLA reported by \citet{fox09} were included in our statistical sample.
\item QSO\,B0102-2931 was observed in a study of the correlation function of the IGM. Therefore, the redshift path was excluded from the statistical sample.
\item QSO\,B0103-260 was targeted in a study of the IGM. Hence, the redshift path was excluded from the statistical sample.
\item QSO\,B0109-353 was observed as part of the ESO Large Program for IGM studies. The redshift path was excluded from the statistical sample. 
\item QSO\,B0112-30 was observed with the aim of measuring the molecular content of known DLAs. The redshift path was included in our statistical sample to search for unknown sub-DLAs.
\item QSO\,J0124+0044 was targeted with the aim of searching for new sub-DLAs. Therefore, the redshift path and sub-DLAs previously reported by \citet{peroux05} were included in the statistical sample.
\item QSO\,B0122-379 was observed as part of the ESO Large Program for IGM studies. The redshift path was excluded from the statistical sample. 
\item QSO\,B0122-005 was targeted in a study of the metal content of $\ion{Mg}{ii}$ absorbers at 0.6 $<$ $z$ $<$ 1.5. The redshift path and an additional sub-DLA reported by \citet{ellison09} were included in our statistical sample. 
\item QSO\,B0128-2150 was observed with the aim of measuring the deuterium content of $z_{\rm abs}=$ 1.85 and 1.64 $\ion{H}{i}$ systems. The redshift path was included in our statistical sample to search for unknown sub-DLAs. The $N_{\ion{H}{i}}$ of the two systems mentioned above are fitted for the first time in paper I . Only one of them is classified as a sub-DLA, the other system has a of lower $\ion{H}{i}$ column density.
\item QSO\,B0130-403 was observed in a study of a nearby galaxy along the quasar sightline. Hence, the redshift path was included in our statistical sample to search for unknown sub-DLAs.
\item QSO\,J0133+0400 was targeted in a search of new sub-DLAs and in a study of the metal content of known DLAs. Therefore, the redshift path and sub-DLAs previously reported by \citet{peroux05} were included in the statistical sample.
\item QSO\,B0135-42 was observed in a search of new sub-DLAs. Therefore, the redshift path and sub-DLAs previously reported by \citet{peroux05} were included in the statistical sample
\item QSO\,J0139-0824 was targeted in a study of the variation of the fine-structure constant in a quasar with known metal lines. Hence, the redshift path wss excluded from the statistical sample. 
\item QSO\,J0143-317 was observed in a study of the IGM. Hence, the redshift path was excluded from the statistical sample.
\item QSO\,J0153-4311 was observed as part of the ESO Large Program for IGM studies. The redshift path was excluded from the statistical sample. 
\item QSO\,B0201+113 was observed in a study of the metal content of a known DLA. Therefore, the redshift path was included in our statistical sample to search for unknown sub-DLAs.
\item QSO\,J0209+0517 was observed in a search of new sub-DLAs. Therefore, the redshift path and a sub-DLA previously reported by \citet{peroux05} were included in the statistical sample.
\item J021741.8-370100 was observed with the aim of measuring the metal content of known DLAs. The redshift path was included in our statistical sample to search for unknown sub-DLAs.
\item QSO\,B0216+0803 was observed with the aim of measuring the molecular and metal content of a known DLA between $z=$ 2--3. The redshift path and an additional sub-DLA at lower redshift reported by \citet{dessauges09} were included in our statistical sample.
\item QSO\,B0227-369 was targeted in a study of the metal content of $\ion{Mg}{ii}$ absorbers at 0.6 $<$ $z$ $<$ 1.5. The redshift path was included to search for sub-DLAs at higher redshift. 
\item QSO\,B0237-2322 was targeted as part of the ESO Large Program for IGM studies. The redshift path and sub-DLAs were excluded from the statistical sample.
\item QSO\,J0242+0049 was observed in a study of its BAL features and therefore the redshift path was included in our statistical sample to search for unknown sub-DLAs.
\item QSO\,J0243-0550 was observed in a study of the metal content of $\ion{Mg}{ii}$ absorbers at 0.6 $<$ $z$ $<$ 1.5. The redshift path was included to search for sub-DLAs at higher redshift. 
\item QSO\,B0241-01 was observed in a search of new sub-DLAs. The redshift path was included in the statistical sample.
\item QSO\,B0244-1249 was targeted in a study of the metal content of $\ion{Mg}{ii}$ absorbers at 0.6 $<$ $z$ $<$ 1.5. The redshift path and an additional sub-DLA reported by \citet{ellison09} were included in the statistical sample. 
\item QSO\,B0254-404 was observed with the aim of measuring the metal content of a known DLA. The redshift path was included in our statistical sample to search for unknown sub-DLAs.
\item J030640.8-301032 was observed in a study of the tomography of the IGM. Therefore, the redshift path was excluded from the statistical sample.
\item J030643.7-301107 was observed in a study of the tomography of the IGM. Hence, the redshift path was excluded from the statistical sample.
\item QSO\,B0307-195A was observed in a study of the tomography of the IGM. Hence, the redshift path was excluded from the statistical sample.
\item QSO\,B0307-195B was observed in a study of the tomography of the IGM. The redshift path and an additional sub-DLA reported by \citet{odorico02} were excluded from the statistical sample.
\item J031856.6-060038 was observed in a study of the quasar outflows. Therefore, the redshift path was included in our statistical sample to search for unknown sub-DLAs.
\item QSO\,B0329-385 was observed as part of the ESO Large Program for IGM studies. The redshift path was excluded from the statistical sample.
\item Q\,B0329-2534 was observed as part of the ESO Large Program for IGM studies. The redshift path was excluded from the statistical sample.
\item QSO\,J0332-4455 was targeted with the aim of measuring the metal content of known DLAs. The redshift path was included in our statistical sample. An additional sub-DLA reported by \citet{fox09} was excluded because it is 3000 km s$^{-1}$ blueward of the quasar emission line.
\item QSO\,B0335-122 was observed with the aim of measuring the metal content of known DLAs. Therefore, the redshift path was included in our statistical sample to search for unknown sub-DLAs.
\item QSO\,J0338+0021 was observed in a study of the intrinsic properties of the quasar. Therefore, the redshift path was included in our statistical sample to search for unknown sub-DLAs. 
\item QSO\,J0338-0005 was observed in a study of the variation of the fine-structure constant in a quasar with known metal lines. Hence, the redshift path was excluded from the statistical sample. 
\item QSO\,B0336-017 was observed in a study of the metal content of a known DLA. Therefore, the redshift path was included in our statistical sample to search for unknown sub-DLAs.
\item QSO\,B0347-383 was observed in a study of the metal content of a known DLA. Therefore, the redshift path was included in our statistical sample to search for unknown sub-DLAs.
\item J035320.2-231418 was targeted in a study of the metal content of $\ion{Mg}{ii}$ absorbers at 0.6 $<$ $z$ $<$ 1.5. The redshift path was included to search for sub-DLAs at higher redshift. 
\item QSO\,J0354-2724 was observed in a study of the metal content of sub-DLAs at 0.7 $<$ $z$ $<$ 1.5. The redshift path was included to search for sub-DLAs at higher redshift.
 \item QSO\,J0403-1703 was observed in a study of narrow absorption lines associated with the quasar. Therefore, the redshift path was included in our statistical sample to search for unknown sub-DLAs.
\item QSO\,J0407-4410 was observed with the aim of measuring the molecular and metal content of known DLAs. The redshift path was included in our statistical sample to search for unknown sub-DLAs.
\item QSO\,J0422-3844 was targeted as part of the ESO Large Program for IGM studies. The redshift path and a sub-DLA were excluded from our statistical sample.
\item QSO\,J0427-1302 was targeted in a study of the metal content of sub-DLAs at 0.7 $<$ $z$ $<$ 1.5. The redshift path was included to search for sub-DLAs at higher redshift.
\item QSO\,J0430-4855 was targeted in a study of the IGM. The redshift path was excluded from the statistical sample.
\item QSO\,B0432-440 was observed with the aim of measure the metal content of a known DLA. Therefore, the redshift path was included in our statistical sample to search for unknown sub-DLAs.
\item QSO\,B0438-43 was observed with the aim of measuring the metal content of a known DLA. Therefore, the redshift path was included in our statistical sample to search for unknown sub-DLAs.
\item QSO\,B0449-1645 was targeted in a study of the metal content of a DLA at $z$ $\leq$ 1.0. The redshift path was included to search for sub-DLAs at higher redshift. 
\item QSO\,B0450-1310B was observed in a study of the metal content of DLAs at $z_{\rm abs}$ $\gtrsim$ 2. The redshift path was included in the statistical sample to search for unknown sub-DLAs.
\item QSO\,J0455-4216 was observed as part of the ESO Large Program for IGM studies. The redshift path was excluded from the statistical sample.
\item 4C-02.19 was observed with the aim of measuring the molecular and metal content of a known DLA.  Later the quasar was re-observed in a study of the variation of the fine-structure constant in known metal lines. To be conservative, the redshift path was excluded from our statistical sample.
\item QSO\,B0515-4414 was targeted in a study of the IGM. Hence, the redshift path and a lower redshift sub-DLA were excluded from the statistical sample.
\item QQSO\,B0528-2505 was observed with the aim of measuring the molecular and metal content of known DLAs and to constrain the variation of ratio of the electron to proton masses. The redshift path was included in our statistical sample to search for unknown sub-DLAs.
\item QSO\,J0530+13 was targeted in a study of the $\ion{Mg}{ii}$ absorber at $z$ = 0.79. The redshift path was included to search for sub-DLAs at higher redshift. 
\item QSO\,B0551-36 was observed with the aim of measuring the molecular content of a known DLA. The redshift path was included in our statistical sample to search for unknown sub-DLAs.
\item J060008.1-504036 was observed in a search of deuterium in a known LLS. The redshift path was included in our statistical sample to search for unknown sub-DLAs.
\item QSO\,B0606-2219 was targeted in a study of the metal content of $\ion{Mg}{ii}$ absorbers at 0.6 $<$ $z$ $<$ 1.5. The redshift path wss included to search for sub-DLAs at higher redshift. 
\item QSO\,B0642-5038 was observed in a search of molecules in a known DLA. The redshift path was included in our statistical sample to search for unknown sub-DLAs.
\item QSO\,B0841+129 was observed in a study of the metal content of DLAs at $z_{\rm abs}$ $\gtrsim$ 2. The redshift path was included in the statistical sample. An additional sub-DLA reported by \citet{fox11} was excluded because it is within 3000 km s$^{-1}$ of the quasar emission line.
\item QSO\,B0908+0603 was observed in a study of the tomography of the IGM. The redshift path wss excluded from our statistical sample. 
\item QSO\,B0913+0715 was targeted with the aim of measuring the metal content and to search for deuterium in a known DLA. Therefore, the redshift path was included in our statistical sample to search for unknown sub-DLAs.
\item QSO\,B0919-260 was targeted in a study of the metal content of $\ion{Mg}{ii}$ absorbers at 0.6 $<$ $z$ $<$ 1.5. The redshift path was included to search for sub-DLAs at higher redshift. 
\item QSO\,B0933-333 was observed with the aim of measuring the metal content of a known DLA. The redshift path was included in our statistical sample to search for unknown sub-DLAs.
\item QSO\,B0951-0405 was targeted with the aim of measuring the metal content of known DLAs. The redshift path and an additional sub-DLA reported by \citet{guimares09} were included in the statistical sample. 
\item QSO\,B0952-0115 was first observed in a study of the tomography of the IGM and later to measure the metal content of a known DLA. To be conservative, the redshift path and an additional sub-DLA reported in paper\,I were excluded from our statistical sample. 
\item QSO\,B1005-333 was targeted in a study of the metal content of $\ion{Mg}{ii}$ absorbers at 0.6 $<$ $z$ $<$ 1.5. The redshift path was included to search for sub-DLAs at higher redshift. 
\item QSO\,B1027-0540 was observed in a study of the epoch of reionization. The redshift path was excluded because the quasar is $z_{\rm em}$ $>$ 5.0. 
\item Q\,1036-272 was observed to constrain the variation of ratio of the electron to proton masses. The redshift path was included in the statistical sample to search for unknown sub-DLAs.
\item QSO\,B1036-2257 was targeted with the aim of measuring the metal content of a known DLA. Therefore, the redshift path and an additional sub-DLA reported in paper\,I were included in our statistical sample.
\item QSO\,J1039-2719 was observed to constrain the variation of ratio of the electron to proton masses in a known sub-DLA and in a study of narrow absorption lines associated with the quasar. The redshift path was included in the statistical sample to search for additional sub-DLAs.
 \item QSO\,B1038-2712 was observed in Visitor Mode as part of a program to search for molecules in known DLAs but the purpose of observing this particular target was to confirm the BAL nature of the quasar. Hence, the redshift path was included in our statistical sample.
 \item QSO\,B1036-268 was targeted in a study of the BAL outflow of the quasar. Hence, the redshift path and an additional sub-DLA reported in paper\,I were included in our statistical sample.
\item J104642.9+053107 was observed in a study of the tomography of the IGM. Therefore, the redshift path was excluded from the statistical sample. 
\item QSO\,B1055-301 was observed with the aim of measuring the metal content of a known DLA. Therefore, the redshift path was included in our statistical sample to search for unknown sub-DLAs.
\item QSO\,B1101-26 was targeted in a study of the variation of the fine-structure constant in a quasar with known metal lines. Hence, the redshift path was excluded from the statistical sample. 
\item QSO\,B1104-181 was observed in a study of the tomography of the IGM. The redshift path was excluded from our statistical sample.
\item QSO\,B1108-07 was targeted with the aim of measuring the metal content and search for deuterium in a known DLA. Hence, the redshift path and an additional sub-DLA reported by \citet{ledoux06} were included in our statistical sample. 
\item QSO\,B1113-1533 was observed in a search of deuterium in a known DLA. The redshift path was included in our statistical sample to search for unknown sub-DLAs. 
\item QSO\,B1114-0822 was observed with the aim of measuring the metal content of a known DLA. The redshift path was included in our statistical sample to search for unknown sub-DLAs. 
\item QSO\,B1122-168 was targeted in a study of the high ionization metal lines in the IGM and broad-line regions. The redshift path was excluded from our statistical sample. 
\item QSO\,J1142+2654 was observed in a study of the $\ion{He}{ii}$ reionization. The redshift path was excluded from our statistical sample. 
\item QSO\,B1151+068 was observed in a study of the metal content in a known DLA. The redshift path was included in our statistical sample to search for unknown sub-DLAs.
\item J115538.6+053050 was observed in a study of the metal content of known DLAs. Therefore, the redshift path was included in our statistical sample to search for unknown sub-DLAs.
 \item QSO\,J1159+1337 was observed in a study of narrow absorption lines associated with the quasar. Therefore, the redshift path and an additional sub-DLA reported by \citet{guimares09} were included in our statistical sample to search for unknown sub-DLAs.
\item QSO\,B1158-1842 was targeted as part of the ESO Large Program for IGM studies. The redshift path was excluded from the statistical sample.
\item QSO\,B1202-074 was observed under the quasar environments, ESO Large Program for IGM studies and DLA abundances programmes. Therefore, the redshift path was not included in our statistical sample. 
\item J120550.2+020131 was observed to confirm the nature of a DLA candidate. The redshift path was included in our statistical sample to search for unknown sub-DLAs.
\item QSO\,B1209+0919 was targeted with the aim of measuring the metal content of a known DLA. The redshift path was included in our statistical sample to search for unknown sub-DLAs.
\item LBQS\,1209+1046 was targeted with the aim of measuring the metal content of a known DLA. The redshift path was included in our statistical sample to search for unknown sub-DLAs.
\item LBQS\,1210+1731 was observed in a study of the metal content of a DLA at $z_{\rm abs}$ $\approx$ 2. The redshift path was included in the statistical sample to search for unknown sub-DLAs.
\item QSO\,B1220-1800 was observed to confirm the nature of a sub-DLA candidate. The redshift path was included in our statistical sample to search for unknown sub-DLAs.
\item LBQS\,1223+1753 was observed in a study of the quasar broad-line region and metal content in a known DLA. The redshift path and an additional sub-DLA reported by \citet{dessauges03} were included in our statistical sample.
\item QSO\,B1228-113 was observed with the aim of measuring the metal content of known DLAs. The redshift path was included in our statistical sample to search for unknown sub-DLAs.
\item QSO\,B1230-101 was targeted in a study of the metal content of $\ion{Mg}{ii}$ absorbers at 0.6 $<$ $z$ $<$ 1.5. The redshift path was included to search for sub-DLAs at higher redshift. 
\item LBQS\,1232+0815 was observed in a study of the molecular content  of a known DLA, quasar outflows and to constrain the variation of ratio of the electron to proton masses. The redshift path and an additional sub-DLA reported in paper\,I  were included in the statistical sample.
\item LBQS\,1242+0006 was observed to confirm the nature of a DLA candidate. The redshift path was included in our statistical sample to search for unknown sub-DLAs.
\item LBQS\,1246-0217 was observed in a study of the metal content of a known DLA. The redshift path was included in our statistical sample to search for unknown sub-DLAs.
\item J124957.2-015929 was observed in a study of the $\ion{He}{ii}$ reionization. The redshift path was excluded from our statistical sample. 
\item QSO\,B1256-177 was targeted in a study of the metal content of $\ion{Mg}{ii}$ absorbers at 0.6 $<$ $z$ $<$ 1.5. The redshift path was included to search for sub-DLAs at higher redshift. 
\item QSO\,B1306+0356 was observed in a study of the epoch of reionization. The redshift path was excluded because the quasar is $z_{\rm em}$ $>$ 5.0. 
\item QSO\,B1317-0507 was observed in a study of the $\ion{He}{ii}$ reionization. The redshift path was excluded from our statistical sample. 
\item QSO\,B1318-263 was targeted in a study of the metal content of $\ion{Mg}{ii}$ absorbers at 0.6 $<$ $z$ $<$ 1.5. The redshift path was included to search for sub-DLAs at higher redshift. 
\item QSO\,B1324-047 was targeted in a study of the metal content of $\ion{Mg}{ii}$ absorbers at 0.6 $<$ $z$ $<$ 1.5. The redshift path was included to search for sub-DLAs at higher redshift. 
\item QSO\,J1330-2522 was observed in a study of the $\ion{He}{ii}$ reionization. The redshift path and three sub-DLAs were excluded from our statistical sample. 
\item QSO\,B1331+170 was targeted in a study of the metal content in a known DLA. Hence, the redshift path was included in our statistical sample to search for known sub-DLAs. 
\item QSO\,J1344-1035 was targeted as part of the ESO Large Program for IGM studies. The redshift path was excluded from our statistical sample.
\item QSO\,J1347-2457 was observed as part of the ESO Large Program for IGM studies. The redshift path was excluded from the statistical sample.
\item QSO\,J1356-1101 was observed with the aim of measuring the metal content of known DLAs. Therefore, the redshift path and an additional sub-DLA reported in paper\,I were included in our statistical sample.
\item QSO\,B1402-012 was targeted in a study of the metal content of $\ion{Mg}{ii}$ absorbers at 0.6 $<$ $z$ $<$ 1.5. The redshift path was included to search for sub-DLAs at higher redshift. 
\item QSO\,B1409+0930 was observed in a study of the metal content of known DLAs. The redshift path and an additional sub-DLA reported by \citet{fox09} were included in our statistical sample.
\item QSO\,B1412-096 was targeted in a study of the metal content of $\ion{Mg}{ii}$ absorbers at 0.6 $<$ $z$ $<$ 1.5. The redshift path was included to search for sub-DLAs at higher redshift. 
\item QSO\,J1421-0643 was observed with the aim of measuring the metal content of a known DLA. Therefore, the redshift path was included in our statistical sample to search for unknown sub-DLAs.
\item QSO\,B1429-008B was targeted in a study of the tomography of the IGM. The redshift was excluded from our statistical sample.
\item QSO\,J1439+1117 was observed with the aim of measuring the molecular content of a known sub-DLA. The redshift path was included in our statistical sample to search for unknown sub-DLAs.
\item QSO\,J1443+2724 was targeted in a search of molecules and measure the CMB temperature in a known DLA. The redshift path was included in our statistical sample to search for unknown sub-DLAs.
\item LBQS\,1444+0126 was observed in a study of the quasar broad-line regions, as well as the metal and molecular content of a known DLA. The redshift path was included in our statistical sample to search for additional unknown sub-DLAs.
\item QSO\,B1448-232 was targeted for the IGM ionization and ESO Large Program for IGM studies. The redshift path was excluded from our statistical sample.
\item J145147.1-151220 was observed as part of the ESO Large Program for IGM studies. The redshift path was not included in our statistical sample.
\item J151352.52+085555.7 was observed in a study of the quasar outflows. Therefore, the redshift path was included in our statistical sample to search for unknown sub-DLAs.
\item QSO\,J1621-0042 was observed in a study of the $\ion{He}{ii}$ reionization. The redshift path was excluded from our statistical sample. 
\item 4C\,12.59 was observed in a study of the metal content of known low-redshift DLA and sub-DLA. The redshift path was included in our statistical sample to search for sub-DLAs at higher redshift.
\item QSO\,J1723+2243 was observed in a search of molecules in a known DLA. The redshift path and an additional sub-DLA reported in paper\,I were included in our statistical sample.
\item QSO\,B1937-1009 was observed in a study of the $\ion{He}{ii}$ reionization. The redshift path was excluded from our statistical sample. 
\item QSO\,B1935-692 was observed in a study of the $\ion{He}{ii}$ reionization. The redshift path was excluded from our statistical sample. 
\item QSO\,B2000-330 was observed as part of the ESO Large Program for IGM studies and in a study of Jeans scale of the IGM. The redshift path and sub-DLAs reported by \citet{omeara07} were not included in the statistical sample.
\item LBQS\,2113-4345 was targeted in a study of the tomography of the IGM. The redshift was excluded from our statistical sample.
\item LBQS\,2114-4347 was targeted in a study of the tomography of the IGM. The redshift and a sub-DLA reported in paper\,I were excluded from our statistical sample.
\item J211739.5-433538 was targeted in a study of the tomography of the IGM. The redshift was excluded from our statistical sample.
\item QSO\,J2119-3536 was targeted in a study of the metal content of a known sub-DLA. Therefore, the redshift path was included in our statistical sample to search for unknown sub-DLAs.
\item QSO\,B2126-15 was observed as part of the ESO Large Program for IGM studies. The redshift path and two sub-DLAs reported by \citet{omeara07} were not included in our statistical sample.
\item QSO\,B2129-4653 was targeted in a study of the variation of the fine-structure constant in a quasar with known metal lines. Hence, the redshift path was excluded from the statistical sample. 
 \item J213314.2-464031 was observed in Visitor Mode as part of a program to search for molecules in known DLAs but the purpose of observing this particular target was to study the tomography of the IGM \citep[see][]{coppolani06}. Hence, the redshift path was excluded from our statistical sample.
\item LBQS\,2132-4321 was targeted in a study of the metal content of a known sub-DLA. The redshift path was included in our statistical sample to search for unknown sub-DLAs.
\item LBQS\,2138-4427 was observed with the aim of measuring the metal content of known DLAs. The redshift path was included in our statistical sample to search for unknown sub-DLAs.
\item QSO\,B2139-4433 was observed in a study of the tomography of the IGM. The redshift path was excluded from our statistical sample.
\item QSO\,B2149-306 was targeted in a study of the metal content of $\ion{Mg}{ii}$ absorbers at 0.6 $<$ $z$ $<$ 1.5. The redshift path was included to search for sub-DLAs at higher redshift. 
\item QSO\,B2204-408 was observed in a study of absorption lines associated with the quasar. Therefore, the redshift path was included in our statistical sample to search for unknown sub-DLAs.
\item LBQS\,2206-1958A was targeted in a study of the metal and molecular content of known DLAs. Therefore, the redshift path was included in our statistical sample to search for unknown sub-DLAs.
\item QSO\,J2220-2803 was observed as part of the ESO Large Program for IGM studies and in a study of broad-line regions of the quasar. The redshift path was not included in our statistical sample.
\item QSO\,B2222-396 was observed with the aim of measuring the metal content of a known DLA. The redshift path was included in our statistical sample to search for unknown sub-DLAs.
\item QSO\,J2227-2243 was observed in a study of the IGM. The redshift path was excluded from the statistical sample.
\item QSO\,B2225-4025 was observed with the aim of measuring the metal content of a known DLA. The redshift path was included in our statistical sample to search for unknown sub-DLAs.
\item LBQS\,2230+0232 was observed in a study of the metal content of a DLA at $z_{\rm abs}$ $\gtrsim$ 2. The redshift path was included in the statistical sample to search for known sub-DLAs.
\item J223851.0-295301 was targeted in a study of the tomography of the IGM. The redshift was excluded from our statistical sample.
\item J223922.9-294947 was targeted in a study of the tomography of the IGM. The redshift was excluded from our statistical sample.
\item J223928.9-295451 was targeted in a study of the tomography of the IGM. The redshift was excluded from our statistical sample.
\item J223941.8-294955 was targeted in a study of the tomography of the IGM. The redshift and a sub-DLA reported in paper\,I were excluded from our statistical sample.
\item J223948.7-294748 was targeted in a study of the tomography of the IGM. The redshift was excluded from our statistical sample.
\item J223951.9-294837 was targeted in a study of the tomography of the IGM. The redshift was excluded from our statistical sample.
\item QSO\,B2237-0607 was targeted in a search of new sub-DLAs. The redshift path was included in the statistical sample.
\item QSO\,J2247-1237 was targeted in a study of the metal content of $\ion{Mg}{ii}$ absorbers at 0.6 $<$ $z$ $<$ 1.5. The redshift path was included to search for sub-DLAs at higher redshift. 
\item QSO\,B2311-373 was observed with the aim of measuring the metal content of a known DLA. Therefore, the redshift path was included in our statistical sample to search for unknown sub-DLAs.
\item J232046.7-294406 was targeted in a study of the tomography of the IGM. The redshift was excluded from our statistical sample.
\item J232059.4-295520 was targeted in a study of the tomography of the IGM. The redshift was excluded from our statistical sample.
\item J232114.3-294725 was observed in a study of the tomography of the IGM. Therefore, the redshift path was excluded from the statistical sample.
\item J232121.2-294350 was targeted in a study of the tomography of the IGM. The redshift was excluded from our statistical sample.
\item QSO\,B2318-1107 was observed in a search of molecules and measure the metal content of known DLAs. The redshift path was included in our statistical sample to search for unknown sub-DLAs.
\item QSO\,B2332-094 was targeted with the aim of measuring the metal and molecular content of a known DLA. The redshift path and an additional sub-DLA reported by \citet{ledoux03} were included in our statistical sample.
\item QSO\,B2342+3417 was observed with the aim of measuring the metal content of a known DLA. The redshift path and an additional sub-DLA reported in paper\,I were included in our statistical sample.
\item QSO\,J2346+1247 was observed in a study of the absorbers associated with known galaxies along the quasar line of sight. The redshift path was excluded from our statistical sample.
\item QSO\,B2343+125 was targeted in a study of the metal content of a known DLA. Hence, the redshift path was included in our statistical sample. 
\item QSO\,B2345+000 was observed in a study of the absorbers associated with known galaxies along the quasar sightline. The redshift path was excluded from our statistical sample.
\item QSO\,B2347-4342 was observed as part of the ESO Large Program and other IGM related programmes. The redshift path was not included in the statistical sample.
\item QSO\,B2348-147 was observed in a study of the metal content of a known DLA. The redshift path was included in our statistical sample to search for unknown sub-DLAs.
\item J235702.5-004824 was observed in a study of its BAL features and therefore the redshift path was included in our statistical sample to search for unknown sub-DLAs.
\end{itemize}

\end{appendix}

\end{document}